\documentclass[iop,apj]{emulateapj}
\usepackage{amsmath,natbib,graphics,graphicx,apjfonts}

\input{buote_new.defs}
\newcommand\mlkband{{M_{\star}/L_K}}
\newcommand\src{\hbox{{NGC~6482}}}

\begin{document}

\title{The Unusually High Halo Concentration of the Fossil Group \src:\\
Evidence for Weak Adiabatic Contraction}
\author{David A.\ Buote}
\affil{Department of Physics and Astronomy, University of California
  at Irvine, 4129 Frederick Reines Hall, Irvine, CA 92697-4575; \\ buote@uci.edu}

\slugcomment{Accepted for Publication in The Astrophysical Journal}

\begin{abstract}
  Although fossil galaxy systems are thought to be very old, and thus
  should possess above-average halo concentrations, typically fossils
  have unexceptional concentrations for their masses. We revisit the
  massive isolated elliptical galaxy / fossil group \src\ for which
  previous X-ray studies of a modest \chandra\ observation obtained a
  very uncertain, but also possibly very high, halo concentration.  We
  present new measurements of the hot gas surface brightness,
  temperature, and iron abundance using the modest \chandra\
  observation and a previously unpublished \suzaku\ observation, the
  latter of which allows the measurements of the gas properties to be
  extended out to $\sim \rtwofiveh$.  By constructing hydrostatic
  equilibrium models of the gas with separate components for the gas,
  BCG stellar mass, and the dark matter (DM), we measure
  $c_{200}=32.2\pm 7.1$ and
  $M_{200}=(4.5\pm 0.6)\times 10^{12}\, M_{\odot}$ using an NFW DM
  profile. For a halo of this mass, the measured concentration
  $c_{200}$ exceeds the mean value (7.1) expected for relaxed \lcdm\
  halos by $3.5\sigma$ in terms of the observational error, and by
  $6\sigma$ considering the intrinsic scatter in the \lcdm\ $c-M$
  relation, which situates \src\ as the most extreme outlier known for
  a fossil system.  We explored several variants of adiabatic
  contraction (AC) models and, while the AC models provide fits of the
  same quality as the un-contracted models, they do have the following
  advantages: (1) smaller $c_{200}$ that is less of an outlier in the
  \lcdm\ $c-M$ relation, and (2) baryon fractions $(f_{\rm b,200})$
  that agree better with the mean cosmic value.  While the standard AC
  prescriptions yield a BCG stellar mass that is uncomfortably small
  compared to results from stellar population synthesis (SPS) models,
  a weaker AC variant that artificially shuts off cooling and star
  formation at $z=2$ yields the same stellar mass as the un-contracted
  models. For these reasons, we believe our X-ray analysis prefers
  this weaker AC variant applied to either an NFW or Einasto DM
  halo. Finally, the stellar mass we measure for the BCG from the
  hydrostatic analysis strongly favors results from SPS models with a
  Chabrier or Kroupa IMF over a Salpeter IMF.
\end{abstract}

\section{Introduction}
\label{intro}

The radial profiles of dark matter (DM) halos formed within the \lcdm\
cosmology are approximately self-similar, where the deviations from
self-similarity are quantified by the relation between halo
concentration $(c)$ and mass ($M$). These deviations are small and
represent a variation of $c$ of a factor of a few over about five
decades in $M$~\citep[e.g.][and references therein]{dutt14a}. X-ray
observations of the hot plasma in massive elliptical galaxies, galaxy
groups, and clusters~\citep[e.g.,][]{buot07a,schm07a,etto10a} obtain
power-law $c-M$ relations with normalizations and slopes consistent
with \lcdm\ after accounting for gas processes in the cosmological
simulations and observational effects like selection
biases~\citep{rasi13a}. The mean and slope of the $c-M$ relation
depends on cosmological parameters~\citep[$\sigma_8,\Omega_{\rm m},w$;
e.g.,][]{eke01a,alam02a,dola05a,kuhl05a,buot07a,macc08a}, while the
scatter about the relation is the same for different variants of
\lcdm\ and is independent of halo mass~\citep[e.g.][and references
therein]{bull01a,macc08a,dutt14a}. Given the robust nature of the
scatter, individual outliers in the $c-M$ relation can provide
valuable tests of the general \lcdm\ galaxy formation paradigm.

Halos classified as ``fossil groups'' are promising locations to
search for such outliers. Fossil groups were originally conceived to
be the end states of group formation~\citep{ponm94a} where all the
large galaxies merged via dynamical friction into a single large
central galaxy (BCG).  The entire merger history of the group was then
contained within the BCG itself, and so the ``group'' could then be
identified with the BCG. The longer cooling time of the hot gas means
the BCG is surrounded by a luminous X-ray halo characteristic of a
group-scale DM halo.  As such, these ``classical'' fossil groups were
presumed to be ancient, highly relaxed, and evolved systems. 

A considerable amount of work has been done over the past 10+ years
characterizing the properties of fossil groups. The conception of a
fossil now is broader than when initially discovered by
\citet{ponm94a} and includes more massive clusters; i.e., a fossil
system is now typically defined as (at least) a 2 magnitude gap
($R$-band) between the BCG and next brightest member and possessing a
bolometric
$L_{\rm x}\ga 5\times 10^{41}$~erg~s$^{-1}$~\citep[e.g.,][]{harr12a}.
While cosmological simulations predict that the magnitude-gap
criterion should reflect older-than-average
systems~\citep[e.g.,][]{dong05a,dari07a,raou16a}, the optical and
X-ray properties of fossil systems are generally consistent with those
of the general cluster
population~\citep[e.g.,][]{gira14a,kund15a}. Moreover, while it might
be expected that such older systems are more dynamically relaxed,
there is no evidence that the dynamical state of the hot gas in
fossils is different from other clusters~\citep{gira14a}.

If fossil systems truly formed earlier than the general population,
they should possess systematically higher halo concentrations. For the
NFW profile~\citep{nfw}, which is the standard function employed to
parameterize \lcdm\ DM halos, the concentration is defined as,
$c\equiv r_{\Delta}/r_s$, where $r_s$ is the scale radius and
$r_{\Delta}$ is the radius within which the mean density is
$\Delta\rho_c$, with $\rho_c$ the critical density of the universe at
the redshift of the halo and typically $\Delta=200$. \citet{nfw}
argued that $c$ is set by the mean density of the universe at the
redshift of halo formation, so that halos with higher concentrations formed
earlier than halos with lower concentrations. Prescriptions have been developed to
associate a formation time for a given $c$ and $M$~\citep[e.g.,][]{zhao09a}.

Using a modest \chandra\ ACIS-S observation for the nearby fossil
group NGC 6482, \citet{khos04a} presented intriguing evidence for an
extremely high concentration value $(c_{200}\sim 60, \, M_{200}\sim 4\times
10^{12}\, M_{\odot})$ compared to the mean \lcdm\ $c-M$
relation~\citep[$c\sim 7$, e.g.,][]{dutt14a}, from which they inferred
a very early formation time ($z\ga 5$). \citet{mamo05a} suggested that
since \citet{khos04a} did not include a separate component for the
stellar mass of the BCG in their hydrostatic equilibrium analysis, the
concentration they measured was likely overestimated. Indeed, when
including the stellar mass in our analysis of the \chandra\ data, we
obtained a much smaller, though very uncertain, value for $c_{200}$ that was
dependent on assumptions about the baryon fraction
profile~\citep[][hereafter H06]{hump06b}.

A proper interpretation of the measured halo concentration must also
account for the halo response to baryons during galaxy formation.  The
key process leading to the steepening of the inner DM halo mass
profile is ``Adiabatic Contraction''~\citep[AC,][]{blum86a}, where the
baryons dissipate rapidly onto the central galaxy, but the DM evolves
more slowly. In particular, a discrepancy between the halo
concentration as given by the mean $c-M$ relation deduced from
cosmological simulations that include only DM~\citep[e.g.,][]{dutt14a}
with the $c$ measured from observations using the standard NFW profile,
in principle can be mitigated if instead the observations are
interpreted with an adiabatically contracted NFW profile.

There are only a few applications of AC models to hydrostatic
equilibrium studies of the hot X-ray--emitting gas in massive
elliptical galaxies, groups, and clusters. H06 and \citet{gast07b}
fitted the modified AC model of \citet{gned04a} to a total of 22
massive elliptical galaxies and group-scale systems. Both of these
studies found the AC model to be modestly disfavored compared to the
pure NFW profile, since the AC model tended to yield BCG stellar
mass-to-light ratios uncomfortably small compared to stellar
population synthesis estimates. AC model fits to the hot gas of the
group/cluster RXCJ2315.7-0222 by ~\citet{demo10a} are compatible with
the data but are not clearly favored over the pure NFW profile because
of degeneracies in the assumed stellar initial mass function (IMF). AC
models are clearly excluded by the X-ray data of the galaxy cluster
A2589~\citep{zapp06a}. A similar lack of evidence supporting AC occurs
from lensing and stellar dynamical studies. While initially several
studies found AC models to be
favored~\citep[e.g.,][]{jian07a,gril12a,sonn12a}, the emerging
consensus~\citep{dutt14b,newm15a} is that for massive elliptical
galaxies and clusters the data favor the pure NFW profile or AC that
is substantially weaker than the standard
prescriptions of~\citet{blum86a} and~\citet{gned04a}.

For our present investigation we revisit the halo concentration of
\src\ for which previous X-ray studies suggested a possibly very high,
but uncertain, value. In 2010 \suzaku\ observed \src\ providing the
opportunity to map the hot gas to larger radius than previously and
thus allow for an improved measurement of the global halo mass profile. In
addition, improvements to the accuracy of the plasma code and our
method of hydrostatic analysis since our study in H06 warrant a
re-analysis of the \chandra\ data. Finally, we take this opportunity
to investigate several variants of AC models including those with
contraction weaker than the standard approaches.

We list several properties of \src\ in Table~\ref{tab.prop}. \src\ is
well-studied at multiple wavelengths and all indications are that it
is an extremely relaxed object; e.g., despite the fact that the
nucleus is a LINER, there is no evidence in the X-ray image or spectra
for AGN emission or cavities~\citep[e.g.,][]{gonz09a,pana14a,shin16a}.
It is also the most relaxed object in the morphological study of a
sample of local X-ray galaxies~\citep{dieh08a}. The combination of
highly relaxed, evolved state and X-ray brightness make \src\ an
optimal low-mass group for hydrostatic equilibrium studies of its hot
gas and mass profile.

The paper is organized as follows. We describe the data reduction in
\S \ref{obs} and spectral analysis in \S \ref{spec}. In \S
\ref{method} and \S \ref{models} we summarize the entropy-based
hydrostatic equilibrium method and the specific parameterized models.
The results and the systematic error budget are discussed in,
respectively, \S \ref{results} and \S \ref{sys}. We discuss many
implications of our results in \S \ref{disc} and provide conclusions
in \S \ref{conc}.  Throughout the paper we assume a flat $\Lambda$CDM
cosmology with $\Omega_{m,0}=0.3$ and
$H_0 = 70$~km~s$^{-1}$~Mpc$^{-1}$.

\begin{table*}[t] \footnotesize
\begin{center}
\caption{Target Properties}
\label{tab.prop}
\begin{tabular}{lcccccc}   \hline\hline\\[-7pt]
& & Distance & $N_{\rm H}$ & $L_{\rm K}$ & $R_e$ & $L_{\rm x, 500}$\\
Name & Redshift & (Mpc) & ($10^{20}$~cm$^{-2}$) & $(10^{11}\, L_{\odot})$  & (kpc)& ($10^{41}$~ergs~s$^{-1}$) \\
\hline \\[-7pt]
\src\ & 0.013129 & 59.2 & 7.7 & 3.3 & $3.65\pm 0.85$ & $5.0 \pm 0.3$\\
\hline \\
\end{tabular}
\tablecomments{The redshift is taken from
  NED\footnote{http://ned.ipac.caltech.edu}. For the distance we use
  the radial velocity corrected for Local Group infall onto Virgo from
  LEDA\footnote{http://leda.univ-lyon1.fr/}. (The angular scale is
  $1\arcsec=0.279$~kpc.)  We calculate the Galactic column density 
  using the HEASARC {\sc w3nh} tool based on the data of~\citet{kalb05a}.  The $K$-band luminosity is from 2MASS while the
  effective radius $(R_e$) represents an average of the 2MASS value
  and that obtained after applying a correction (see \S
  \ref{stars}). $L_{\rm x, 500}$ is the 0.1-10.0~keV luminosity
  computed for the fiducial hydrostatic model within a
  three-dimensional volume with radius $r_{500}$ (\S \ref{results}).}
\end{center}
\end{table*}

\section{Observations and Data Preparation}
\label{obs}

\begin{table}[t] \footnotesize
\begin{center}
\caption{Observations}
\label{tab.obs}
\begin{tabular}{ccccc}   \hline\hline\\[-7pt]
& & & & Exposure\\
Telescope & Obs.\ ID & Obs.\ Date & Instrument & (ks)\\
\hline \\[-7pt]
\chandra\ & 3218 & 2002, May 20 & ACIS-S & 17.5\\
\suzaku\ & 804050010 & 2010, Feb.\ 11 & XIS & 43.4\\
\hline \\
\end{tabular}
\tablecomments{The exposure times refer to those obtained after
  filtering the light curves (\S \ref{obs}). The exposures for
  the \chandra\ and \suzaku\ observations before filtering were 19.6~ks
  and 43.7~ks respectively.}
\end{center}
\end{table}

\begin{figure*}
\parbox{0.49\textwidth}{
\centerline{\includegraphics[scale=0.35,angle=0]{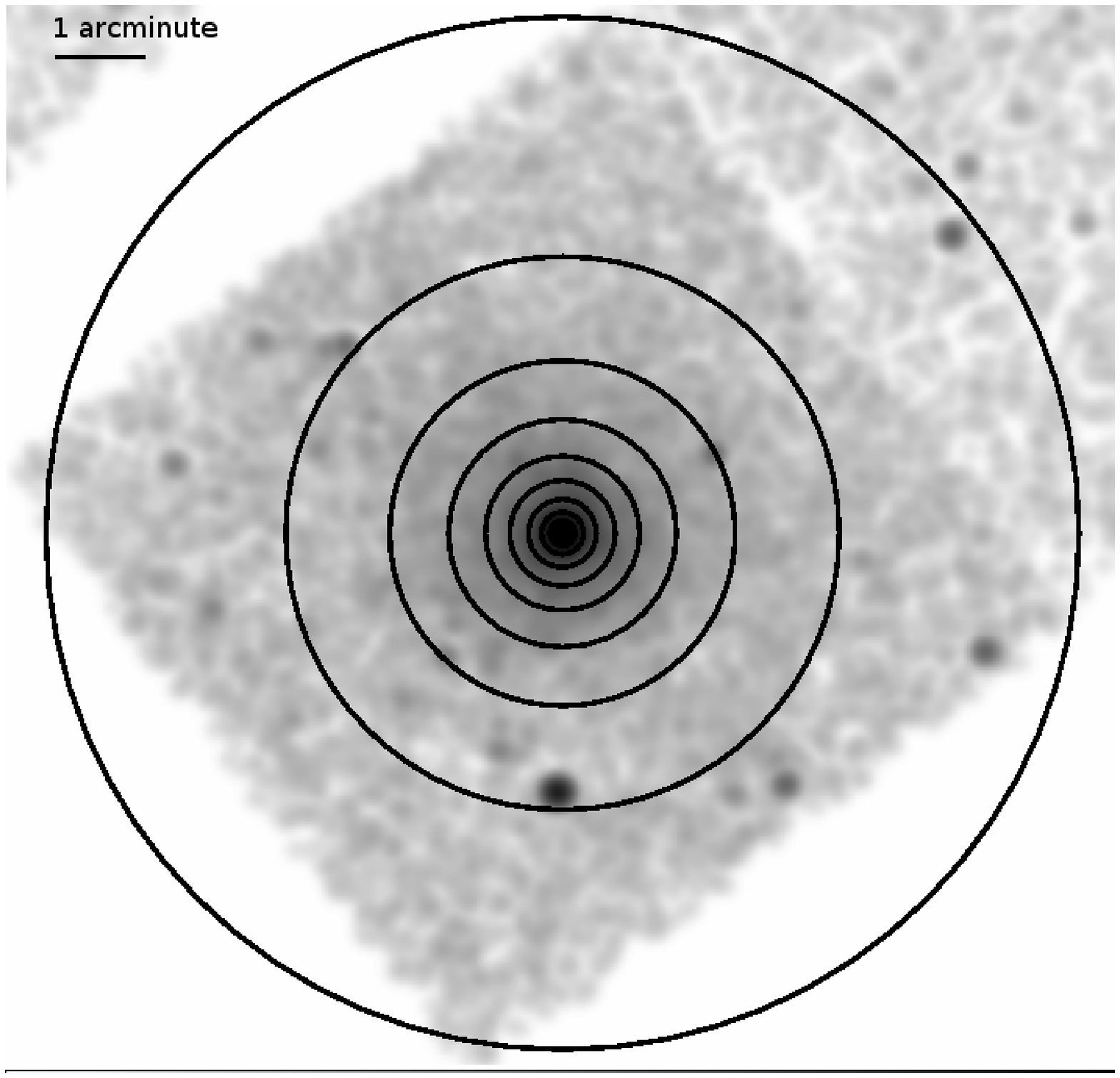}}}
\parbox{0.49\textwidth}{
\centerline{\includegraphics[scale=0.35,angle=0]{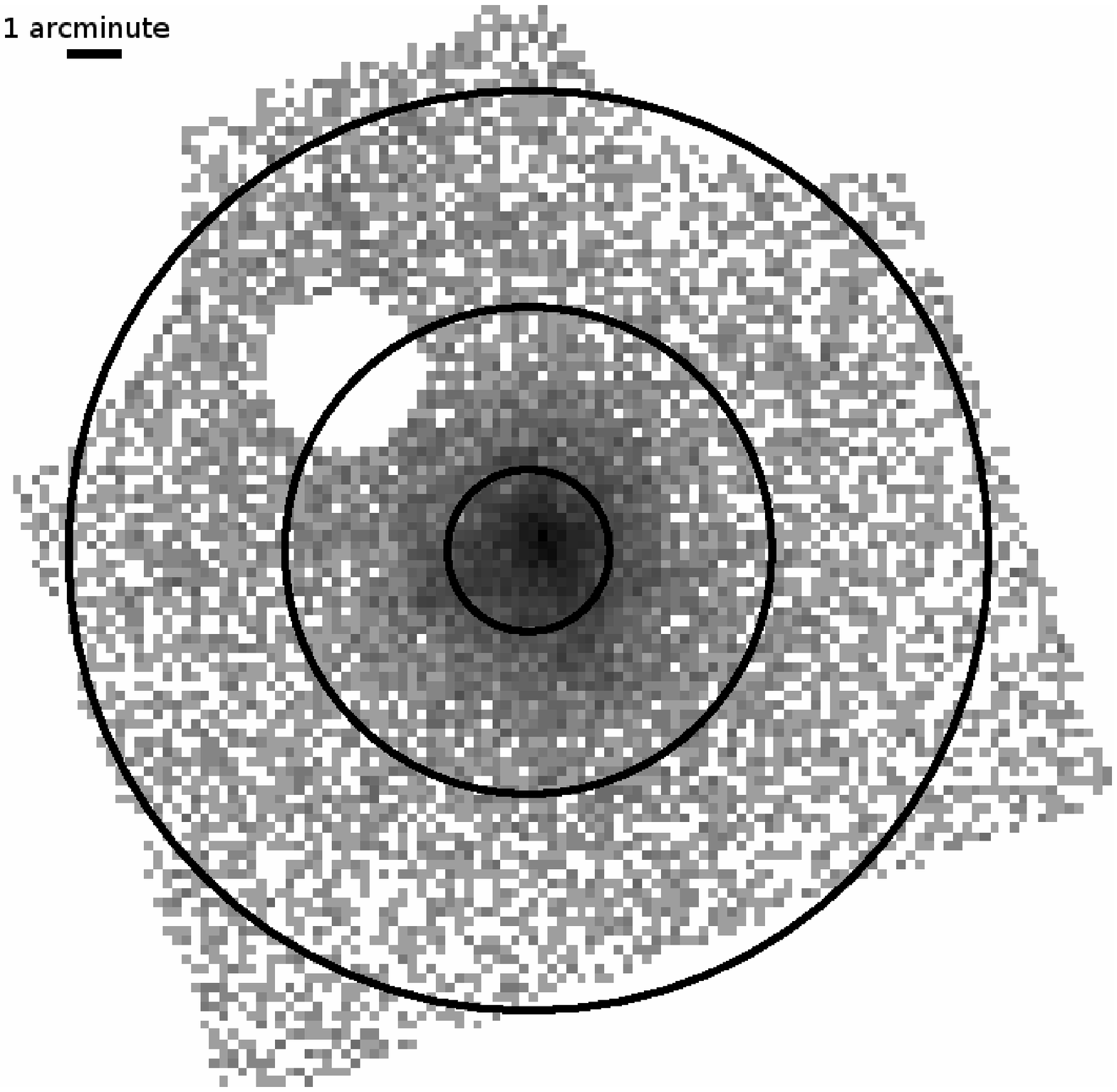}}}
\caption{\label{fig.regions} Regions adopted for spectral analysis.
  ({\sl Left Panel} ) \chandra\ spectral extraction regions overlaid
  on the raw image (without point sources masked out). The outermost
  annulus (i.e., Annulus 11) used only for background modeling is not
  shown.  ({\sl Right Panel}) \suzaku\ spectral extraction apertures
  overlaid on the processed XIS1 image with sources (point and
  calibration) masked out.}
\end{figure*}

Table \ref{tab.obs} lists details of the \chandra\ and \suzaku\
observations of NGC~6482. Below we describe how the data were
prepared for spectral analysis (\S \ref{spec}). We defer discussion of
possible Solar Wind Charge Exchange (SWCX) emission to \S \ref{swcx}. 

\subsection{Chandra}
\label{chandra}

\chandra\ observed \src\ on-axis with the ACIS instrument in the
ACIS-S configuration for $\approx 20$~ks in 2002 as part of observing
cycle AO3.  We prepared the data for spatially resolved spectral
analysis using the \ciao\ (v4.8) and \heasoft\ (v6.18) software suites
along with version 4.7.1 of the \chandra\ calibration database. First,
we followed the standard \chandra\ data-reduction
threads\footnote{http://cxc.harvard.edu/ciao/threads/index.html} and
reprocessed the level 1 events file using the latest calibration. Next,
we extracted a light curve from regions free of obvious point sources
and the main emission of NGC~6482. From visual inspection of the light
curve we removed periods of high background which resulted in the
exclusion of $\approx 2$~ks giving a final, cleaned exposure of
17.5~ks. From the cleaned events list we constructed an image in the
0.3-7.0~keV band to detect point sources using the \ciao\ {\sc
  wavdetect} tool. We set the detection threshold to $10^{-6}$ and
used a 1.7~keV monochromatic exposure map. All point sources detected
by {\sc wavdetect} were verified by eye.

We extracted spectra in a series of concentric, circular annuli
positioned at the optical center of the elliptical galaxy (Fig.\
\ref{fig.regions}). Point sources, chip gaps, and any off-chip regions
were masked out. We adjusted the widths of the annuli so that each
annulus contained $\approx 1000$
background-subtracted counts, which resulted in 9 annuli within a
radius of 184.5\arcsec (3.1\arcmin). These annuli are located entirely
on the ACIS-S3 CCD.  We also included two additional annuli
(Annulus 10: 3.1\arcmin - 5.7\arcmin, Annulus 11: 5.7\arcmin - 12.3\arcmin), with
smaller numbers of source counts to facilitate simultaneous
determination of the background. Annulus 10 is located mostly on the
S3 chip, but with its outermost regions largely off any active ACIS CCD except for
some parts that fall on the S2. (This is the largest annulus shown in Fig.\
\ref{fig.regions}.) Annulus 11 (not shown) partly overlaps the active
CCDs S2, I2, and I3. 

\begin{figure*}
\parbox{0.49\textwidth}{
\centerline{\includegraphics[scale=0.35,angle=-90]{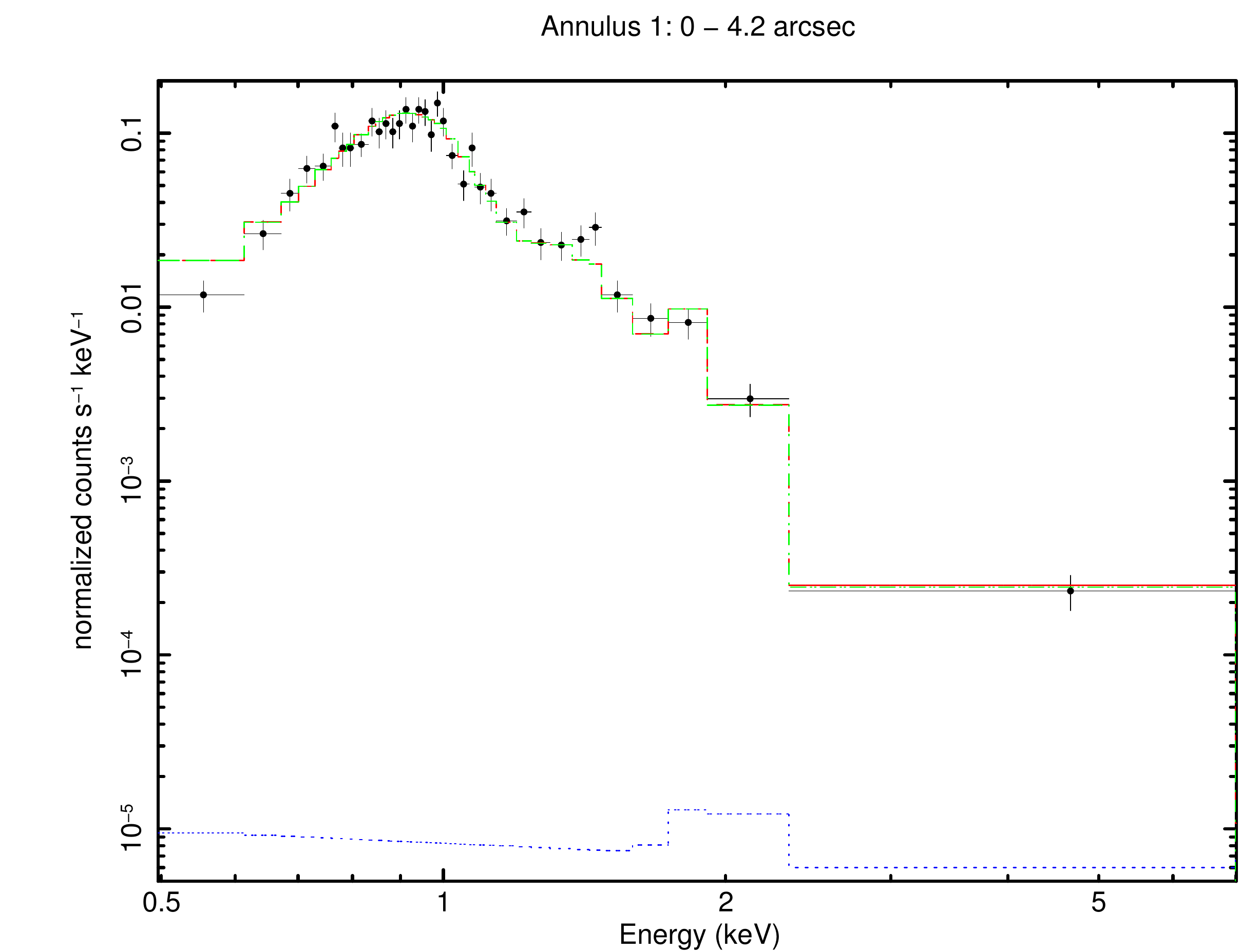}}}
\parbox{0.49\textwidth}{
\centerline{\includegraphics[scale=0.35,angle=-90]{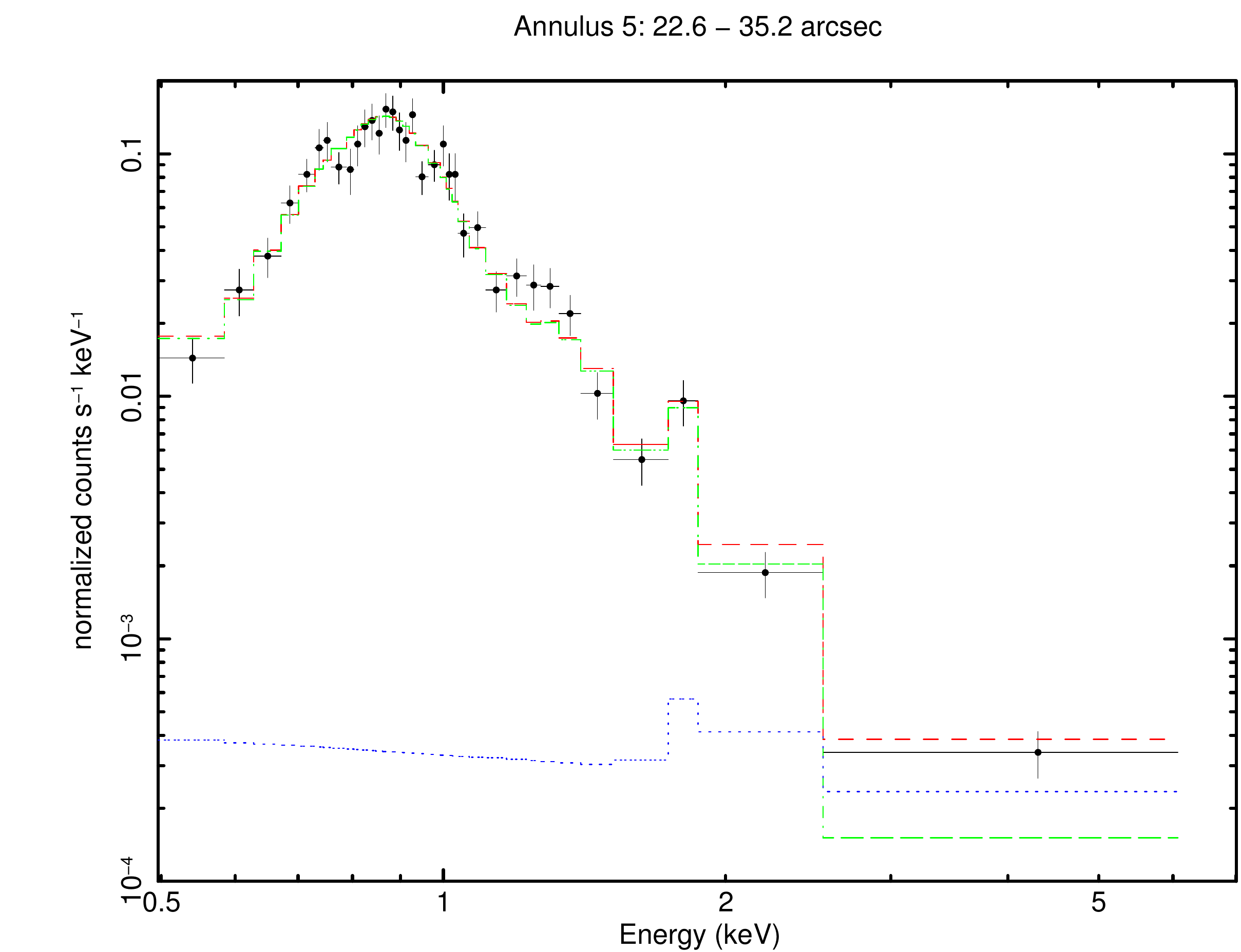}}}

\vskip 0.2cm

\parbox{0.49\textwidth}{
\centerline{\includegraphics[scale=0.35,angle=-90]{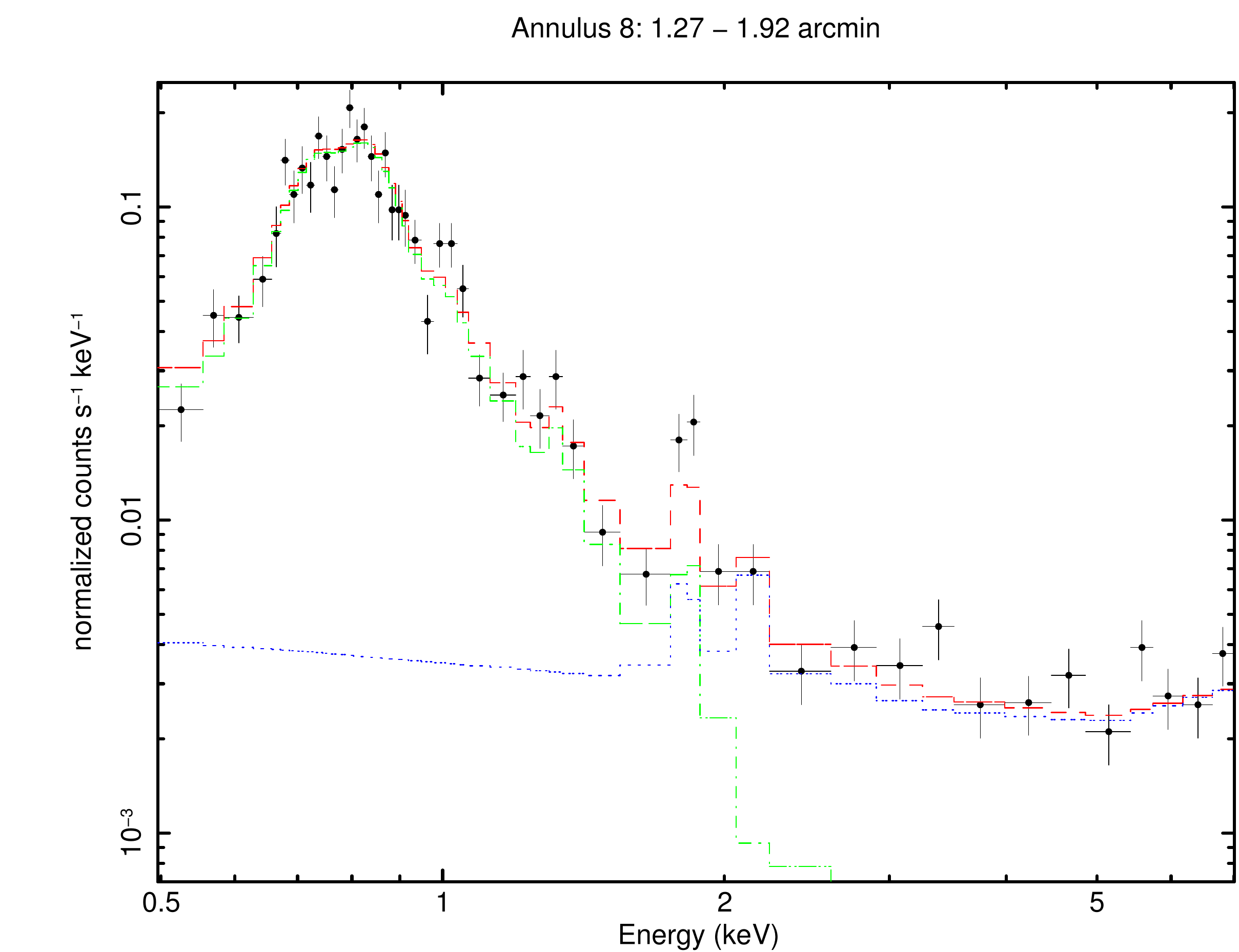}}}
\parbox{0.49\textwidth}{
\centerline{\includegraphics[scale=0.35,angle=-90]{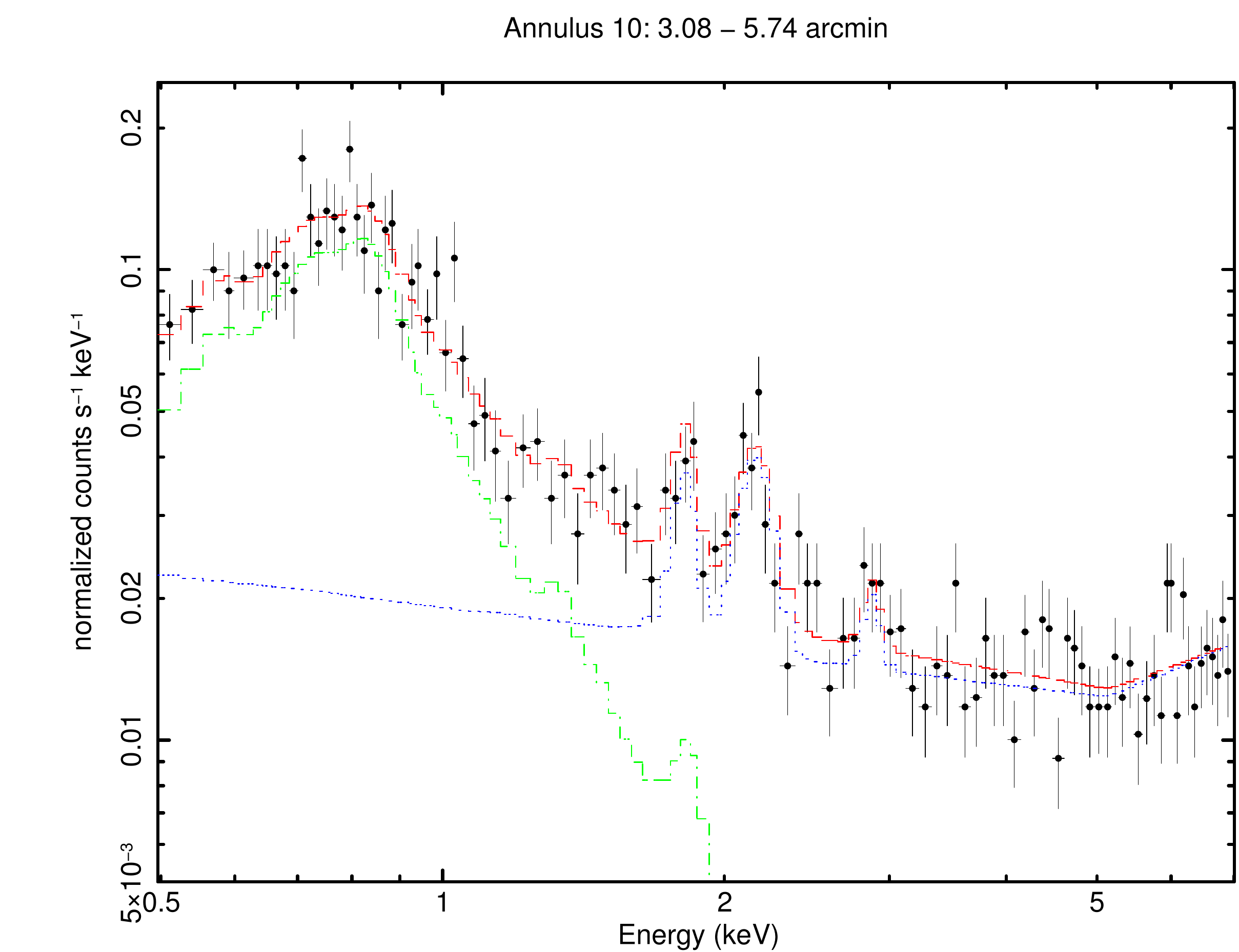}}}

\caption{\label{fig.chandra} Representative \chandra\ spectra in the
  0.5-7.0~keV band without any background subtraction. Also plotted
  are the best-fitting models (red dashed) broken down into the separate contributions
  from the following:  (1) hot gas and unresolved LMXBs from \src\
  along with the CXB (green dot-dash), and (2) particle background
  (blue dotted).}
\end{figure*}

For each annulus we constructed counts-weighted response files (ARFs
and RMFs) using the \ciao\ tools {\sc mkwarf} and {\sc mkacisrmf}. In
Fig.\ \ref{fig.chandra} we plot representative spectra. Note that we
do not subtract any background beforehand. Instead, we model it and
fit it simultaneously with the hot gas emission below (\S
\ref{spec}). For all spectral fitting with \chandra\ we include only
data between 0.5-7.0~keV.

\subsection{Suzaku}
\label{suzaku}

As part of observing cycle AO4, in 2010 \suzaku\ observed \src\ with
the XIS instrument for a nominal exposure of 46.5~ks at a location
slightly offset ($\approx 0.9\arcmin$) from the optical position. We
prepared the data for spatially resolved spectral analysis using the
\heasoft\ (v6.18) software suite along with version 20160204 of the
\suzaku\ calibration database. We followed the standard \suzaku\
data-reduction
threads~\footnote{http://heasarc.gsfc.nasa.gov/docs/suzaku/analysis/abc/}
and used the {\sc aepipeline} tool to reprocess the events using the
latest calibration. Data taken in $5\times 5$ readout mode were
converted to $3\times 3$ mode and then were merged with the original
$3\times 3$ mode data.  We also applied an additional screening on
geomagnetic cutoff rigidity (i.e., COR2$>6$) to reduce the particle
background as recommended in the standard data-reduction threads. We
visually inspected the light curve to remove any periods of high
background. Since the observation was very quiescent, only about
0.3~ks were removed resulting in a cleaned exposure of 43.4~ks.  We
masked out regions near the calibration sources and surrounding one
faint point source located $\sim 5\arcmin$ to the NE of
\src. For the point source we used a circular region of radius
$1.5\arcmin$. While this radius encloses only $\sim 70\%$ of the point
source counts, larger regions would unacceptably exclude more diffuse
emission of \src.

We extracted spectra for the XIS0, XIS1, and XIS3 in three concentric
annuli
$(0\arcmin-1.5\arcmin,1.5\arcmin-4.5\arcmin,4.5\arcmin-8.5\arcmin)$
centered on \src\ (Fig.\ \ref{fig.regions}). For each spectrum we
created an associated RMF file using the task {\sc xisrmfgen}. Unlike
with \chandra, it is necessary to account for spatial mixing between
different annuli because of the large point spread function of
\suzaku\ ($\sim 2\arcmin$ HPD).  Consequently, we created special
mixing ARF files following~\citet{hump11a} using the task {\sc
  xissimarfgen}~\citep{ishi07a}. Since the {\sc xissimarfgen} task,
which performs ray-tracing calculations, requires a model image for the
true spatial source profile, we constructed one from the \chandra\
image using two $\beta$ models with the same $\beta$ for each
component. The parameters were obtained by fitting the 0.5-7.0~keV
\chandra\ surface brightness profile in conjunction with appropriate
background terms. We found the model to be a good fit with
best-fitting parameters, $\beta=0.57$, $r_{c,1}=3.8\arcsec$,
$r_{c,2}=29\arcsec$, and relative normalization of component 2 to 1 of
0.99. Finally, for each spectrum we also generated corresponding
Non-X-ray (i.e., particle) Background (NXB) spectral files using the
task {\sc xisnxbgen}.

For the spectral fitting below in \S \ref{spec}, we initially
considered data between 0.5-7.0~keV for all detectors like we did with
\chandra. From visual inspection of the spectra and comparison to the
spectral models, we decided to restrict further the fitted energy
ranges of the \suzaku\ data. The data at high energies contain little
contribution from \src\ and are instead dominated by the NXB and
CXB. To reduce any sensitivity of our results on the accuracy of the
NXB model we lowered the upper bandpass limit from 7.0 to 5.25~keV. As
for the lower energy limit, we noticed that the two front-illuminated (FI)
CCDs (XIS0, XIS3) generally exhibited large fit residuals near the
lower bandpass limit. Consequently, we excluded energies below 0.7~keV
for the FI CDDs. Therefore, we used the following energy ranges in the
spectral fitting: XIS0, XIS3: 0.7-5.25~keV, XIS1: 0.5-5.25~keV.

\begin{figure*}
\parbox{0.49\textwidth}{
\centerline{\includegraphics[scale=0.35,angle=-90]{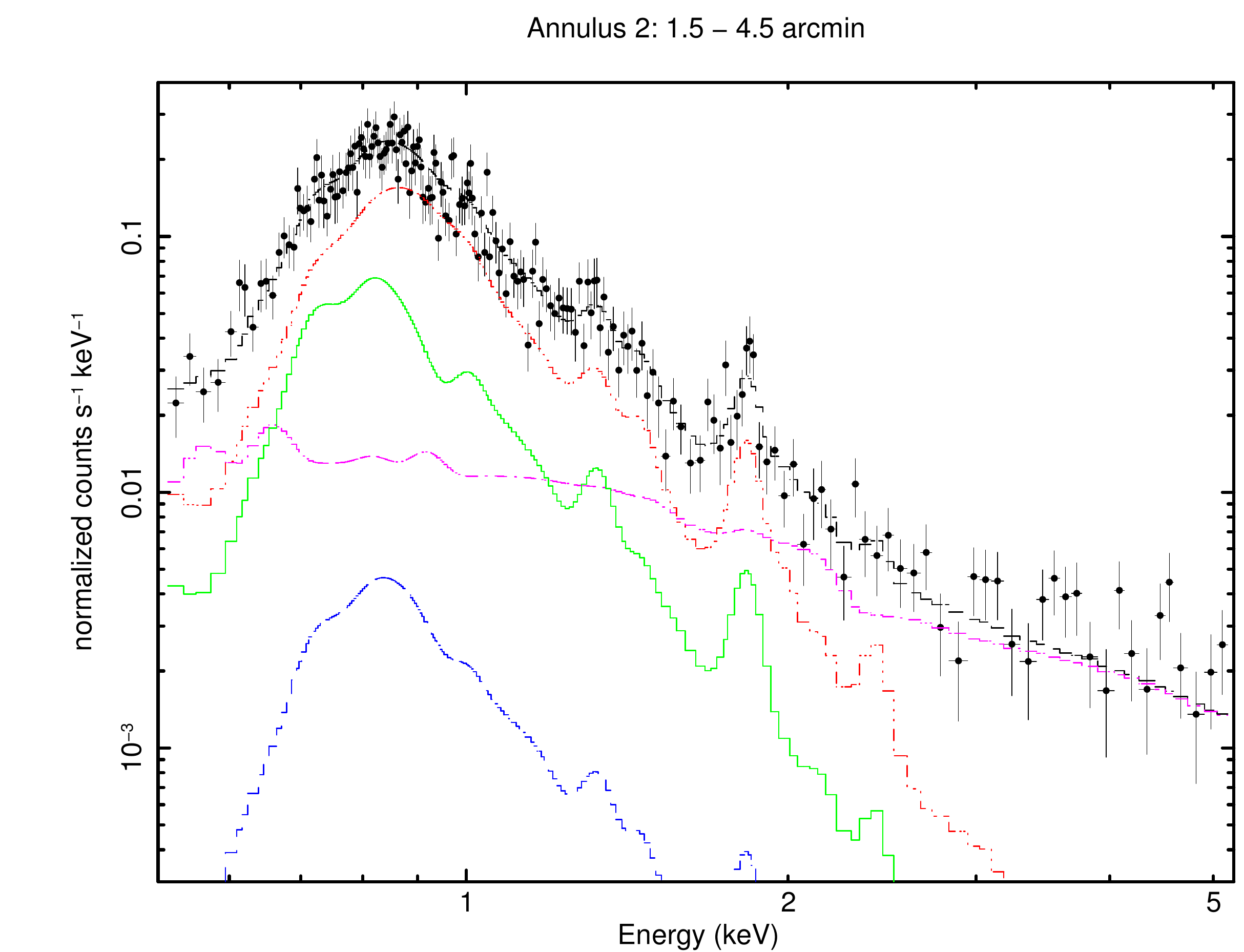}}}
\parbox{0.49\textwidth}{
\centerline{\includegraphics[scale=0.35,angle=-90]{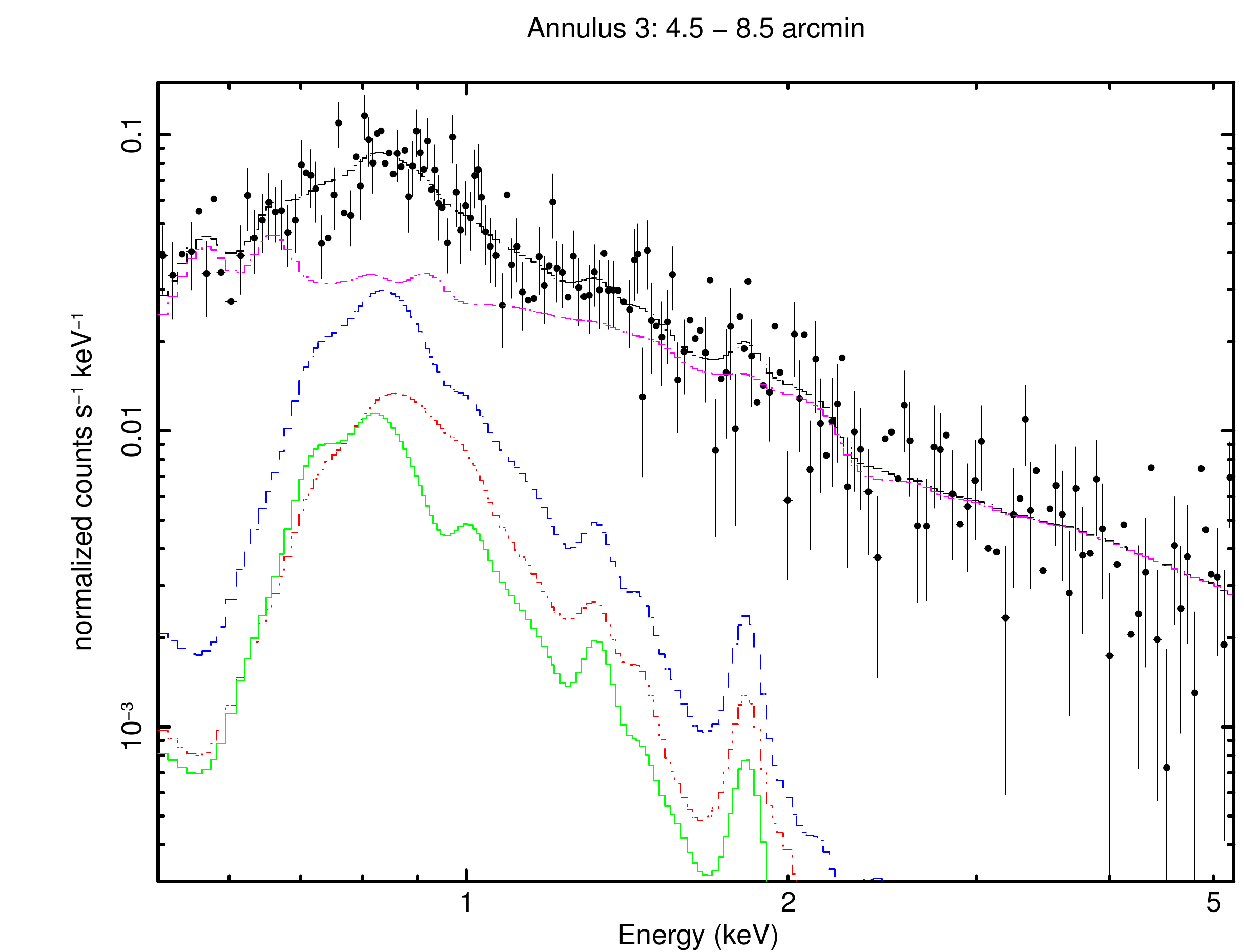}}}
\caption{\label{fig.suzaku} \suzaku\ spectra in the 0.5-7.0~keV band
  with the NXB subtracted for annulus 2 ({\sl Left Panel}) and annulus
  3 ({\sl Right Panel}). Also plotted are the best-fitting models
  (black solid) broken down into the separate contributions from the
  following: hot gas from \src\ from annulus 1 (green solid), annulus
  2 (red dot-dash), and annulus 3 (blue dash) along with CXB (magenta
  dash). Notice in particular that hot gas emission is clearly visible
  near $\la 1$~keV in both annuli.}
\end{figure*}

We plot the XIS1 spectra of Annulus~2 and Annulus~3 in
Figure~\ref{fig.suzaku}.

\section{Spectral Analysis}
\label{spec}

\begin{table*}[t] \footnotesize
\begin{center}
\caption{Hot Gas Properties}
\label{tab.gas}
\begin{tabular}{lcccccccc}   \hline\hline\\[-7pt]
& & $R_{\rm in}$ & $R_{\rm out}$ & $\Sigma_{\rm x}$ (0.5-7.0~keV) &
                                                                    $k_BT$ & $Z_{\rm Fe}$\\
Telescope & Annulus & (kpc) & (kpc) & (ergs cm$^2$ s$^{-1}$ arcmin$^{-2}$) & (keV) & (solar)\\
\hline \\[-7pt]
\\ \chandra\\
&   1 & 0.00 & 1.17 &   1.83e-11 $\pm$   3.21e-12 & $  0.958 \pm   0.019$ &  $   0.68 \pm    0.16$ \\
&   2 & 1.17 & 2.34 &   5.94e-12 $\pm$   1.13e-12 & $  0.889 \pm   0.021$ &  $   1.06 \pm    0.24$ \\
&   3 & 2.34 & 3.92 &   2.49e-12 $\pm$   4.53e-13 & $  0.830 \pm   0.018$ &  tied \\
&   4 & 3.92 & 6.32 &   9.40e-13 $\pm$   1.38e-13 & $  0.828 \pm   0.018$ &  $   0.67 \pm    0.10$ \\
&   5 & 6.32 & 9.83 &   3.99e-13 $\pm$   5.78e-14 & $  0.801 \pm   0.019$ &  tied \\
&   6 & 9.83 & 14.36 &   1.97e-13 $\pm$   3.14e-14 & $  0.711 \pm   0.019$ &  $   0.66 \pm    0.13$ \\
&   7 & 14.36 & 21.23 &   9.16e-14 $\pm$   1.51e-14 & $  0.621 \pm   0.017$ &  tied \\
&   8 & 21.23 & 32.16 &   3.78e-14 $\pm$   6.44e-15 & $  0.575 \pm   0.017$ &  tied \\
&   9 & 32.16 & 51.53 &   1.42e-14 $\pm$   4.05e-15 & $  0.578 \pm   0.024$ &  $   0.26 \pm    0.08$ \\
&  10 & 51.53 & 96.19 &   5.45e-15 $\pm$   8.77e-16 & $  0.588 \pm   0.035$ &  tied \\
\\ \suzaku\\
&   2 & 25.14 & 75.42 &   8.20e-15 $\pm$   3.78e-15 & $  0.572 \pm   0.042$ &  $   0.58 \pm    0.43$ \\
&   3 & 75.42 & 142.45 &   1.44e-15 $\pm$   7.64e-16 & $  0.647 \pm   0.072$ &  tied \\
\hline \\
\end{tabular}
\tablecomments{1~kpc = 3.58\arcsec. Annuli where the iron abundance is
  linked to the value in the previous annulus are indicated as
  ``tied.'' See \S \ref{spec_suzaku} for discussion of the central
  \suzaku\ annulus.}
\end{center}
\end{table*}

\begin{figure*}
\parbox{0.49\textwidth}{
\centerline{\includegraphics[scale=0.43,angle=0]{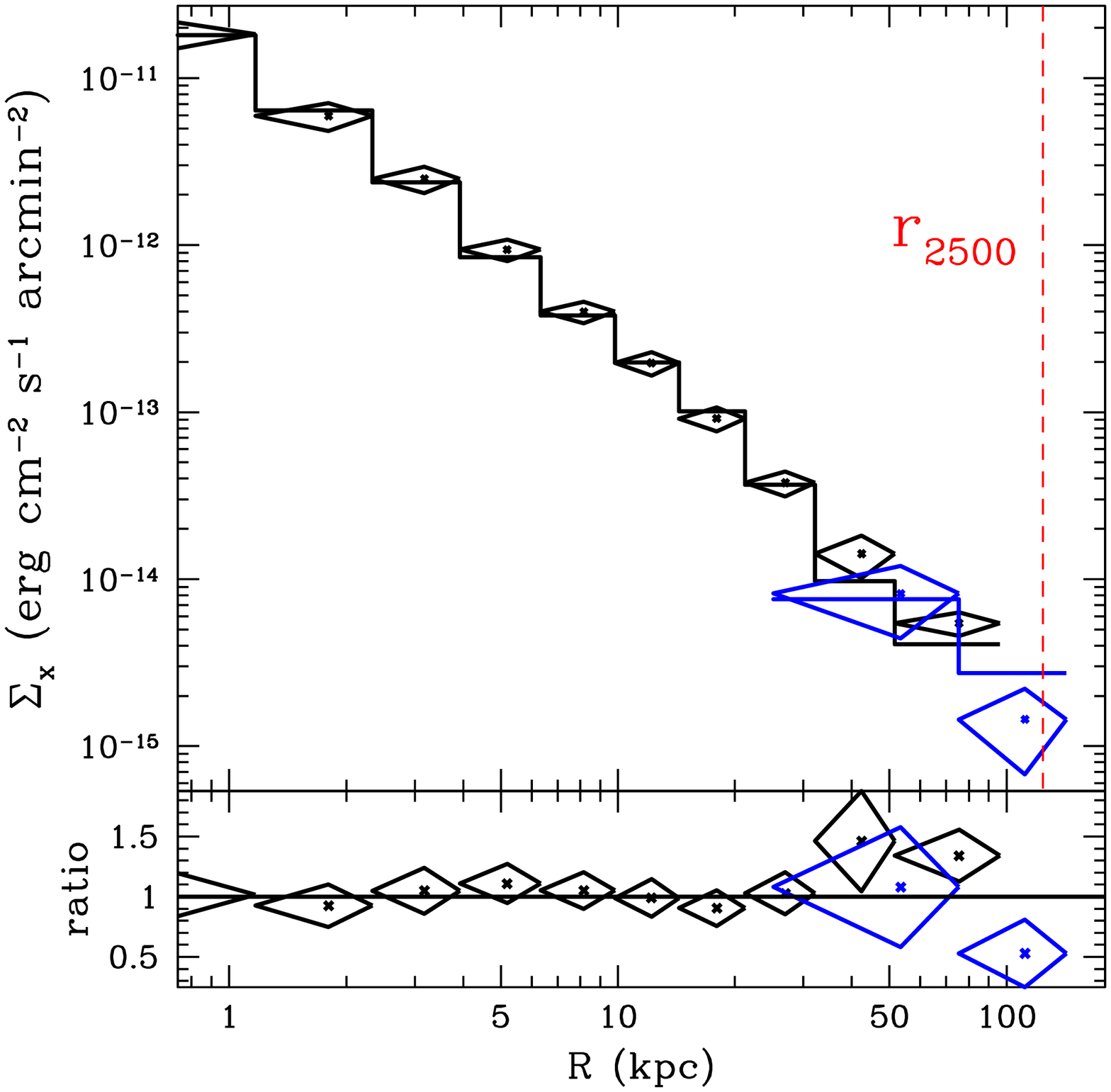}}}
\parbox{0.49\textwidth}{
\centerline{\includegraphics[scale=0.43,angle=0]{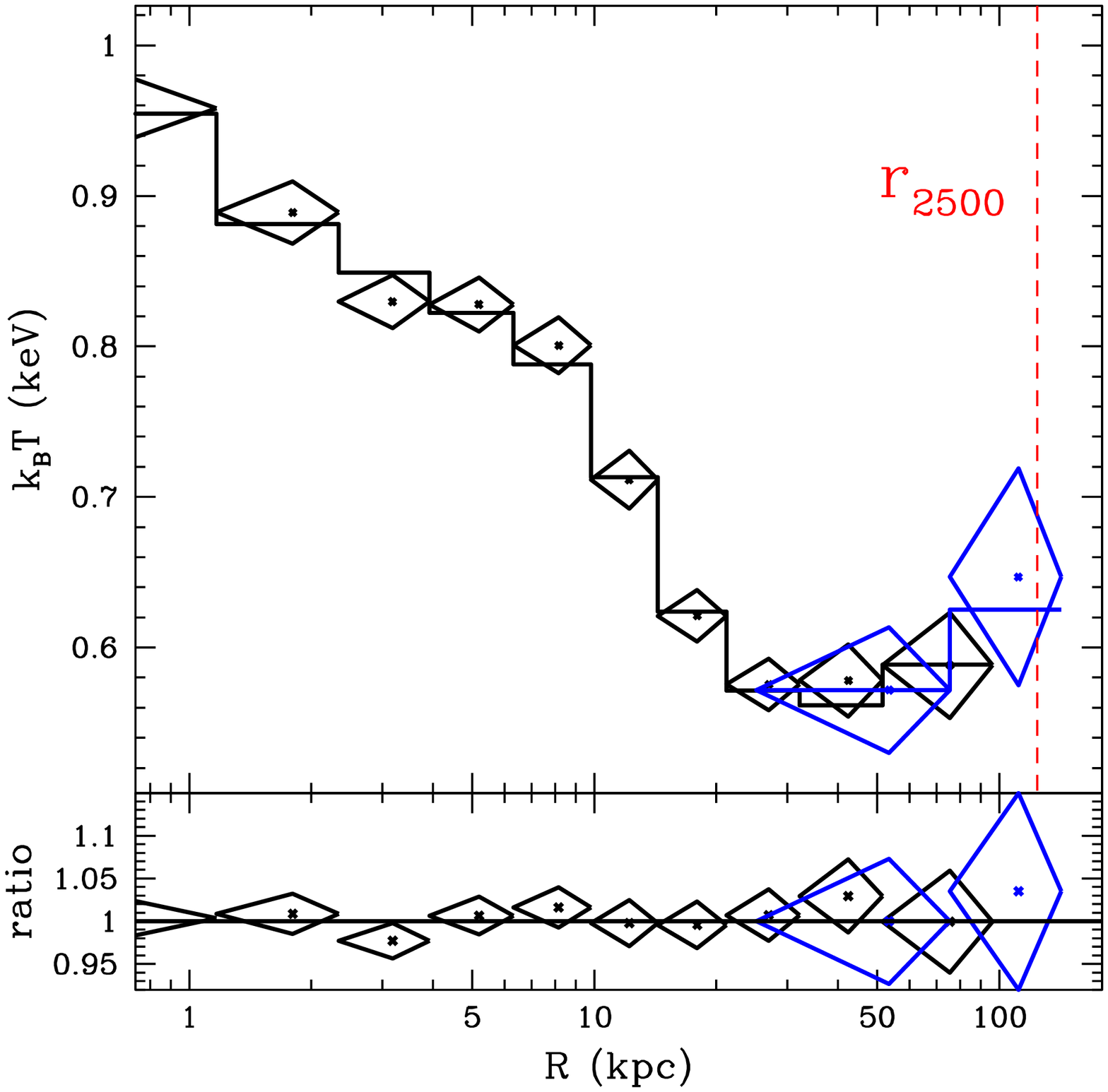}}}
\caption{\label{fig.best} \chandra\
  (black) and \suzaku\ (blue) data, $1\sigma$ errors, and the best-fitting
  fiducial hydrostatic model in 
  each circular annulus on the sky.  ({\sl Left
    Panel}) Surface brightness (0.5-7.0~keV). ({\sl Right Panel})
  Projected emission-weighted temperature ($k_BT$).
 The bottom panels plot the data/model ratios.}
\end{figure*}

We fitted models to the spectra in a standard frequentist approach
using \xspec\ v12.9.0k~\citep{xspec} by minimizing the
C-statistic~\citep{cstat} to mitigate biases associated with the
$\chi^2$ statistic for Poisson-distributed data even in the
high-counts regime~\citep{hump09a}. While not strictly necessary when
using the C-statistic, we rebinned each spectrum so that each data bin
contained a minimum of 20 counts.  \citet{hump09a} note that, like
with the $\chi^2$ statistic, rebinning the data can improve model
selection by emphasizing differences between the model and data. In
addition, fewer bins typically translates to fewer model evaluations
and therefore speedier convergence. In \S \ref{misc.spec} we compare
results using the C-statistic to those obtained using standard
$\chi^2$.

Unless stated otherwise, all quoted parameter errors represent 68\%
confidence limits on one interesting parameter obtained using the {\sc
  error} command in \xspec. For the C-statistic this standard
procedure begins with the model evaluated with the best-fitting values
of all the parameters. Then the parameter of interest is stepped
to lower and higher values (while all other free parameters are varied
to minimize C) until the C-statistic increases by 1 from its
best-fitting  value.

\subsection{Spectral Models}
\label{specmod}

The procedure we adopt to interpret the X-ray spectra closely follows
that used originally for \xmm\ data of the galaxy group
NGC~5044~\citep{buot04b} and has been refined modestly in subsequent
studies, the most recent of which~\citep{su15a} examined a
combination of \chandra\ and \suzaku\ as in our present study.  Here
we summarize the models that we fitted separately to the \chandra\ and
\suzaku\ spectral data.  Unless stated otherwise, all the emission
models described below were modified by foreground Galactic absorption
with the {\sc phabs} model in \xspec\ using the photoelectric absorption
cross sections of \citet{phabs} and a hydrogen column density, $N_{\rm
  H} = 7.7\times 10^{20}$~cm$^{-2}$~\citep{kalb05a}.

\subsubsection{Galactic Emission Models}
\label{galmod}

The principal source of X-ray emission from \src\ is the hot, diffuse
interstellar medium (ISM) / intragroup medium (IGrM). We account for
this emission with the {\vapec} coronal plasma model in \xspec. We do
not perform onion-peeling--type deprojection (e.g., using the {\sc
  projct} mixing model in \xspec) because it amplifies noise
particularly in the challenging background-dominated outer regions. In
addition, the ISM emission outside the bounding annulus is not easily
accounted for by the standard deprojection algorithms, which can be a
sizable source of systematic
error~\citep[e.g.,][]{nuls95a,mcla99a,buot00c}.

As is readily apparent from Figures~\ref{fig.chandra} and
\ref{fig.suzaku}, the broad ``hump'' near 1~keV composed of an
unresolved forest of emission lines from iron L-shell transitions is
the dominant spectral line feature. After iron, for
$k_{\rm B}T\sim 1$~keV the strongest line features arise from
$K$-shell transitions of O, Ne, Mg, Si, and S. For both \chandra\ and
\suzaku, the O, Ne, and Mg lines are heavily blended with the strong,
highly temperature-sensitive Fe-L lines. Consequently, in our spectral
model these weaker, unresolved lines may attempt to partially
compensate for inaccuracies in the Fe-L lines due to, e.g., plasma
code deficiencies or the assumed temperature structure that gives rise
to the Fe Bias~(\citealt{buot00a}; see
also~\citealt{buot94,buot98c,buot99a,buot00c,mole01a,buot03b}); i.e., the
underestimate of the iron abundance when assuming a single plasma
temperature when in fact multiple temperature components are present
arising from, e.g., a range of temperatures due to the projection
along the line of sight of a radial temperature gradient. The Si and S
lines do not suffer from blending with the Fe L shell lines, but they
are contaminated by strong lines in the particle
background. Therefore, by default we assume a single temperature
component for each spectrum and allow the iron abundance to be fitted
freely while restricting the other metal abundances to vary in their
solar ratios with respect to iron; e.g., for silicon, the quantity
$Z_{\rm Si}/Z_{\rm Fe}$ is fixed at the solar ratio. For completeness,
however, we also examine the effects of allowing the non-Fe abundances
to be varied separately and adding additional temperature
components. By default we employ the solar abundance standard
of~\citet{aspl}.

A weaker source of emission arises from unresolved discrete sources in
the galaxy, especially from Low-Mass X-Ray binaries (LMXBs). We
represent the spectrum of these unresolved point sources by a thermal
bremsstrahlung component with $k_{\rm B}T=7.3$~keV and assume it
follows the stellar mass by requiring the relative fluxes in the
different annuli to obey the $K$-band light profile (\S \ref{stars}).
We only include this model interior to $D_{25}$ (i.e., $\la 1\arcmin$)
which corresponds to annuli 1-6 for \chandra\ and the inner annulus
$(<1.5\arcmin)$ for \suzaku. Based on the $K$-band luminosity, the
expected 0.3-7.0~keV luminosity within $D_{25}$ from the LMXBs is,
$L_{\rm LMXB}=(3.5\pm 2.1)\times 10^{40}$~erg~s$^{-1}$, using the
scaling relation of~\citet[][i.e., their equation 8]{hump08b}.

For the \suzaku\ data we obtain a fitted value consistent with the
scaling relation. For the \chandra\ data our best-fitting value is
$L_{\rm LMXB}\approx 5.0 \times 10^{40}$~erg~s$^{-1}$; i.e., about
$1\sigma$ above the predicted value.  Given the nature of our complex,
multi-component spectral model with source and background emission,
the LMXB component is somewhat degenerate with the CXB and particle
packground models, which likely accounts for the possible
discrepancy. Consequently, by default for the \chandra\ spectral
fitting we fixed $L_{\rm LMXB}$ at the predicted value from the
scaling relation and treated as a systematic error the case where
$L_{\rm LMXB}$ is allowed to be freely fitted (\S \ref{misc.spec}).

\subsubsection{Background Emission Models}

We account for the cosmic X-ray background (CXB) with several model
components. The soft portion of the CXB consists of emission from the
Local Hot Bubble (LHB) and hot plasma associated with the disk and
halo of the Milky Way~\citep[e.g.,][]{kunt00a}. Since the temperatures
associated with these components $(k_{\rm B}T\approx 0.1-0.2$~keV) are
well below the band-pass of our observations, we find that at most two
separate soft components are needed, as in, e.g., \citet{lumb02a}. As
we did in \citet{su15a}, we include separate components for the LHB
(unabsorbed {\sc apec} component with $k_{\rm B}T=0.08$~keV and solar
abundances) and the halo gas (absorbed {\sc apec} component with
$k_{\rm B}T=0.20$~keV and solar abundances). The hard portion of the
CXB is dominated by unresolved AGN which we model as a power-law with
index $\Gamma=1.41$~\citep[e.g.,][]{delu04a}.

As mentioned above in \S \ref{suzaku}, we adopt a standard model for
the particle background and subtact it from the \suzaku\ spectra
before fitting. For \chandra, however, we parameterize the particle
background with a simple broken power-law model and several Gaussian
components -- none of which are folded through the ARF (i.e., the
particle background model components are ``un-vignetted'') -- which we
find is sufficient to provide an acceptable description of the data. In
particular, to obtain a good fit for the particle background for
annuli 1-10 (which lie mostly on the BI chip S3) we added three
gaussian components with best-fitting central energies, 1.81, 2.17,
and 2.88~keV. Their emission is easily seen in the spectrum of
annulus~10 shown in Figure~\ref{fig.chandra}. For annulus~11, which
spans three FI chips, the most important gaussian component we added
has energy 2.17~keV, though we also added minor components at 1.49~keV
and 2.51~keV.

We required the CXB to be uniform across the field. Therefore, we tied
all CXB parameters for all annuli and insured that the fluxes of the
CXB model components were the same in each annuli when rescaled to the
aperture area. The same was done for the \chandra\ particle background
model. We experimented with allowing the particle background
normalizations to vary separately for each annulus and found the
results were not changed significantly.

\subsection{Chandra}
\label{spec_chan}

In addition to background and LMXB components, the fiducial spectral
model described above (\S \ref{specmod}) consists of a single \vapec\
coronal plasma component for each annulus where only the temperature,
normalization, and iron abundance (all other elements tied to iron in
their solar ratios) are allowed to vary. We find this fiducial model
provides a good overall fit to the data with a minimum C-statistic of
631.85 for 595 degrees of freedom (dof). This best fit also translates
to a reduced $\chi^2=1.07$ with a null hypothesis probability of 12\%,
indicating a formally acceptable fit. The good quality of the global
fit is on display in Figure~\ref{fig.chandra} where we show the
best-fitting model over-plotted on the spectra in four of the annular
regions.

For the hot gas we list the surface brightness ($\xsurf$), temperature
($\ktemp$), and $\ziron$ for each annulus in Table~\ref{tab.gas} and
plot the radial profiles of $\xsurf$ and $\ktemp$ in
Figure~\ref{fig.best}. We define $\xsurf$ to be the 0.5-7.0~keV flux
(erg cm$^2$ s$^{-1}$) of the gas emission in the aperture divided by
the aperture solid angle taken to be $\pi\theta^2$ (arcmin$^2$).  The
temperature profile peaks at the center ($\ktemp\approx 0.96$~keV) and
decreases with radius until flattening out in the last two bins (9-10)
with a slight hint of an upturn that is more pronounced in the
\suzaku\ spectra (\S \ref{spec_suzaku}).

The shape of the declining temperature profile agrees well with our
previous study of \src\ by H06 using similar models and
data reduction methods. However, the temperatures we have measured
here systematically exceed those reported by H06 by
$\approx 20\%$, while our iron abundances are much smaller.  In our
present study the \vapec\ plasma emission code uses the current version (2.0.2)
of the atomic database {\sc AtomDB}\footnote{http://www.atomdb.org},
whereas H06 used version 1.3.1. We find that if we force
\xspec\ to use {\sc AtomDB} version 1.3.1, then we obtain results that
are overall very consistent with those of H06.

Even though we obtain a fit that is formally acceptable, we examined
whether the fit could be improved further.  Perhaps the most notable
fit residuals occur between 0.5-0.6~keV for annulus 1, and also annuli
3-4 (not shown), where the model over-predicts the data by $\sim 60\%$
with 2-$3\, \sigma$ significance. (Excluding these bins does not affect
the fitted parameters much.) Other notable, but less significant,
residuals occur in the Fe-L region (i.e., 0.7-1.4~keV) at
approximately the $\pm 25\%$ level for annuli 1-6. These features are
characteristic of the Fe Bias (\S \ref{galmod}), but since the annular
widths are fairly small, the effect is not highly significant; for a
more detailed study of the effect of aperture width on the Fe Bias,
see, e.g., the case of NGC~5044~\citep{buot03a,buot03b}. Indeed, we
find that adding a second temperature component leads to only a
marginal improvement in the fit while leading to much more poorly
constrained gas parameters.

We also find that essentially the same reduction in the C-statistic
obtained by adding a second temperature component can be achieved
instead by allowing some other elemental abundances to be varied
separately from iron.  Since the abundances of these other elements
are not as well constrained as iron, we tied their values for all annuli.
When doing so, the largest fit improvement occurs for Ne, Mg and
Si respectively for which we obtain the following abundance ratios (in solar
units): $\zneon/\ziron=2.5^{+0.4}_{-0.3}$, $\zmag/\ziron=1.6\pm 0.1$,
and $\zsil/\ziron=1.3^{+0.4}_{-0.3}$. Since we expect that these
non-solar abundance ratios reflect to a large extent the Fe Bias, we
treat this variable abundance case as a systematic error for the mass
models (\S \ref{misc.spec}).

\subsection{Suzaku}
\label{spec_suzaku}

The \suzaku\ data from all annuli and all instruments (XIS0, XIS1,
XIS3) are fitted simultaneously. To account for any relative
calibration uncertainties between the detectors, we allow the relative
normalizations of the model components for the XIS1 and XIS3 to vary
with respect to the XIS0 through the use of additional multiplicative
constants.

The strong temperature gradient measured by \chandra\ within the
region of annulus~1 (i.e., $<1.5\arcmin$, see Table~\ref{tab.gas})
compels us to consider multiple temperatures there so that the
spatial mixing between annuli due to the large PSF is sufficiently
accurate in our model for all annuli. Like our previous study of
RXJ~1159+5531~\citep{hump12a,su15a}, for this purpose we adopted a
simple two-temperature model (2T), since we have shown previously that
such a simple model provides a good description at CCD resolution of
the integrated spectrum in a region containing a radial temperature
gradient~\citep{buot99a,buot03a}. The 2T model improves the global fit
modestly (i.e., $\Delta\rm C = 11.0$, $\Delta\chi^2=12.6$ for 1136 dof
and 3 additional parameters), so that the gas parameters in the other
annuli are also modestly affected. For example, in annulus~2, $\ktemp$
rises from 0.47~keV (1T) to 0.57~keV (2T), the difference of which is
$\approx 2.5\sigma$ significant.  We therefore employ the 2T model in
annulus~1 (and 1T models in annuli~2 and 3) for the fiducial model of the
\suzaku\ spectra and treat the case of using the 1T model in annulus~1
as a systematic error (\S \ref{misc.spec}).

Whereas we obtain tight constraints on the gas parameters in annulus~1
for a 1T model (e.g., $\ktemp=0.766\pm 0.006$~keV,
$\ziron=0.70\pm 0.08\, \zsolar$), for the 2T model the temperatures
for each component are not very well constrained. In fact, the
emission-weighted parameters of the 2T model (e.g.,
$\ktemp=0.73\pm 0.12$~keV) have much larger uncertainties than the
corresponding \chandra\ measurements in the same region. Consequently,
we find that annulus~1 does not provide tangible constraining power
for the hydrostatic models (\S \ref{models}) above that provided by
the \chandra\ data, and thus we do not include annulus~1 in our
subsequent analysis.

From visual inspection of Figure~\ref{fig.suzaku}, where we show the
best-fitting fiducial model to the spectra in annuli~2-3, it is
apparent that the fiducial model provides a good description of the
data there, and the strong spatial mixing of the gas components is
readily seen.  The best fit yields a C-statistic of 1349.35 for
1136 dof and reduced $\chi^2=1.20$. Unlike the \chandra\ data, the
\suzaku\ fit is not formally acceptable, and we were unable to greatly
improve the fit with small alterations to the fiducial model (e.g.,
varying other abundances, see below). The most prominent model
residuals occur for annulus~1 (not shown), providing additional
justification for not including the parameters of annulus~1 in our
subsequent analysis.

We list the parameters for the hot gas for annuli~2-3 in
Table~\ref{tab.gas} and plot them in Figure~\ref{fig.best}. The
parameters of annulus~2 measured by \suzaku\ agree very well with
those measured by \chandra\ in their overlapping region (i.e.,
\chandra\ annuli~8-10) but with larger statistical error. For
annulus~3, which is mostly exterior to the \chandra\ data, the
temperature appears to rise. Though the rise is not very significant
($\sim 1\sigma$), it is consistent with the hint of a rise in
\chandra\ annuli~8-10. The isolated elliptical galaxy
NGC~1521~\citep{hump12b}, which has a halo mass similar to \src,
displays a similar rise in temperature near $\rtwofiveh$.

Allowing the Ne abundance to be free yields a super-solar value
$(\zneon/\ziron=2.2\pm 0.3)$ very consistent with that obtained from
\chandra\ (\S \ref{spec_chan}). However, for Mg and Si we obtain
ratios, $\zmag/\ziron=1.09\pm 0.08$, and $\zsil/\ziron=1.00\pm 0.07$,
very consistent with solar and, particularly for Mg, significantly
less than obtained with \chandra.  As we do for \chandra, we treat
this variable abundance case as a systematic error for the mass models
(\S \ref{misc.spec}).

\section{Entropy-Based Method}
\label{method}

To measure the mass profile from the X-ray observations we adopt the
entropy-based approach of \citet{hump08a}; see \citet{buot12a} for a
discussion of the benefits of this approach and a review of other
hydrostatic methods. The equation of hydrostatic equilibrium may be
written in terms of the variable $\xi \equiv P^{2/5}$,  with $P$ the
total thermal pressure, so that
\begin{equation}
\frac{d\xi}{dr} = -\frac{2}{5}\frac{GM(<r)}{r^2}S_{\rho}^{-3/5},
\label{eqn.he}
\end{equation}
where $M(<r)$ is the total gravitating mass enclosed within radius
$r$, and
\begin{equation}
  S_{\rho} \equiv \frac{k_{\rm B}T}{\mu m_{\rm a}\rho_{\rm
  gas}^{2/3}} = \frac{S}{\mu m_{\rm
    a}}\left(\frac{2+\mu}{5\mu}\frac{1}{m_{\rm a}}\right)^{2/3}.
\end{equation}
The quantity $S \equiv k_{\rm B}Tn_e^{-2/3}$ is the conventional
entropy proxy expressed in units of keV~cm$^{2}$, $m_a$ is the atomic
mass unit, and $\mu$ is the mean atomic mass of the hot ISM. We take
$\mu=0.62$ for a fully ionized cosmic plasma.

We assume parameterized functions for $S$, BCG stellar mass, and DM,
insert them into Equation~\ref{eqn.he}, and solve for $\xi$ using an
iterative procedure to ensure the gas mass is treated
self-consistently~\citep{buot16a}. For the boundary condition on $\xi$
we choose the value of the pressure at a radius of 1~kpc which we
designate as the ``reference pressure,'' $P_{\rm ref}$. Given a
solution for $\xi$ we compute profiles of gas density,
$\rho_{\rm gas}= (P/S_{\rho})^{3/5} = S_{\rho}^{-3/5}\xi^{3/2}$, and
temperature,
$k_{\rm B}T/\mu m_{\rm a} = S_{\rho}^{3/5}P^{2/5} =
S_{\rho}^{3/5}\xi$.

We compare this hydrostatic equilibrium model to the observations
as follows. Using the three-dimensional profiles of $\rhog$ and
$\ktemp$ we construct the X-ray volume emissivity,
$\epsilon_{\nu}\propto\rho_{\rm gas}^2 \Lambda_{\nu}(T,Z)$, where
$\Lambda_{\nu}(T,Z)$ is the plasma emissivity of the \vapec\ model in
\xspec. We project $\epsilon_{\nu}$ and the emission-weighted
temperature along the line of sight using equations (64) and (67) of
\citet{buot12b} for a spherical system but with the aperture
luminosity replaced by the flux and the aperture area expressed in
square arcseconds~\citep[see also Appendix B of ][]{gast07b}. Then
we compare the resulting surface brightness $\xsurf$ and projected
temperature map to the observations.

To fully specify $\Lambda_{\nu}(T,Z)$ also requires the radial
elemental abundance profiles to be specified. In our previous studies
of elliptical galaxies and galaxy clusters we have found that simply
using the measured projected elemental abundance profiles for the true
three-dimensional profiles for the models yields satisfactory
results. In \S \ref{mischm} we examine an alternative procedure to
account for the metal abundance profile in the hydrostatic models. 

\section{Models}
\label{models}

\subsection{Entropy}
\label{models.entropy}

To represent the profile of the entropy proxy we use the simple function,
\begin{equation}
S(r) = s_0 + s_1f(r), \label{eqn.entropy}
\end{equation}
where $s_0$ is a spatially uniform constant, $s_1=S(1\, {\rm kpc}) -
s_0,$ and $f(r)$ is a dimensionless function with $r$ expressed in kpc, 
\begin{equation*}
f(r) = \left \{
\begin{array}{l}
r^{\alpha_1}  \hfill r \le r_{\rm b,1}\\
f_1r^{\alpha_2}  \qquad\qquad \hfill r_{\rm b,1}< r \le r_{\rm b,2}\\
f_2r^{\alpha_3}   \hfill r > r_{\rm b,2}
\end{array}  \right .\
\end{equation*}
The two break radii are denoted by $r_{\rm b,1}$ and $r_{\rm b,2}$ and
the coefficients $f_1$ and $f_2$ are,
\begin{equation*}
f_n  =  f_{n-1} r_{{\rm b},n}^{\alpha_{n}-\alpha_{n+1}},
\end{equation*}
with $f_0\equiv 1.$ We restrict $\alpha_1,\alpha_2,\alpha_3\ge 0$ to
enforce convective stability consistent with the condition of
hydrostatic equilibrium. In addition, we ensure that beyond some
radius $r_{\rm  baseline}$ the logarithmic entropy slope matches that of the
baseline entropy profile (i.e., $\approx 1.1$) derived from
cosmological simulations with only
gravity~\citep[e.g.,][]{tozz01a,voit05a}.  We find our results are not
very sensitive to the choice of $r_{\rm  baseline}$. By default we
set $r_{\rm baseline}=200$~kpc, which lies between $\rtwofiveh$ and
$\rfiveh$ and is not far outside the measured data extent. We treat a
larger value of $r_{\rm baseline}$ as a systematic error (\S
\ref{sys.entropy}).

\subsection{Stellar Mass}
\label{stars}

Since the $K$-band light should trace stellar mass more closely than
that of shorter wavelengths, we represent the stellar mass of the BCG
using the $K$-band light profile from the Two Micron All-Sky
Survey (2MASS) as listed in the Extended Source
Catalog~\citep{jarr00a}. We convert the 2MASS total $K$-band magnitude
($\rm k\_m\_ext = 8.372$) to absolute magnitude following equation~(1) of
\citet{ma14a} using Galactic extinction $A_V=0.277$ and distance
$\rm D=59.2$~Mpc to yield, $M_K=-25.52$ and thus
$L_K=3.3\times 10^{11}\, L_{\odot}$ (Table~\ref{tab.prop}). Despite
the relatively shallow 2MASS photometry, \citet{ma14a} find no
evidence that the total $K$-band magnitudes in their local sample of
elliptical galaxies have been systematically underestimated.

To assign the 2MASS effective radius to our spherical models we follow
the procedure of \citet{capp11a} and take the median effective
semi-major axes from the $J$-, $H$-, and $K$-band photometry and
convert it to a spherical radius using the geometric mean of the 2MASS
effective minor and major axes to yield, $R_e=10.1\arcsec=2.8$~kpc.
There is widespread evidence, however, that the typical effective
radius obtained by the 2MASS photometry is smaller than obtained from
analysis of photometry at shorter wavelengths. For example,
\citet{ma14a}'s equation~(4) provides a conversion between the typical
effective radius measured by 2MASS and the NASA-Sloan Atlas
(NSA). Using this conversion for \src, we obtain $R_e=4.5$~kpc which
is $\approx 60\%$ larger than the 2MASS value.  Although the $K$-band
is preferred for representing the stellar mass, given the large
differences in the effective radius just noted, we adopt for our
fiducial model the mean of the 2MASS and NAS values (i.e.,
$R_e=3.65$~kpc) and consider the full range 2.8-4.5~kpc as a
systematic error (\S \ref{sys.stars}).

In our models we treat the stellar mass-to-light ratio $(\mlkband)$ of
the BCG as a free parameter. To provide a check on the accuracy of our
hydrostatic model as well as provide a constraint on the stellar
initial mass function (IMF), it is interesting to compare $\mlkband$
determined from the hydrostatic analysis to that obtained from
single-burst stellar population synthesis (SPS) models.  In
H06 we provided SPS estimates of $\mlkband$ for \src\
based on updated versions of the models of~\citet{mara98a} and an age
and metallicity derived from Lick indices. Since no uncertainty was
used for the age in H06, we have updated the SSP fits using
the same procedure as H06 but with 10\% error on the age
of \src\ as found by~\citet{sanc06a}.  For an age $12\pm 1.2$~Gyr and
metallicity $[\rm Z/H\rm] = 0.06 +/- 0.15$, we obtain $\mlkband$
values of $1.58\pm 0.12$ (Salpeter IMF), $1.05\pm 0.08$ \citep[Kroupa
IMF,][]{krou01a}, and $0.89\pm 0.07$ \citep[Chabrier
IMF,][]{chab03a}. Fully consistent values are obtained if instead we
use the age and metallicity found by \citet{sanc06a}, i.e., $\log(\rm
Age)= 10.058 \pm 0.044$ and $[\rm Z/H\rm] = 0.116\pm 0.032$.

We treat as a systematic error in \S \ref{sys.stars} the additional
(less well constrained) stellar mass contributions from non-central
galaxies and intracluster light.

\subsection{Dark Matter}
\label{dm}

We consider the following models for the distribution of DM.  All the
models possess two free parameters which may be expressed as a
concentration $c_{\Delta}$, and mass $M_{\Delta}$, evaluated with
respect to an overdensity $\Delta$ times the critical density of the
universe.

\begin{itemize}
\item{\bf NFW} We adopt the NFW profile~\citep{nfw} for our fiducial
  DM model since it is still the current standard both for modeling
  observations and simulated clusters. 

\item{\bf Einasto} A more accurate model for \lcdm\ halos is the
  Einasto profile~\citep{eina65}. Our implementation for the Einasto
  profile follows~\citet{merr06a}, but we use \citet{reta12a}'s
  approximation for $d_n$.  We fix $n=5.9\, (\alpha=0.17)$ appropriate for a
  halo of mass similar to \src~\citep[e.g., see figure 13 of ][]{dutt14a}. 

\item{\bf CORELOG} Finally, we also investigate a pseudo-isothermal
  logarithmic potential with a core to provide a strong contrast to
  the NFW and Einasto models; i.e., a model having a constant density
  core but with a density that approaches $r^{-2}$ at large
  radius. See \S 2.1.2 of \citet{buot12c} for this model expressed in
  terms of $c_{\Delta}$ and $M_{\Delta}$.

\end{itemize}

We evaluate all virial parameters (i.e., $c_{\Delta}$,
$M_{\Delta}$, and $r_{\Delta}$) at the redshift of \src. Note that the
concentration values we quote below for the total mass profile (i.e.,
stars+gas+DM) employ the scale radius of the DM profile; i.e.,
$c_{\Delta}^{\rm tot}\equiv r_{\Delta}^{\rm tot}/r_s^{\rm DM}$, where
$r_s^{\rm DM}$ is the DM scale radius and $r_{\Delta}^{\rm tot}$ is
the virial radius of the total mass profile. To obtain
$r_{\Delta}^{\rm tot}$ we begin with the DM interior to $r_{\Delta}^{\rm DM}$,
add in the baryonic components, and then recompute the virial radius. We
iterate until the change in $r_{\Delta}^{\rm tot}$ is sufficiently
small.

\subsection{Adiabatic Contraction}
\label{ac}

\begin{table*}[t] \footnotesize
\begin{center}
\caption{Adiabatic Contraction Model Definitions}
\label{tab.ac.defs}
\begin{tabular}{ccccccc}   \hline\hline\\[-7pt]
Model & a & b & c & d & Description & Ref.\\
\hline \\[-7pt]
AC1 & 0 & 1 & 0 & 1 & Standard & \citet{blum86a}\\
AC2 & 0 & 1 & 0 & 0.8 & Modified & \citet{gned04a}\\
AC3 & 1  & -0.52 & -1 & 2 & Standard Hydro & \citet{dutt15a}\\
AC4 & 0.75  & 0.25 & 0 & 2 & Forced Quenching & \citet{dutt15a}\\
\hline \\
\end{tabular}
\tablecomments{Parameters refer to Equation~\ref{eqn.ac}. AC2
  approximates \citet{gned04a}'s modified AC model following
  \citet{dutt15a}. Both the Standard Hydro and Forced Quenching models
  refer to the results of cosmological hydrodynamical simulations at
  $z=0.$ The Forced Quenching model arbitrarily turns off cooling and
  star formation at $z=2.$. The AC models we compute in this paper use
  the NFW and Einasto profiles.} 
\end{center}
\end{table*}

We consider ``adiabatic contraction'' (AC) models that modify the
shape of the DM halo due to the presence of the assembled stars in the
central galaxy. The standard analytic AC model presented by
\citet{blum86a} uses an adiabatic invariant, $r_{\rm f}M_{\rm f} =
r_{\rm i}M_{\rm i}$, to connect the initial total mass profile
(assumed to follow the unperturbed DM profile) and the final mass
profile consisting of the central galaxy's stellar mass and the
perturbed DM profile. \citet{blum86a} assumed spherical symmetry and
circular orbits. To allow for more complex AC models, we generalize
the contraction factor $r_{\rm f}/r_{\rm i}$ following \cite{dutt15a},
\begin{equation}
\frac{r_{\rm f}}{r_{\rm i}} = a + b\left(\frac{M_{\rm i}}{M_{\rm f}} +
  \rm c\right)^{\rm d}, \label{eqn.ac}
\end{equation}
where $a$, $b$, $c$, and $d$ are constants. In Table~\ref{tab.ac.defs}
we list the parameters of the AC models we consider in this paper. In
addition to the standard analytic model of \citet[][hereafter
AC1]{blum86a}, we investigate the analytic modification by
\citet[][hereafter AC2]{gned04a} that allows for non-circular
orbits. Our implementation of AC2 uses the approximation suggested by
\citet{dutt15a}. We denote by ``AC3'' the results of nominally
standard cosmological hydrodynamical simulations by \citet{dutt15a}
that include cooling and star formation but not AGN feedback. Finally, we
examine the ``Forced Quenching'' model (AC4) of  \citet{dutt15a}
based on the same simulations as used for AC3 except that cooling
and star formation are arbitrarily turned off at $z=2$. 

We implement the AC models by employing the iterative procedure
described by \citet{blum86a} but using the generalized contraction
factor in Equation~(\ref{eqn.ac}). We neglect the hot gas for
computing the contracted DM profile, which should be a good
approximation since the gas mass is sub-dominant everywhere. In our
present study we apply the AC models to the NFW and Einasto profiles.
We quote results for the concentration parameters of these AC models
using the scale radius of the initial unperturbed mass
profile. Consequently, the quoted AC concentrations can be directly
compared to the results of cosmological simulations with only
DM~\citep[e.g.,][]{dutt14a}.

\section{Results}
\label{results}

\subsection{Overview}
\label{overview}

To fit the hydrostatic equilibrium model to the data we employ
a Bayesian ``nested sampling'' analysis using the MultiNest code
v2.18~\citep{multinest}. We use a $\chi^2$ likelihood function
consisting of the surface brightness and temperature data in
Table~\ref{tab.gas}. We adopt flat priors on the free
parameters except for $P_{\rm ref}$, stellar mass ($M_{\star}$), and DM
($M_{\rm DM}$) for which we use flat priors on their logarithms.  For
each parameter we quote two ``best'' values: (1) the mean parameter
value of the posterior, which we refer to as the ``Best Fit'', (2) the
parameter that maximizes the likelihood, which we refer to as the
``Max Like''.  Unless stated otherwise, errors quoted are the standard
deviation ($1\sigma$) of the posterior.  Other than
Figure~\ref{fig.best}, which evaluates the fiducial model using the
Max Like parameters, all figures for the hydrostatic models plot the
mean and standard deviation of the posterior as a function of radius
for the quantity of interest; e.g., entropy, pressure, etc.

The fiducial model has 11 free parameters, which are as follows (see
\S \ref{models}).  Pressure boundary condition: 1 parameter
($P_{\rm ref}$).  Entropy profile: 7 parameters ($s_0$, $s_1$,
$r_{\rm b,1}$, $r_{\rm b,2}$, $\alpha_1$, $\alpha_2$,
$\alpha_3$). Stellar mass of the central galaxy: 1 parameter
($\mlkband$).  DM profile: 2 parameters ($c_{\Delta}$,
$M_{\Delta}$). We display the best-fitting fiducial model to the
surface brightness and temperature profiles along with fractional
residuals in Figure~\ref{fig.best}. The fit is generally excellent
with almost all residuals within $\approx 1\sigma$ of the model
values. The most deviant data points are the last annuli for $\xsurf$
for each data set (i.e., \chandra\ annulus~10 and \suzaku\ annulus~3),
each of which lie $\approx 1.6\sigma$ away from the best-fitting
model.  To provide a straightforward quantitative measure of
goodness-of-fit we also perform a standard frequentist $\chi^2$
analysis. We obtain $\chi^2=9.7$ for 13 dof, which is formally
acceptable. (For reference, if the stellar mass component of the
central galaxy is omitted, $\chi^2=20.0$; i.e., the data require it at
$3\sigma$ according to the F-test. If instead the DM profile is
omitted, $\chi^2=285.2$, clearly demonstrating the failure of a
mass-follows-BCG light model.)

Since we shall often refer to the ``Best Fit'' virial radii of the
fiducial model, we list their values here: $\rtwofiveh=124\pm 4$~kpc,
$\rfiveh=236\pm 10$~kpc, and $\rtwoh=340\pm 15$~kpc. The extent of the
data is $\approx\rtwofiveh$ which we indicate in Figure~\ref{fig.best}.

\subsection{Entropy}
\label{entropy}

\begin{figure*}
\parbox{0.49\textwidth}{
\centerline{\includegraphics[scale=0.43,angle=0]{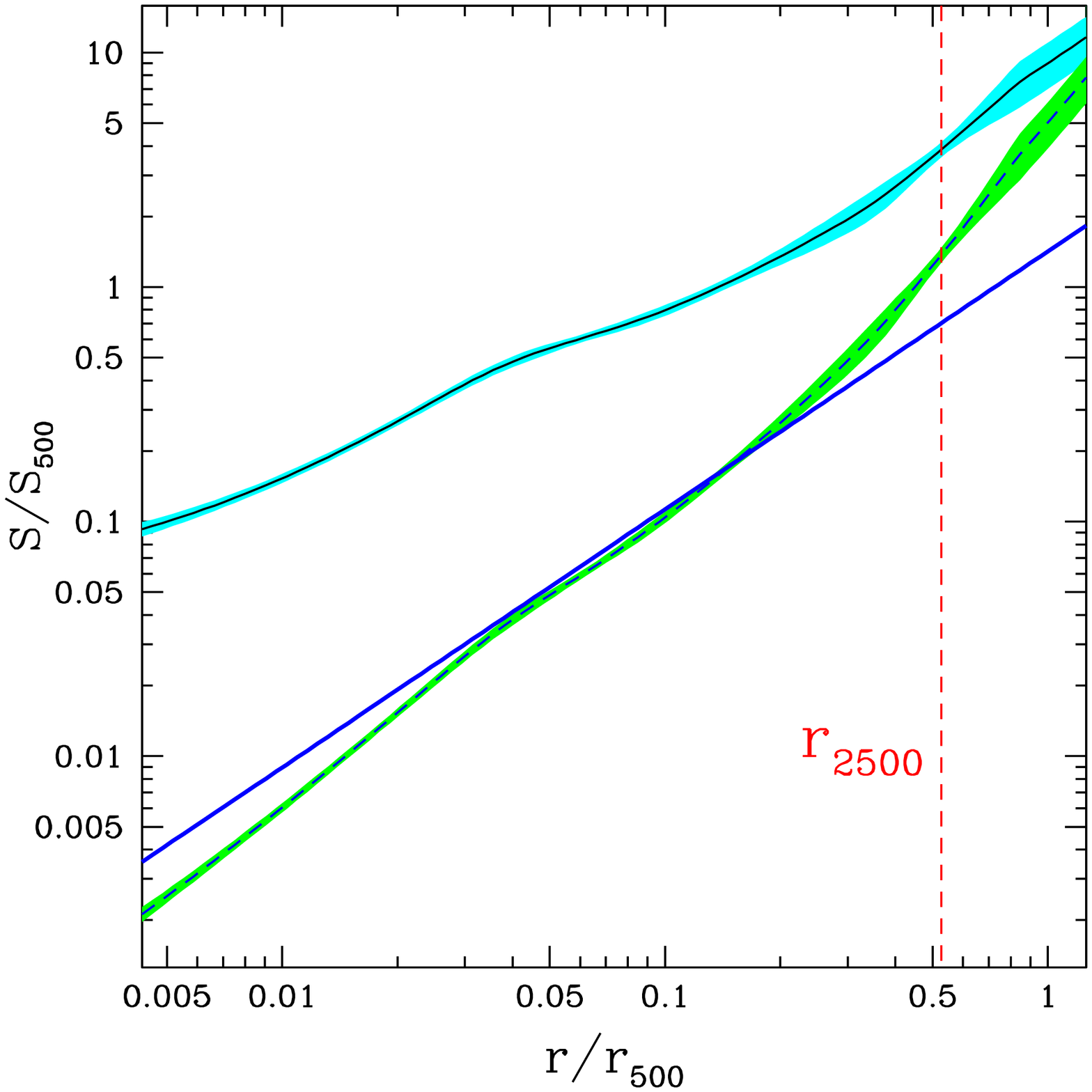}}}
\parbox{0.49\textwidth}{
\centerline{\includegraphics[scale=0.43,angle=0]{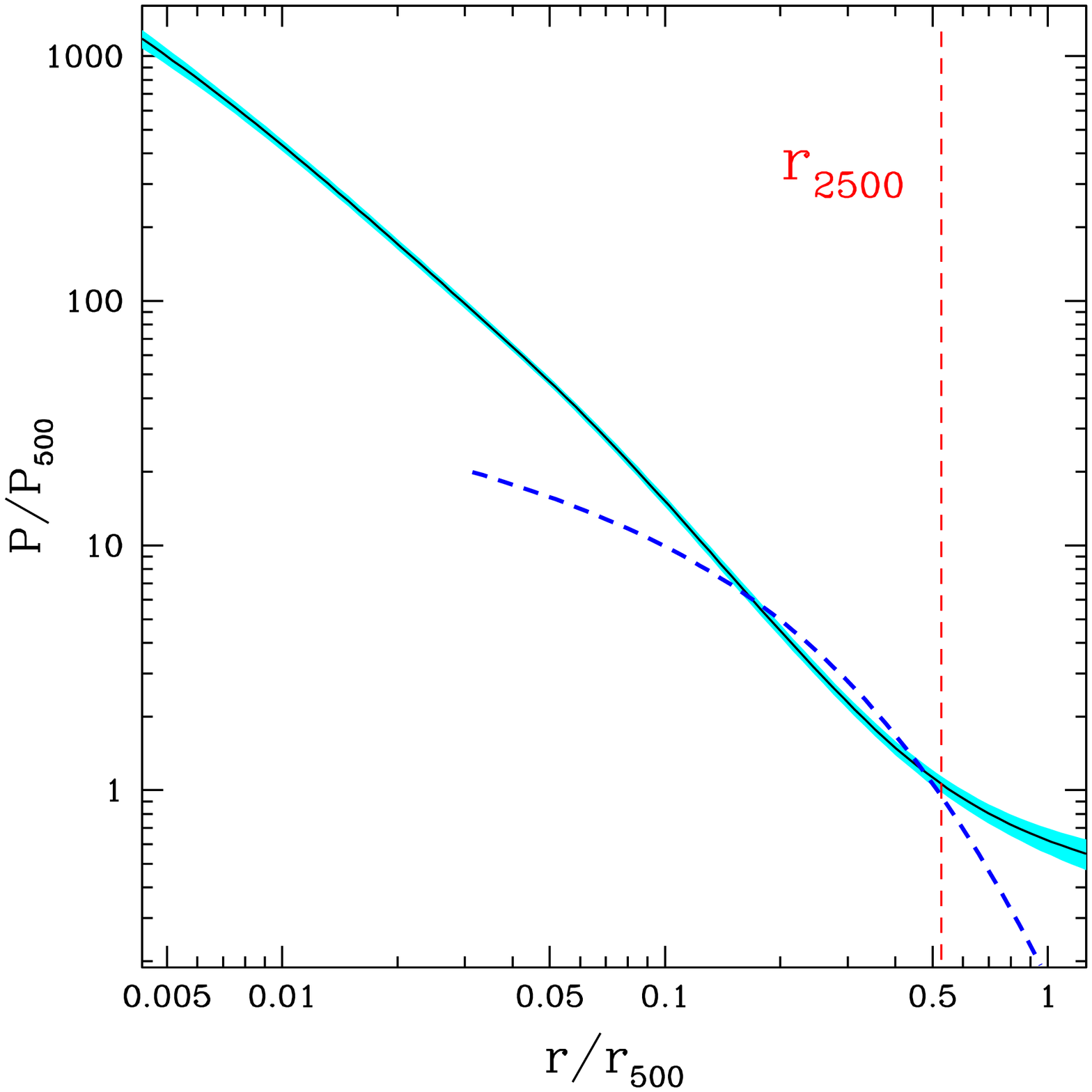}}}
\caption{\label{fig.entropy} ({\sl Left Panel}) Radial profile of the
  entropy (black) and $1\sigma$ error region (cyan) for the fiducial
  hydrostatic model rescaled by $S_{500}=44.7$~keV~cm$^2$. The
  baseline $r^{1.1}$ profile obtained by cosmological
  simulations~\citep{voit05a} with only gravity is shown as a blue
  line. The result of rescaling the entropy profile by $\propto f_{\rm
    gas}^{2/3}$ \citet{prat10a} is shown by the black dashed line (and
  green $1\sigma$ region). ({\sl Right Panel}) Radial profile of the
  gas pressure (black) and $1\sigma$ error region (cyan) rescaled by
  $P_{500} = 1.80\times 10^{-4}$~keV~cm$^{-3}$. For comparison we also
  show the universal profile of \citet{arna10a} derived for galaxy clusters.}
\end{figure*}

\begin{table*}[t] \footnotesize
\begin{center}
\caption{Pressure and Entropy}
\label{tab.entropy}
\begin{tabular}{lc|ccccccc}  \hline\hline\\[-7pt]
 & $P_{\rm ref}$ & $s_0$ & $s_1$ & $r_{\rm b,1}$ & $r_{\rm b,2}$ & $\alpha_1$ & $\alpha_2$ & $\alpha_3$\\ 
& $(10^{-1}$ keV cm$^{-3}$) & (keV cm$^2$) & (keV cm$^2$) & (kpc) & (kpc) \\ 
\hline \\[-7pt]
Best Fit & $2.18 \pm 0.19$ & $2.33 \pm 0.70$ & $1.76 \pm 0.55$ & $8.81 \pm 1.87$ & $  52 \pm   27$ & $1.11 \pm 0.16$ & $0.55 \pm 0.17$ & $1.38 \pm 0.41$ \\
(Max Like) & $(2.29)$ & $(1.77)$ & $(2.08)$ & $(10.24)$ & $(  23)$ & $(1.02)$ & $(0.34)$ & $(1.02)$ \\
\hline \\
\end{tabular}
\tablecomments{Best values and error estimates (see \S \ref{overview})
  for the free parameters of the pressure and entropy components of
  the fiducial hydrostatic equilibrium model. $P_{\rm ref}$ refers to
  the total gas pressure evaluated at the reference radius $r=1$~kpc and
  serves as the boundary condition for the hydrostatic model. The
  entropy parameters are defined in \S \ref{models.entropy}.}
\end{center}
\end{table*}

We show the entropy profile of the best-fitting fiducial hydrostatic
model in Figure~\ref{fig.entropy} and list the parameter measurements
in Table~\ref{tab.entropy}. In the figure we plot the entropy scaled
in terms of the quantity, $S_{500}=44.7$~keV~cm$^{2}$~\citep[see eqn.\
3 of][]{prat10a}. The two breaks in the entropy profile occur at the
locations of changes in slope of the temperature profile. The first
break occurs near $r=9$~kpc where $\ktemp$ drops sharply, while the
second break is near $r=50$~kpc where the temperature profile
inverts. Except for the region between the two break radii, where the entropy
slope flattens somewhat, the slope of the entropy profile is
consistent with that of the $\sim r^{1.1}$ baseline model. 

The entropy profile lies well above the baseline gravity-only model
illustrating the role of energy injection from feedback processes. For
comparison, we perform the scaling suggested by \citet{prat10a} where
the entropy profile is multiplied by $(f_{\rm gas}/f_{b,U})^{2/3}$,
where $f_{\rm gas}$ is the cumulative gas fraction and $f_{b,U}=0.155$
is the cosmic baryon fraction. The result is shown in
Figure~\ref{fig.entropy} and indicates much better agreement between
the rescaled entropy profile and the baseline profile interior to
$\approx\rtwofiveh$ and suggests that for $r\la \rtwofiveh$ the
feedback energy injected into \src\ has not raised the gas temperature
but instead spatially redistributed the gas. These basic results are
very consistent with those we have obtained previously for the massive
isolated elliptical galaxies NGC~720~\citep{hump11a} and
NGC~1521~\citep{hump12b} and results for galaxy
clusters~\citep[e.g.,][]{prat10a}. In \S \ref{entropy.disc} we discuss
the increasing deviation of the rescaled entropy profile from the
baseline model exterior to $\rtwofiveh.$

Finally, for reference we note that the second break in the entropy
profile is required statistically. Using a frequentist analysis, a
one-break model yields a minimum $\chi^2=24.2$ for 15 dof, so that the
second break is required at a significance of $\approx 3\sigma$
according to the F-test.

\subsection{Pressure}
\label{pressure}

We list the constraints on $P_{\rm ref}$ in Table~\ref{tab.entropy}
and plot the pressure profile for the fiducial hydrostatic model in
Figure~\ref{fig.entropy} scaled in terms of $P_{500}=1.80\times
10^{-4}$~keV~cm$^{-3}$~\citep[see eqn.\ 5 of][]{arna10a} expressed as
a total gas pressure (not electron pressure). For comparison we also
display the ``universal'' pressure profile of \citet{arna10a}
determined from an analysis of galaxy clusters having
$M_{500}>10^{14}\, M_{\odot}$.  The pressure profile of \src\ is
similar to (considering the intrinsic scatter) the universal profile
of clusters for radii approximately 0.1-0.7~$\rfiveh$, but exceeds it
elsewhere. Despite significant differences, the broad
agreement over a sizable range in radius is noteworthy since the
universal profile was calibrated for systems about 100 times more
massive than \src. The increasing deviations of the observed profile
from the universal one outside 0.1-0.7~$\rfiveh$, however, demonstrate a
significant breakdown of the mass scaling underlying the universal
profile and points to the increasing importance of non-gravitational
energy for group-scale halos.

\subsection{Mass}
\label{mass}

\begin{figure*}
\parbox{0.49\textwidth}{
\centerline{\includegraphics[scale=0.43,angle=0]{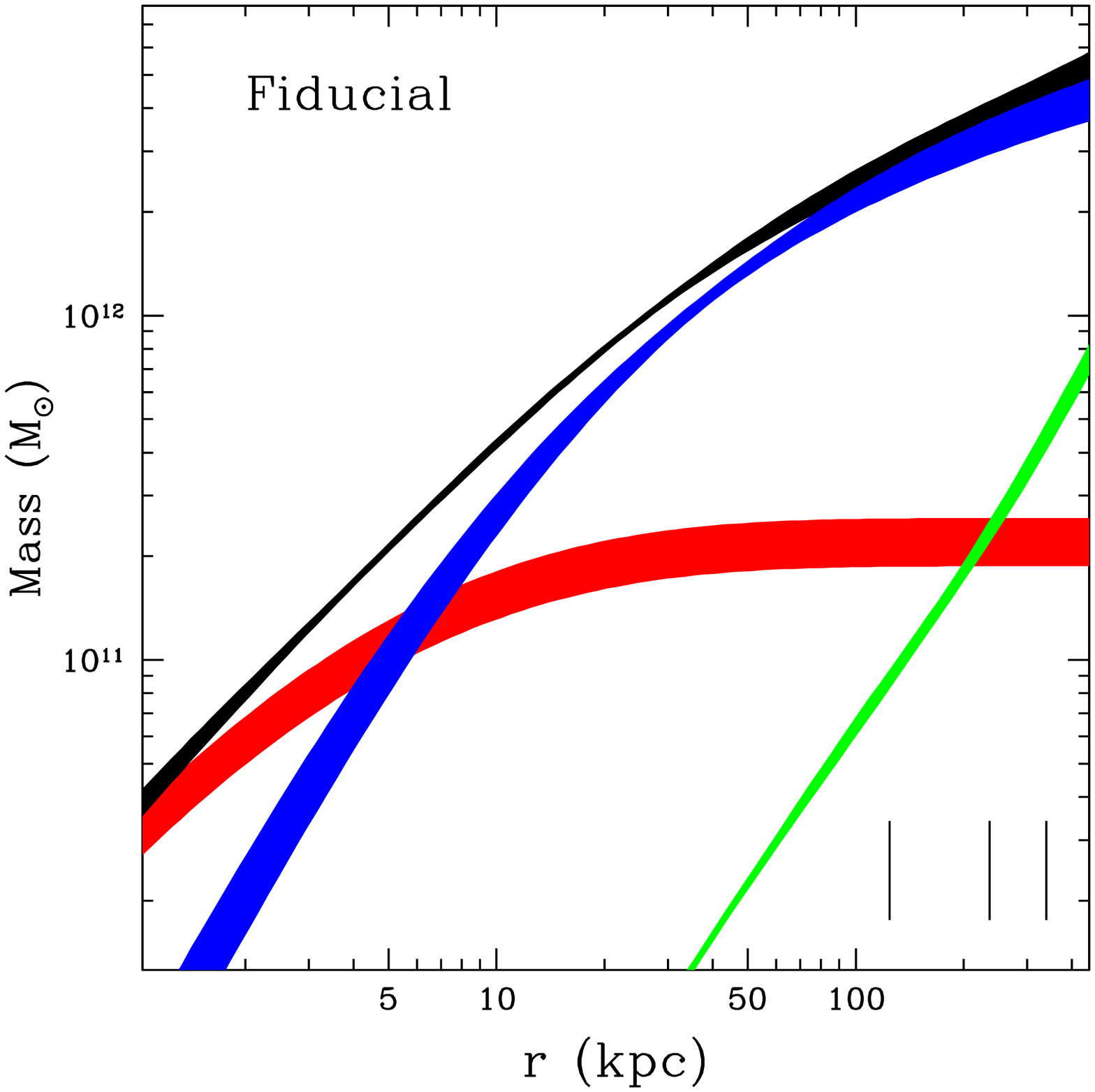}}}
\parbox{0.49\textwidth}{
\centerline{\includegraphics[scale=0.43,angle=0]{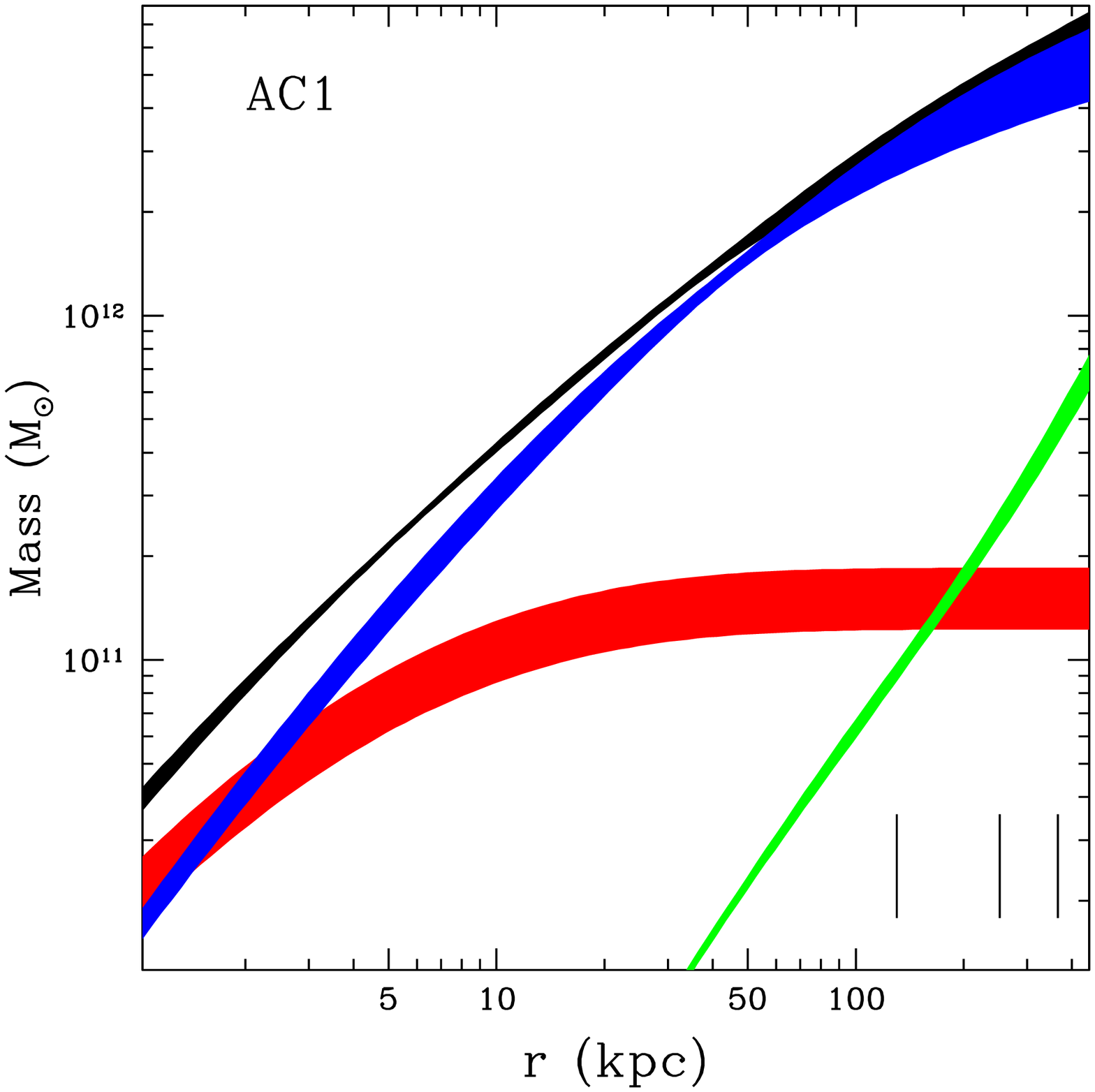}}}
\caption{\label{fig.mass}  ({\sl Left Panel}) Radial profiles of the total mass (black) and
individual mass components  of the fiducial hydrostatic model: total
 NFW DM (blue), stars (red), hot gas (green).  The black  vertical
 lines indicate the virial radii; i.e., from left to right: 
  $r_{2500}$, $r_{500}$, and $r_{200}$. ({\sl Right Panel}) Same
  quantities plotted as in the {\sl Left Panel} except for a
  hydrostatic model with an AC1 DM halo;
  i.e., an NFW DM halo modified by standard adiabatic contraction
  according to \citet{blum86a}.}
\end{figure*}

\begin{figure*}
\parbox{0.49\textwidth}{
\centerline{\includegraphics[scale=0.43,angle=0]{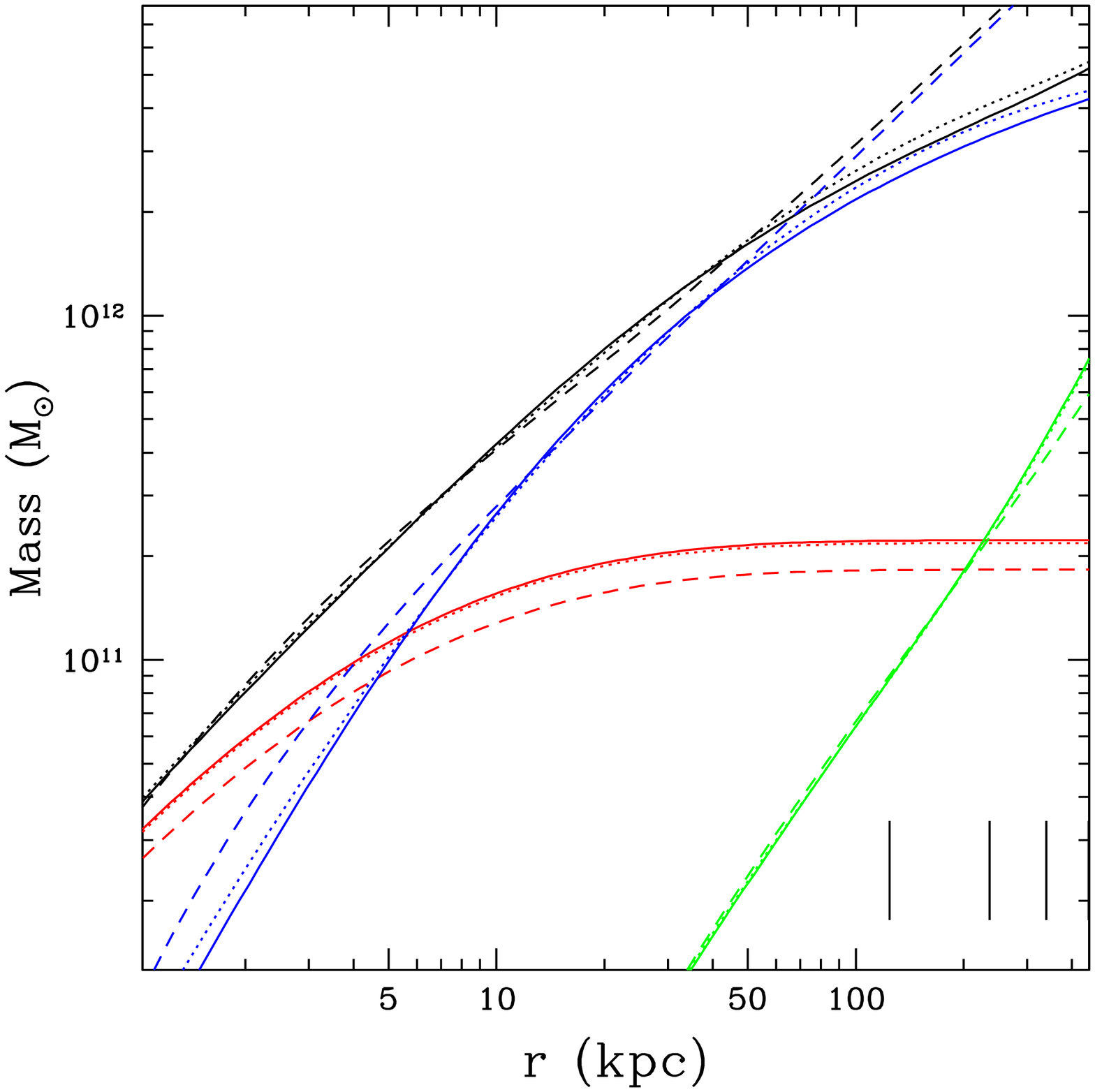}}}
\parbox{0.49\textwidth}{
\centerline{\includegraphics[scale=0.43,angle=0]{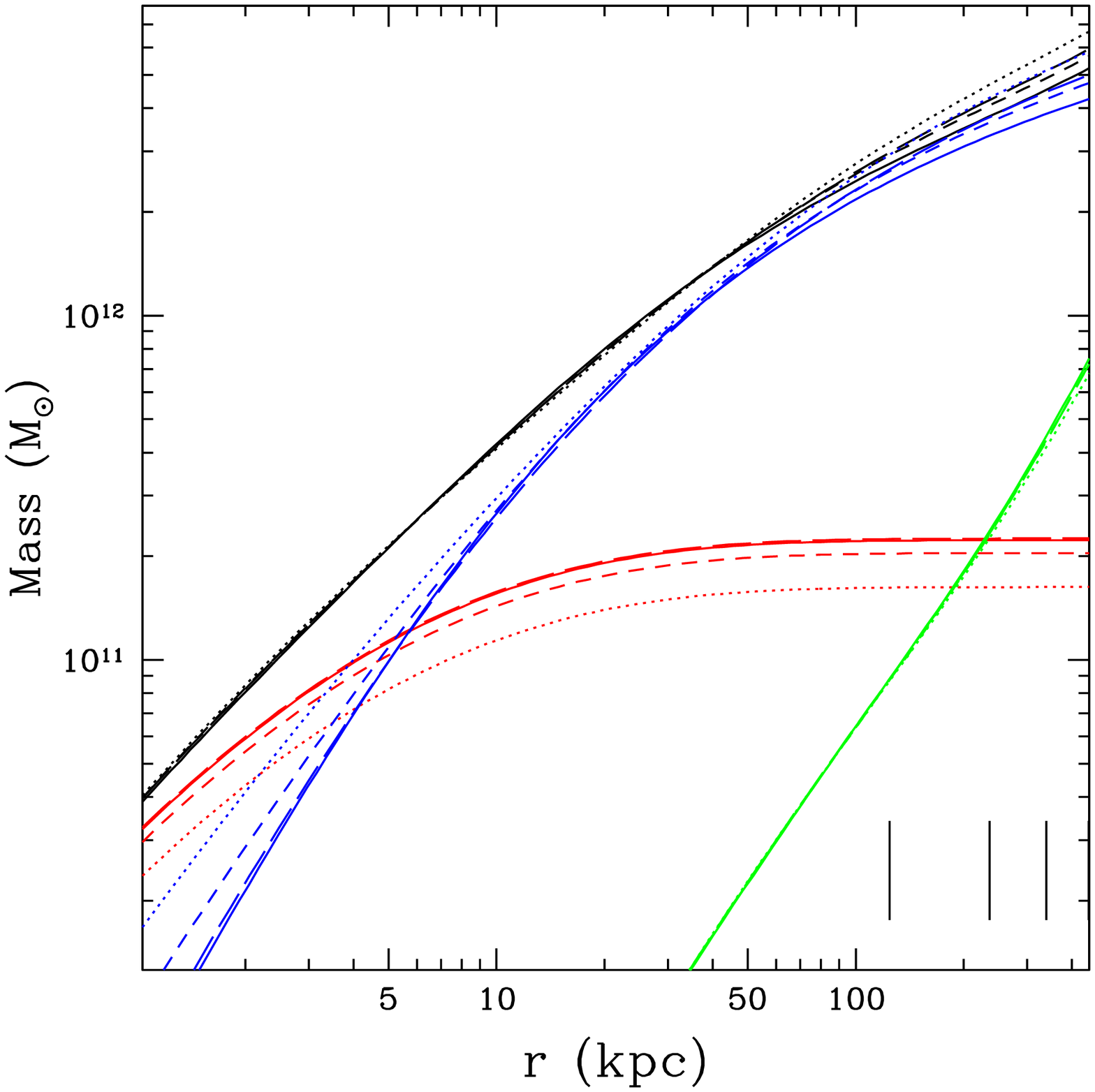}}}
\caption{\label{fig.mass_compare} Radial mass profiles for hydrostatic models having different DM profiles:
  NFW (solid lines), Einasto (dotted lines), CORELOG (dashed
  lines). We show only the best-fitting models for clarity. The color scheme is
  the same as for Figure~\ref{fig.mass}. The gas mass profiles are almost
  indistinguishable between the models. ({\sl Right Panel})  Same
  quantities plotted as in the {\sl Left Panel} except for hydrostatic
  models having adiabatically contracted DM halos: AC1 (dotted), AC3
  (short dash), AC4 (long dash). AC2 (not shown) is very
  similar to AC1. }
\end{figure*}

\begin{table*}[t] \footnotesize
\begin{center}
\caption{Stellar and Total Mass}
\label{tab.mass}
\begin{tabular}{lc|cc|cc|cc}  \hline\hline\\[-7pt]
& $M_{\star}/L_K$ & $c_{2500}$ & $M_{2500}$  & $c_{500}$ & $M_{500}$  & $c_{200}$ & $M_{200}$ \\ 
& ($M_{\odot} L_{\odot}^{-1}$) &  & $(10^{12}\, M_{\odot})$ &  & $(10^{12}\, M_{\odot})$ &  & $(10^{12}\, M_{\odot})$\\
\hline \\[-7pt]
Best Fit & $0.68 \pm 0.11$  & $11.8 \pm  2.7$ & $ 2.8 \pm  0.3$ & $22.4 \pm  5.0$ & $ 3.8 \pm  0.5$ & $32.2 \pm  7.1$ & $ 4.5 \pm  0.6$\\ 
(Max Like) & $(0.75)$ & $(10.3)$ & $( 2.9)$ & $(19.6)$ & $( 4.0)$ & $(28.2)$ & $( 4.8)$\\ 
\hline \\[-7pt]
Spherical & $\cdots$ & $\cdots$ & $\cdots$ & $_{-0.9}^{+0.4}$ & $_{-0.02}^{+0.07}$ & $\cdots$ & $\cdots$\\
Einasto & $-0.01$ & $-1.2$ & $ 0.3$ & $-2.4$ & $ 0.4$ & $-3.7$ & $ 0.4$ \\ 
CORELOG & $-0.12$ & $98.0$ & $ 1.8$ & $222.1$ & $ 6.2$ & $355.1$ & $11.4$ \\ 
AC & $-0.21$ & $-3.9$ & $ 0.4$ & $-7.2$ & $ 0.9$ & $-10.3$ & $ 1.2$ \\ 
BCG & $^{+0.21}_{-0.15}$ & $^{+ 1.0}_{-1.8}$ & $^{+ 0.1}_{-0.0}$ & $^{+ 1.9}_{-3.3}$ & $^{+ 0.2}_{-0.1}$ & $^{+ 2.8}_{-4.8}$ & $^{+ 0.3}_{-0.1}$ \\ 
Entropy & $0.05$ & $-1.4$ & $ 0.1$ & $-2.6$ & $ 0.2$ & $-3.7$ & $ 0.3$ \\ 
Proj. Limit & $0.03$ & $-0.7$ & $ 0.1$ & $-1.4$ & $ 0.1$ & $-2.0$ & $ 0.2$ \\ 
Distance & $-0.06$ & $-0.4$ & $ 0.1$ & $-0.7$ & $ 0.2$ & $-1.0$ & $ 0.2$ \\ 
Frequentist & $0.04$ & $-1.0$ & $ 0.1$ & $-1.9$ & $ 0.1$ & $-2.6$ & $ 0.2$ \\ 
$\Lambda_{\nu}(T,Z)$ & $0.04$ & $ 2.0$ & $-0.4$ & $ 3.7$ & $-0.6$ & $ 5.6$ & $-0.7$ \\ 
No Suzaku & $-0.08$ & $ 4.3$ & $-0.5$ & $ 8.0$ & $-0.8$ & $11.9$ & $-0.9$ \\ 
Exclude Last Bin & $0.03$ & $-1.3$ & $ 0.5$ & $-2.5$ & $ 0.8$ & $-3.6$ & $ 0.9$ \\ 
SWCX & $0.01$ & $ 0.1$ & $-0.1$ & $ 0.3$ & $-0.1$ & $ 0.5$ & $-0.1$ \\ 
CXBSLOPE & $^{+0.07}_{-0.02}$ & $^{+ 1.3}_{-2.0}$ & $^{+ 0.4}_{-0.3}$ & $^{+ 2.4}_{-3.7}$ & $^{+ 0.5}_{-0.4}$ & $^{+ 3.6}_{-5.4}$ & $^{+ 0.6}_{-0.4}$ \\ 
NXB & $-0.03$ & $ 1.4$ & $-0.2$ & $ 2.6$ & $-0.3$ & $ 3.7$ & $-0.4$ \\ 
$\chi^2$ & $-0.03$ & $ 1.1$ & $-0.2$ & $ 2.0$ & $-0.3$ & $ 2.8$ & $-0.3$ \\ 
$N_{\rm H}$ & $0.02$ & $-0.3$ & $^{+ 0.0}_{-0.0}$ & $-0.5$ & $^{+ 0.0}_{-0.0}$ & $-0.7$ & $^{+ 0.1}_{-0.0}$ \\ 
LMXBs & $0.06$ & $-1.8$ & $ 0.1$ & $-3.3$ & $ 0.2$ & $-4.7$ & $ 0.3$ \\ 
Solar Abun. & $^{+0.01}_{-0.01}$ & $^{+ 0.3}_{-0.3}$ & $^{+ 0.0}_{-0.1}$ & $^{+ 0.5}_{-0.6}$ & $^{+ 0.1}_{-0.1}$ & $^{+ 0.7}_{-0.9}$ & $^{+ 0.1}_{-0.1}$ \\ 
Other Abun & $0.06$ & $-1.6$ & $-0.0$ & $-3.0$ & $ 0.0$ & $-4.3$ & $ 0.0$ \\ 
$1T$ Ann~1 Suzaku & $-0.01$ & $ 0.9$ & $-0.2$ & $ 1.6$ & $-0.3$ & $ 2.2$ & $-0.4$ \\ 
\\ 
\hline \\
\end{tabular}
\tablecomments{Best values and error estimates (see \S \ref{overview})
  for the free parameters of the mass components of the fiducial
  hydrostatic equilibrium model; i.e., stellar mass-to-light ratio
  $(M_{\star}/L_K)$, concentration, and enclosed total mass
  (stars+gas+DM). We show the concentration and mass results obtained
  within radii $r_{\Delta}$ for overdensities $\Delta=200, 500, 2500.$
  In addition, we give a detailed budget of systematic errors (\S
  \ref{sys}). For each column we quote values with the same
  precision. In a few cases, an error has a value smaller than the
  quoted precision, and thus it is listed as a zero; e.g., ``0.0'' or
  ``-0.0''. Briefly, the various systematic errors are as follows,
  \\
  (``Spherical'', \S \ref{sys.sphere}): spherical symmetry assumption\\
  (``Einasto'', ``CORELOG'', ``AC''\S \ref{sys.dm}): Using different
  DM profiles: Einasto, CORELOG, or adiabatically contracted models\\
  (``BCG'', \S \ref{sys.stars}): Allowing for a range of $R_e$ for the BCG\\
  (``Entropy'', \S \ref{sys.entropy}): Variations of the entropy model\\
  (``Proj.\ Limit''), \S \ref{mischm}): Varying the adopted outer
  radius for computing model projections\\
  (``Distance'',  \S \ref{mischm}): Using a different distance\\
  (``Frequentist'', \S \ref{mischm}): Fit the hydrostatic equilibrium
  models using a standard $\chi^2$ frequentist approach\\
  (``$\Lambda_{\nu}(T,Z)$'', \S \ref{mischm}): Using a parameterized
  model for the plasma emissivity\\
  (``No Suzaku'', \S \ref{mischm}): Omitting the \suzaku\ data from
  the hydrostatic equilibrium model fits \\
  (``Exclude Last Bin'', \S \ref{mischm}): Omitting the outer spectral
  bin from both the \chandra\ and \suzaku\ data in
  the hydrostatic equilibrium model fits \\
  (``SWCX'', \S \ref{swcx}): Solar Wind Charge Exchange emission\\
  (``CXBSLOPE'', \S \ref{sys.bkg}): Varying the CXB power-law slope\\
  (``NXB'', \S \ref{sys.bkg}): Varying the normalization of the  non-X-ray background model for the \suzaku\ data\\
  (``$\chi^2$'', \S \ref{misc.spec}): Minimizing $\chi^2$ rather than
  the C-statistic for the spectral fitting\\
  (``$N_{\rm H}$'', \ref{misc.spec}): Varying the Galactic hydrogen
  column density\\
  (``LMXB'', \ref{misc.spec}): Varying the amount of emission from
  LMXBs in the spectral fitting\\
  (``Solar Abun.\", \S \ref{misc.spec}):  Using different solar abundance tables\\
  (``Other Abun.\", \S \ref{misc.spec}): Allowing some abundance to
  vary with a non-solar ratio to iron\\
  (``$1T$ Ann~1 Suzaku", \S \ref{misc.spec}): Using only a single
  temperature to model the spectrum of the central \suzaku\ annulus}
\end{center}
\end{table*}
\renewcommand{\arraystretch}{1}

\begin{table*}[t] \footnotesize
\begin{center}
\caption{Results for Adiabatic Contraction Models}
\label{tab.custom}
\begin{tabular}{r|cc|cc|cc|cc|cc}   \hline\hline\\[-7pt]
  & \multicolumn{2}{c}{$M_{\star}/L_{\rm K}$} & \multicolumn{2}{c}{$c_{ 200}$} & \multicolumn{2}{c}{$M_{200}$} & \multicolumn{2}{c}{$f_{\rm gas, 200}$} & \multicolumn{2}{c}{$f_{\rm b, 200}$}\\
  & \multicolumn{2}{c}{$(M_{\odot}L_{\odot}^{-1})$} & \multicolumn{2}{c}{} & \multicolumn{2}{c}{$(10^{12}\, M_{\odot})$} & \multicolumn{2}{c}{} & \multicolumn{2}{c}{}\\
 Model  & Best & 99\% & Best & 99\% & Best & 99\% & Best & 99\% & Best & 99\%\\
\hline \\[-7pt]
              Fiducial   &   0.68    & $(- 0.28,+ 0.22)$ &  32.2    & $(- 13.9,+ 18.7)$ &   4.5    & $(-  1.1,+  2.0)$ & 0.099    & $(-0.030,+0.028)$ & 0.149    & $(-0.038,+0.033)$\\ 
         AC1   &   0.49    & $(- 0.17,+ 0.22)$ &  19.6    & $(- 12.8,+ 13.9)$ &   6.1    & $(-  2.1,+  7.2)$ & 0.082    & $(-0.043,+0.030)$ & 0.110    & $(-0.053,+0.034)$\\ 
         AC2   &   0.51    & $(- 0.21,+ 0.25)$ &  22.6    & $(- 15.2,+ 18.5)$ &   5.7    & $(-  1.9,+  6.8)$ & 0.086    & $(-0.044,+0.030)$ & 0.117    & $(-0.055,+0.033)$\\ 
         AC3   &   0.62    & $(- 0.20,+ 0.25)$ &  28.4    & $(- 15.9,+ 15.9)$ &   5.0    & $(-  1.3,+  3.8)$ & 0.093    & $(-0.037,+0.027)$ & 0.134    & $(-0.048,+0.032)$\\ 
         AC4   &   0.68    & $(- 0.25,+ 0.26)$ &  23.0    & $(- 12.9,+ 14.8)$ &   5.3    & $(-  1.5,+  4.1)$ & 0.091    & $(-0.038,+0.028)$ & 0.135    & $(-0.050,+0.035)$\\ 
\hline \\
      Einasto   &   0.66    & $(- 0.28,+ 0.26)$ &  28.5    & $(- 16.0,+ 18.4)$ &   4.9    & $(-  1.6,+  4.3)$ & 0.095    & $(-0.040,+0.031)$ & 0.140    & $(-0.053,+0.038)$\\ 
 AC1   &   0.48    & $(- 0.17,+ 0.20)$ &  16.9    & $(- 11.1,+ 14.3)$ &   7.3    & $(-  3.2,+  8.2)$ & 0.077    & $(-0.042,+0.034)$ & 0.100    & $(-0.052,+0.040)$\\ 
 AC2   &   0.52    & $(- 0.18,+ 0.22)$ &  18.0    & $(- 12.1,+ 14.2)$ &   7.0    & $(-  3.0,+  8.4)$ & 0.079    & $(-0.044,+0.033)$ & 0.104    & $(-0.054,+0.039)$\\ 
 AC3   &   0.60    & $(- 0.22,+ 0.27)$ &  26.1    & $(- 17.2,+ 17.8)$ &   5.4    & $(-  2.0,+  6.8)$ & 0.090    & $(-0.046,+0.031)$ & 0.127    & $(-0.061,+0.038)$\\ 
 AC4   &   0.66    & $(- 0.25,+ 0.27)$ &  20.2    & $(- 12.6,+ 14.7)$ &   6.1    & $(-  2.2,+  6.4)$ & 0.085    & $(-0.041,+0.032)$ & 0.122    & $(-0.055,+0.039)$\\ 
\hline \\
\end{tabular}
\tablecomments{Best-fitting results and 99\% confidence limits for
  selected mass parameters of the fiducial hydrostatic model, the
  model with an Einasto DM halo, and the corresponding AC models. The
  definitions of the AC models are given in Table~\ref{tab.ac.defs}.}
\end{center}
\end{table*}

We list in Table~\ref{tab.mass} the constraints on the mass parameters
$(\mlkband, c_{\Delta}, M_{\Delta})$ obtained for the fiducial
hydrostatic model along with estimates for systematic errors (\S
\ref{sys}). We also plot the mass profiles for all sub-components in
Figure~\ref{fig.mass}.  We quote the DM halo parameters evaluated for
three overdensities, $\Delta=200,500,2500$.  Our hydrostatic models
self-consistently evaluate the projection of gas for the entire
system, and thus the data do constrain the gas for $r\ga \rtwofiveh$
albeit with less statistical precision and increased model dependence.

Considering only the statistical errors, the most weakly constrained
mass parameter is the halo concentration (e.g., $\pm 22-23\%$),
followed by the stellar mass $(\pm 16\%)$, and finally the total halo
mass ($\pm 11-13\%$). The gas mass profile is the most tightly
constrained ($\pm 6-12\%$).  Most of the systematic errors induce
parameter changes of the same size or smaller than the $1\sigma$
statistical error and are therefore insignificant. The most
significant changes in the parameters occur for the following
systematic tests: $\mlkband$ (BCG), concentration (No Suzaku), total
mass (No Suzaku, Exclude Last Bin, $\Lambda_{\nu}(T,Z)$,
CXBSLOPE). Since these largest effects lead to parameter changes of
only $\la 1.5\sigma$, we conclude that systematic errors, though
important in some cases, do not dominate the error budget.

The stellar mass-to-light ratio of the BCG that we measure
($\mlkband=0.68 \pm 0.11$, in solar units) agrees very well with our
previous X-ray hydrostatic measurement~(H06). However, this value of
$\mlkband$ lies below the estimates from the SPS models (\S
\ref{stars}) by $1.9\sigma$ (Chabrier IMF), $3.4\sigma$ (Kroupa IMF),
and $8.2\sigma$ (Salpeter) considering only the statistical error on
the hydrostatic measurement. If the systematic error associated with
the choice of the BCG effective radius is considered (i.e., BCG in
Table~\ref{tab.mass}), Then the best-fitting value rises to
$\mlkband=0.89$ solar, which agrees very well with the SPS value for a
Chabrier IMF and reasonably well for the Kroupa IMF. Hence, the
hydrostatic analysis strongly favors SPS models with a Chabrier or
Kroupa IMF over Salpeter, an issue to which we return in \S \ref{he}.

The halo concentration and mass were previously measured from
hydrostatic X-ray studies by \citet{khos04a} and H06. \citet{khos04a}
estimated $M_{200}\approx 4\times 10^{12}\, M_{\odot}$, in good
agreement with our result. But their estimate of $c_{200}\sim 60$ far
exceeds our value, as expected, because they neglected the mass of the
BCG. In our previous analysis of H06 we did include the BCG but placed
{\it ad hoc} restrictions on $f_{\rm b,200}$. For the case where the
baryon fraction was restricted to the range
$0.032\le f_{\rm b,200} \le 0.16$, the fitted baryon fraction pegged
at the upper limit $(f_{\rm b,200}=0.16^{+0.00}_{-0.10})$ so that
$c_{\rm vir}=38_{-24}^{+76}$ and
$M_{\rm vir} = 3.6_{-1.5}^{+5.5}\times 10^{12}\, M_{\odot}$ computed
within $r_{103}$ (90\% conf; see Table 4 of H06). If in our present
study we use a similar radius (actually $r_{102.3}$) we obtain
$c_{\rm vir}=42.2_{ -13.9}^{+16.7}$ and
$M_{\rm vir} = 5.2_{-1.0}^{+1.4} \times 10^{12}\, M_{\odot}$ (also
90\% conf.) fully consistent with the very uncertain results of
H06. However, unlike these previous measurements, which were extremely
uncertain (no error bar quoted by \citealt{khos04a}), here we have
made a precise measurement revealing a high concentration that
includes the BCG stellar mass and does not place {\it ad hoc}
restrictions on the baryon fraction.

While our new measurement of $c_{200}=32.2\pm 7.1$ is about half the
value obtained by \citet{khos04a} that neglected the stellar mass of
the BCG, it is nevertheless very large when compared to simulated
\lcdm\ halos.  For a ``relaxed'' DM halo with
$M_{200}=4.5\times 10^{12}\, M_{\odot}$ in the \planck\ cosmology
\citet{dutt14a} obtain $c_{200}=7.1$ with a log-normal intrinsic
scatter of 0.11 dex.  (In \S \ref{c-m} we discuss the values obtained
from other theoretical $c_{200}-M_{200}$ relations.)  Our measurement
lies $6\sigma$ above the mean \lcdm\ relation in terms of the
intrinsic scatter and $3.5\sigma$ above it in terms of the statistical
error on the hydrostatic measurement. Such an extreme outlier would be
difficult to reconcile with the \lcdm\ simulations, and thus we
examine whether modifications of the fiducial NFW DM halo can reduce
significantly the inferred concentration.  If instead we use the
Einasto profile for the DM halo, we obtain $c_{200}=28.5\pm 7.1$ and
$M_{200}=(4.9\pm 1.1)\times 10^{12}\, M_{\odot}$. For this mass and an
Einasto profile, \citet{dutt14a} obtain $c_{200}=8.1$ with an
intrinsic scatter of 0.13 dex for \lcdm\ halos.  While somewhat less
extreme of an outlier, our measurement using the Einasto profile still
lies $4.2\sigma$ above the mean \lcdm\ relation in terms of the
intrinsic scatter and $2.9\sigma$ above it in terms of the statistical
measurement error. Below we consider the AC models and especially
their effect on the concentration and discuss the implications of the
high concentration in \S \ref{highc}.

We compare results for the fiducial (NFW), Einasto, and
pseudo-isothermal CORELOG models in Figure~\ref{fig.mass_compare}. For
all mass components over all radii plotted the NFW and Einasto models
are extremely similar. The largest deviations between the two models
occur only at the very smallest and largest radii plotted. The CORELOG
model, however, displays substantial differences from the fiducial
model. First, the stellar mass for CORELOG is lower (i.e., about
$1\sigma$) than NFW. Second, the DM exceeds the NFW profile both at
small ($\la 10$~kpc) and large radii ($\ga 50$~kpc). Finally, the
CORELOG total mass also exceeds the fiducial model for radii $\ga
50$~kpc. In good agreement with what we found previously for the
fossil cluster RXJ~1159+5531~\citep{buot16a}, the gas mass is very
similar for all the models and the total mass ``pinches'' near 1-2
$R_e$ where the DM crosses over the stellar mass of the BCG so that
the total mass is nearly identical for all the models there.  Despite
some large differences in the profiles of the different mass
components, when we perform frequentist fits we obtain minimum
$\chi^2=9.7,9.7,9.4$ for 13 dof respectively for the NFW, Einasto, and
CORELOG models; i.e., the X-ray data do not statistically distinguish
the quality of the fits between the models.

Like the Einasto and CORELOG models, frequentist fits of the AC models
with an NFW DM halo give minimum $\chi^2$ values virtually the same as
the fiducial model; i.e., the quality of the fits of the AC models is
also statistically indistinguishable from the fiducial model. In the
right panel of Figure~\ref{fig.mass} we plot the mass profiles of the
various components for AC1. The most notable features are the higher
mass of DM within 1-2~$R_e$ and lower overall stellar mass compared to
the fiducial model. These features are strongest for AC1 and AC2
compared to the other AC models.  We compare the best-fitting mass
profiles of the fiducial and AC models in
Figure~\ref{fig.mass_compare}.  Notice in particular that AC4 yields
stellar mass and central DM mass profiles very similar to the fiducial
model, but the DM and total mass of AC4 exceed those of the fiducial
model at large radius.

In Table~\ref{tab.custom} we list the best-fitting results for
$\mlkband$, $c_{200}$, and $M_{200}$ from the bayesian analysis along
with their 99\% confidence limits defined as the 0.5\% and 99.5\%
values of the cumulative distributions obtained from the
posterior. Generally, compared to the fiducial un-contracted model,
the AC models give smaller $\mlkband$, smaller $c_{200}$, and larger
$M_{200}$.  The smaller concentrations are expected since they refer
to the scale radius of the un-contracted NFW profile~(\S \ref{ac});
i.e., AC converts an initially lower concentration halo into a higher
concentration halo, the latter of which is similar to what we measured
above using the pure NFW profile.  The AC1 and AC2 models give
$c_{200}$ values that are $\approx 10$ less than the fiducial model
but still $>4\sigma$ larger than expected for \lcdm\ halos in terms of
the intrinsic scatter. Moreover, their much smaller $\mlkband$ values
are difficult to reconcile with the SPS models for any IMF even
considering reasonable systematic error associated with $R_e$. AC3
yields a more intermediate value for $\mlkband$, but its $c_{200} $ is
close to the fiducial value. AC4, however, gives $\mlkband$ the same
as the fiducial value while also yielding a smaller $c_{200}$ that is
a little less extreme of an outlier from the mean \lcdm\ relation.  As
shown in Table~\ref{tab.custom}, these results for the NFW DM halo
(un-modified and AC) agree extremely well with those obtained for the
Einasto DM halo (un-modified and AC). We discuss further the
implications of the high concentration and AC models in \S
\ref{highc}.

\subsection{Mass and Density Slopes}
\label{slope}

\begin{figure}[t]
\begin{center}
\includegraphics[scale=0.42]{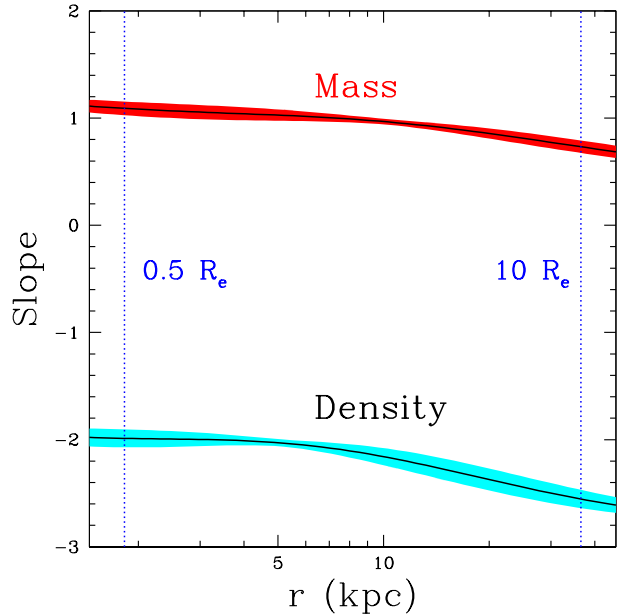}
\end{center}
\caption{\footnotesize The radial profiles of the radial logarithmic
  derivatives (i.e., slopes) and $1\sigma$ error regions of the total
  mass and total mass density for the fiducial hydrostatic model. }
\label{fig.slope}
\end{figure}

\begin{table}[t] \footnotesize
\begin{center}
\caption{Mass-Weighted Total Density Slope}
\label{tab.slope}
\begin{tabular}{rrc}   \hline\hline\\[-7pt]
Radius & Radius\\
(kpc) & ($R_e$) & $\langle\alpha\rangle$ \\
\hline \\[-7pt]
  1.8 &   0.5 & $ 1.91 \pm  0.06$ \\
  3.6 &   1.0 & $ 1.95 \pm  0.06$ \\
  7.3 &   2.0 & $ 2.00 \pm  0.04$ \\
 18.2 &   5.0 & $ 2.13 \pm  0.04$ \\
 36.5 &  10.0 & $ 2.27 \pm  0.06$ \\
\hline \\
\end{tabular}
\tablecomments{These values refer to the application of
  Equation~(\ref{eqn.slope}) to the fiducial hydrostatic model.}
\end{center}
\end{table}

The total mass density profiles of massive elliptical galaxies are
described accurately by $\rho\sim r^{-\alpha}$ with $\alpha\approx 2$
over a wide radial range; see, e.g., reviews of the evidence from
X-ray hydrostatic equilibrium studies by \citet{hump10a} and from
stellar dynamics and lensing by \citet{cour14a}. It is now known
this power-law relation extends to central BCGs in groups and clusters
but with smaller $\alpha$~\citep[e.g.,][and references
therein]{hump10a,newm13a,cour14a,capp15a}. The slope correlates with
various structural properties including the stellar half-light radius
of the BCG.  Using a toy model of a power-law total mass profile
decomposed approximately into separate BCG stellar and NFW DM halos,
\citet{hump10a} found that $\alpha$ and $R_e$ obey the following
relation over radii 0.2-10$~R_e$, $\alpha=2.31 - 0.54\log (R_e/\rm
kpc),$ in agreement with the relation obtained by~\citet{auge10a}
for a larger galaxy sample analyzed with stellar dynamics and
gravitational lensing.

We show in Figure \ref{fig.slope} the slopes (i.e., radial logarithmic
derivatives) of the radial profiles of the total mass and total mass
density.  Within $\approx 2R_e$ the slopes of the mass and
density are very slowly varying, and then they steepen more quickly at
larger radii.  Over radii 0.5-10~$R_e$ the slopes range from $\approx
0.73-1.09$ for the mass to $\approx$ -2.0 to -2.55 for the density,
representing a radial density variation of almost $30\%$, with most of
the variation occurring for radii $\ga 2R_e$. Rather than fitting a
power-law to the mass profile to obtain an average value for the
slope, it is convenient to compute the mass-weighted total density
slope $\langle\alpha\rangle$ of our fiducial model following equation (2) of
\citet{dutt14b},
\begin{equation}
\langle\alpha\rangle = 3 - \frac{d\ln M}{d\ln r}, \hskip 0.4cm
\alpha\equiv -\frac{d\ln \rho}{d\ln r} \label{eqn.slope},
\end{equation}
where $M$ is the total mass enclosed within radius $r$. We list the
mass-weighted slopes for selected radii between
0.5-10~$R_e$ in Table~\ref{tab.slope}. The mass-weighted slope within
$10R_e$, $\langle\alpha\rangle = 2.27\pm 0.06$, is $\approx 12\%$
larger than the mean value $\alpha=2.02$ obtained using the
$\alpha-R_e$ scaling relation quoted above but is consistent within the
observed scatter~\citep{auge10a}.

\subsection{Gas and Baryon Fraction}
\label{baryfrac}

\begin{figure}[t]
\begin{center}
\includegraphics[scale=0.42]{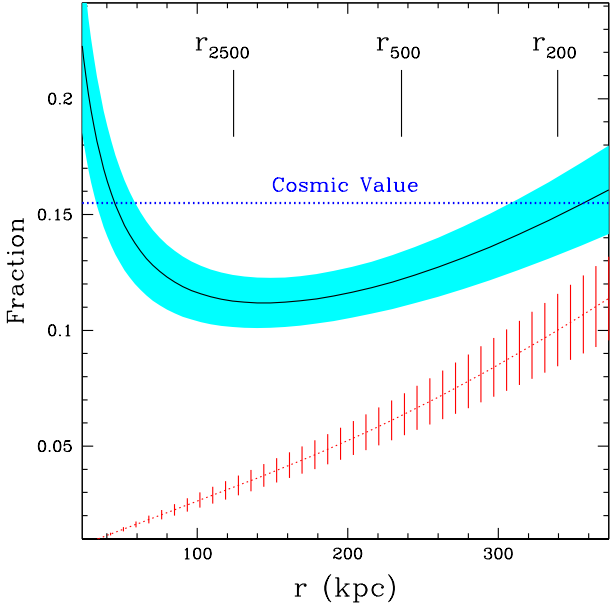}
\end{center}
\caption{\footnotesize Baryon fraction (solid black line, shaded cyan
  $1\sigma$ error region) and gas fraction (dotted red line, shaded
  red $1\sigma$ error region) of the fiducial hydrostatic model.}
\label{fig.baryfrac}
\end{figure}

\begin{table*}[t] \footnotesize
\begin{center}
\caption{Gas and Baryon Fraction}
\label{tab.fb}
\begin{tabular}{lcc|cc|cc}   \hline\hline\\[-7pt]
& $f_{\rm gas, 2500}$ & $f_{\rm b, 2500}$ & $f_{\rm gas, 500}$ & $f_{\rm b, 500}$ & $f_{\rm gas, 200}$ & $f_{\rm b, 200}$\\
\hline \\[-7pt]
Best Fit 
 & $0.032 \pm 0.003$ & $0.112 \pm 0.011$ & $0.063 \pm 0.006$ & $0.122 \pm 0.011$ & $0.099 \pm 0.011$ & $0.149 \pm 0.014$\\ 
(Max Like) 
 & $(0.030)$ & $(0.115)$ & $(0.062)$ & $(0.124)$ & $(0.099)$ & $(0.151)$\\ 
\hline \\[-7pt]
$M_{\rm other}^{\rm stellar}$ & $\cdots$ & $^{+0.046}$ & $\cdots$ & $^{+0.045}$ & $\cdots$ & $^{+0.043}$\\
Spherical & $\cdots$ & $\cdots$ & $_{-0.0003}^{+0.0003}$ & $\cdots$ & $\cdots$ & $\cdots$\\
Einasto & $-0.002$ & $-0.010$ & $-0.003$ & $-0.009$ & $-0.004$ & $-0.008$ \\ 
CORELOG & $-0.007$ & $-0.046$ & $-0.026$ & $-0.066$ & $-0.052$ & $-0.090$ \\ 
AC & $-0.003$ & $-0.035$ & $-0.008$ & $-0.033$ & $-0.014$ & $-0.036$ \\ 
BCG & $^{+0.000}_{-0.001}$ & $^{+0.020}_{-0.018}$ & $^{+0.000}_{-0.002}$ & $^{+0.013}_{-0.013}$ & $^{+0.001}_{-0.003}$ & $^{+0.009}_{-0.010}$ \\ 
Entropy & $^{+0.000}_{-0.001}$ & $0.001$ & $^{+0.000}_{-0.004}$ & $^{+0.001}_{-0.003}$ & $-0.007$ & $-0.006$ \\ 
Proj. Limit & $^{+0.000}_{-0.001}$ & $0.001$ & $^{+0.000}_{-0.001}$ & $^{+0.001}_{-0.001}$ & $^{+0.001}_{-0.003}$ & $^{+0.001}_{-0.003}$ \\ 
Distance & $0.001$ & $-0.001$ & $0.002$ & $-0.000$ & $0.003$ & $0.001$ \\ 
Frequentist & $-0.002$ & $0.001$ & $0.001$ & $0.002$ & $0.003$ & $0.004$ \\ 
$\Lambda_{\nu}(T,Z)$ & $0.006$ & $0.028$ & $0.025$ & $0.041$ & $0.046$ & $0.058$ \\ 
No Suzaku & $0.011$ & $0.018$ & $0.032$ & $0.037$ & $0.056$ & $0.060$ \\ 
Exclude Last Bin & $-0.006$ & $-0.011$ & $-0.012$ & $-0.017$ & $-0.019$ & $-0.023$ \\ 
SWCX & $0.002$ & $0.005$ & $0.007$ & $0.009$ & $0.014$ & $0.015$ \\ 
CXBSLOPE & $^{+0.005}_{-0.005}$ & $^{+0.011}_{-0.006}$ & $^{+0.014}_{-0.014}$ & $^{+0.018}_{-0.016}$ & $^{+0.024}_{-0.025}$ & $^{+0.027}_{-0.026}$ \\ 
NXB & $0.003$ & $0.005$ & $^{+0.004}_{-0.002}$ & $0.006$ & $^{+0.004}_{-0.005}$ & $^{+0.007}_{-0.003}$ \\ 
$\chi^2$ & $0.002$ & $0.004$ & $0.001$ & $0.003$ & $-0.000$ & $0.002$ \\ 
$N_{\rm H}$ & $0.001$ & $0.002$ & $0.000$ & $0.001$ & $^{+0.000}_{-0.000}$ & $0.001$ \\ 
LMXBs & $^{+0.000}_{-0.002}$ & $^{+0.006}_{-0.003}$ & $-0.006$ & $^{+0.002}_{-0.007}$ & $-0.011$ & $^{+0.001}_{-0.012}$ \\ 
Solar Abun. & $-0.003$ & $^{+0.000}_{-0.001}$ & $-0.005$ & $-0.004$ & $-0.007$ & $-0.006$ \\ 
Other Abun & $-0.005$ & $0.004$ & $-0.009$ & $-0.003$ & $-0.013$ & $-0.008$ \\ 
$1T$ Ann~1 Suzaku & $0.004$ & $0.009$ & $0.005$ & $0.009$ & $0.006$ & $0.009$ \\ 
\\ 
\hline \\
\end{tabular}
\tablecomments{Best values and error estimates (see \S \ref{overview})
  for the gas and baryon fractions of the fiducial hydrostatic
  equilibrium model quoted for several over-densities. $M_{\rm
    stellar}^{\rm other}$ refers to the addition of non-central
  stellar bayons (\S \ref{sys.stars}).  See the notes to
  Table~\ref{tab.mass} regarding the other systematic error tests.}
\end{center}
\end{table*}
\renewcommand{\arraystretch}{1}

In Figure \ref{fig.baryfrac} we show the radial profiles of the gas
and baryon fractions for the fiducial hydrostatic model and quote the
derived values and errors within $\rtwofiveh$, $\rfiveh$, and $\rtwoh$ in
Table \ref{tab.fb} along with the detailed systematic error budget.
Similar to what we find for the mass profile (\S \ref{mass}), most of
the systematic errors for the gas and baryon fractions are
insignificant, and the most important effects occur for the systematic
tests, No Suzaku, Exclude Last Bin, $\Lambda_{\nu}(T,Z)$, CXBSLOPE.
These largest effects testify to the sensitivity of the gas and baryon
fractions to the accuracy of the surface brightness and temperature at
the largest radii in our study (i.e., near $\rtwofiveh$) determined
primarily by the \suzaku\ measurements of the background-dominated
group X-ray emission. (See \S \ref{sys} and \S \ref{baryfrac.disc}.)

For $\rtwofiveh$ and $\rfiveh$ the baryon fraction of the fiducial
hydrostatic model yields, $f_{\rm b}<f_{\rm b,U}$, where $f_{\rm
  b,U}=0.155$ is the mean baryon fraction of the universe as
determined by \planck\ \citep{plan14a}.  At $\rtwoh$, we obtain
$f_{\rm b, 200}=0.149\pm 0.014$, consistent with $f_{\rm b,U}$, and
where the hot gas comprises 67\% of the total baryons. The Einasto
model yields results very similar to the fiducial NFW model, whereas
CORELOG has a much smaller value, $f_{\rm b, 200}=0.06\pm
0.01$, owing to its larger total mass. Similarly, since the AC models also yield
larger total masses than the fiducial model, they also produce smaller
gas and baryon fractions (Table~\ref{tab.custom}), where the largest
effect is for AC1;  e.g., $f_{\rm b, 200}=0.110\pm
0.017$.

Thus far we have only considered the stellar baryons of the BCG as
indicated by the fitted $M_{\star}/L_K$. In \S \ref{sys.stars} we
estimate the amount of non-central stellar baryons and list the
results ($M_{\rm stellar}^{\rm other}$) as a systematic error in
Table~\ref{tab.fb}. The addition of these non-central stellar baryons
increases the baryon fraction to $\approx f_{\rm b,U}$ for
$\rtwofiveh$ and $\rfiveh$.  However, at $r_{200}$ the value becomes
$f_{\rm b}\approx 0.19$ which exceeds the cosmic value with $\approx
2.5\sigma$ statistical significance. Although the excess is marginally
significant (and other systematic effects do mitigate the difference
further), if our estimates for the contributions from the
non-central baryons are accurate, then the super-cosmic baryon fraction at
$r_{200}$ for the fiducial model provides evidence in support of the AC
models and their smaller baryon fractions.

\subsection{MOND}
\label{mond}

\begin{figure*}[t]
\parbox{0.49\textwidth}{
\centerline{\includegraphics[scale=0.42,angle=0]{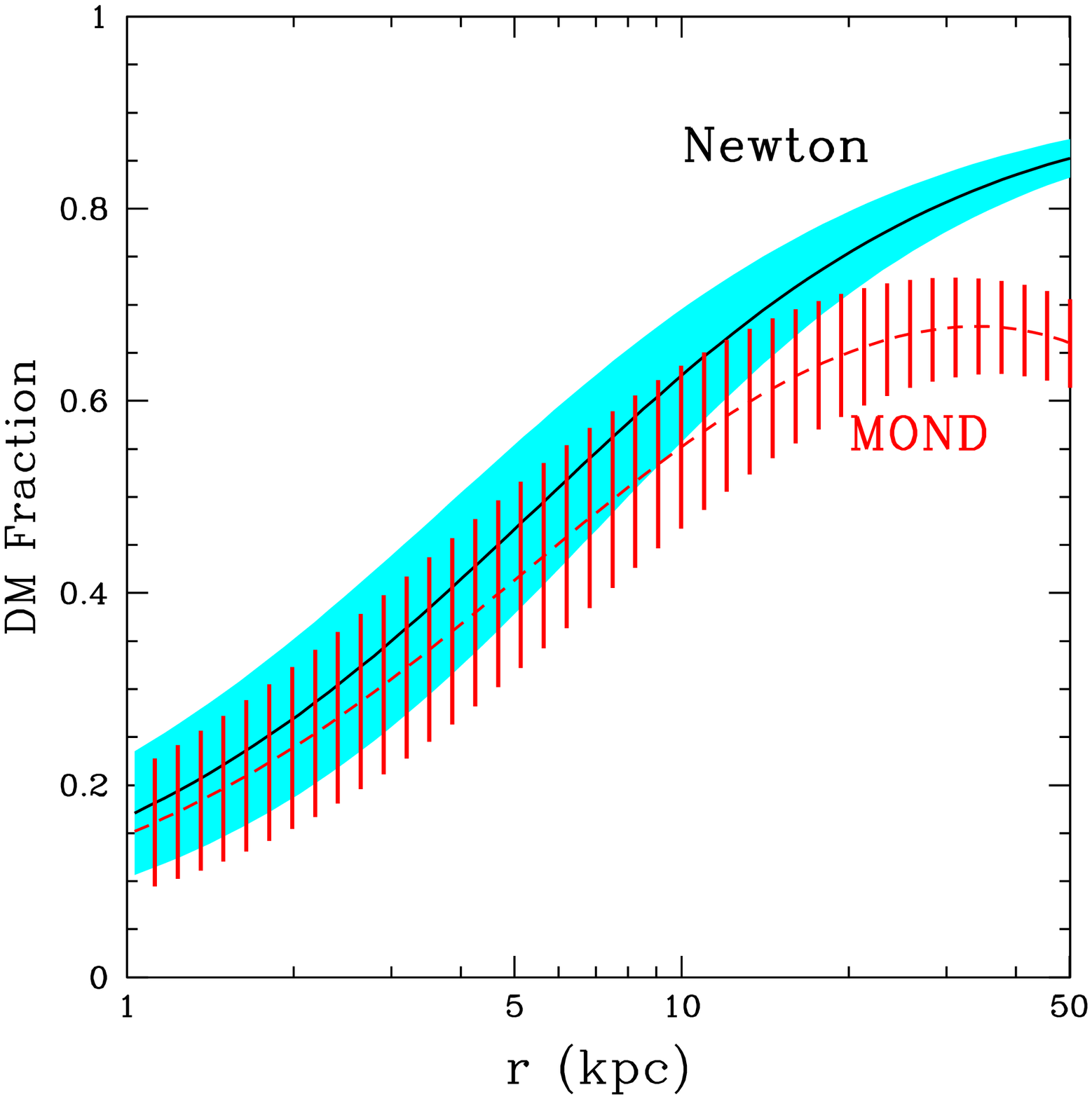}}}
\parbox{0.49\textwidth}{
\centerline{\includegraphics[scale=0.42,angle=0]{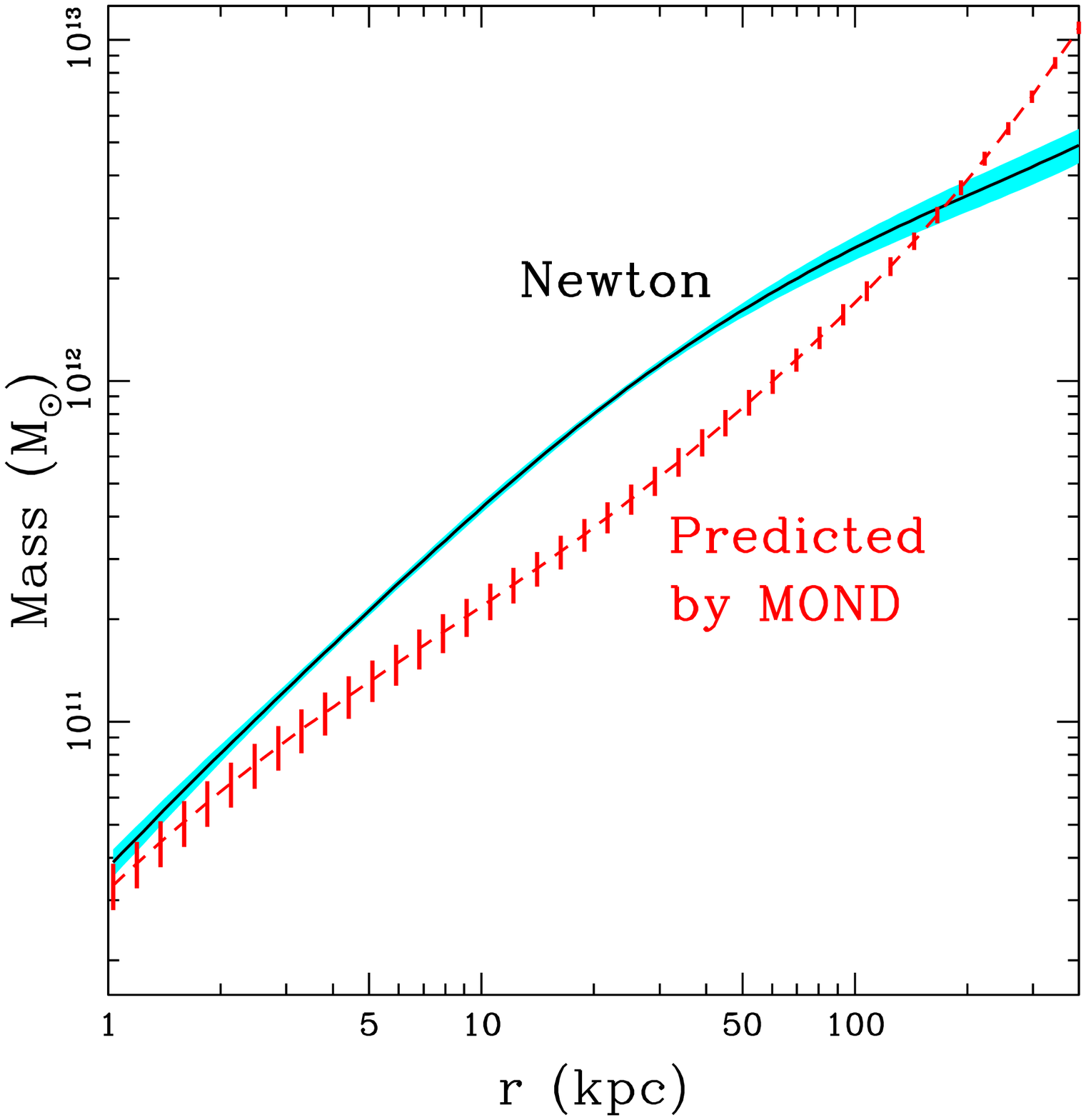}}}
\caption{\footnotesize  ({\sl Left Panel}) Radial profiles of the DM fraction for the
  fiducial hydrostatic model for the Newtonian (black and cyan) and
  MOND (red) cases. The shaded and hashed regions represent $1\sigma$
  errors. ({\sl Right Panel}) Total Newtonian mass profiles of (1) the fiducial
  hydrostatic model (black and cyan, same as in Fig.\ref{fig.mass}),
  and (2) that predicted by MOND without DM (red) from eqn.\ \ref{eqn.mond.other}.}
\label{fig.mond}
\end{figure*}

For comparison to our standard Newtonian analysis with a DM halo, we
also consider the most widely investigated
and successful modified gravity theory, MOND~\citep{milg83a}. Despite
its many successes, MOND is presently unable to explain the mass
profiles of galaxy clusters without
DM~\citep[e.g.,][]{sand99a,poin05a,angu08a,milg15a}. To determine
whether this also applies to \src,  we use the approach of \citet{angu08a}
that is based on \citet{sand99a}. 

Following \citet{sand99a}, we assume spherical symmetry and associate
the gravitational acceleration derived from the equation of
hydrostatic equilibrium,
\begin{equation}
  \vec{g}_{\rm HE} = \frac{\hat{r}}{\rho}\frac{dP}{dr},
\end{equation}
with the general acceleration independent of the gravity theory. (Note
-- in \citealt{buot16a} we referred to the magnitude of this quantity
as $g_{\rm N}$.)  For a spherical system, the MONDian gravitational
acceleration ($\vec{g}_{\rm M}$) is related to the enclosed MONDian
mass ($M_{\rm M}(<r)$) by the relation~\citep{milg83a,beke84a},
\begin{equation}
  -GM_{\rm M}\frac{\hat{r}}{r^2} = \mu(g_{\rm M}/a_0) \vec{g}_{\rm M}, \label{eqn.mond.fundamental}
\end{equation}
where $g_{\rm M} \equiv |\vec{g}_{\rm M}|$,
$a_0\approx 1.2\times 10^{-8}$~cm~s$^{-2}$ is the MOND acceleration
constant, and $\mu(g_{\rm M}/a_0)$ is some unspecified function
interpolating between the Newtonian and MONDian regimes.  Replacing
$\vec{g}_{\rm M}$ with $\vec{g}_{\rm HE}$ in
Equation~(\ref{eqn.mond.fundamental}), and using the simple
interpolating function, $\mu(x) = x/(1+x)$, we have,
\begin{equation}
\frac{GM_{\rm M}}{r^2} = \frac{g_{\rm HE}}{1 + a_0/g_{\rm HE}}, \label{eqn.mond}
\end{equation}
where $g_{\rm HE}\equiv |\vec{g}_{\rm HE}|$ and we have made use of
the relation, $\vec{g}_{\rm M} = -g_{\rm M}\hat{r}$.

Since we have already computed $g_{\rm HE}(r)$ in our Newtonian
analysis, we may immediately compute $M_{\rm M}(<r)$ without any
additional fitting. The use of the simple interpolating function $\mu$
in Equation~(\ref{eqn.mond}) means our solutions have the undesirable
feature that $M_{\rm M}(<r)$ attains a maximum value at some radius
and then decreases~\citep[for additional discussion of this issue
see][]{angu08a}. Presently, we discuss results interior to the radius
roughly where the MONDian mass profile attains its maximum.

We plot the cumulative DM fractions $(M_{\rm DM}/M_{\rm total})$ in
Figure \ref{fig.mond} for the fiducial hydrostatic model computed in
the context of both Newtonian and MONDian gravity. The Newtonian DM
fraction rises continuously with radius whereas the MONDian DM
fraction attains its maximum value for $r\approx 35$~kpc. At this
radius the DM dominates the MONDian mass profile with a DM fraction
$0.68 \pm 0.05$. Using an estimate for the contribution from
non-central baryons (\S \ref{sys.stars}) reduces the MOND DM fraction
to $\approx 0.59$.  The MOND DM fraction is reduced an additional 0.09
if we use the upper limit for $R_e$ for the BCG (\S \ref{stars});
i.e., MOND still needs as much DM as baryonic matter.

It is instructive to examine what a Newtonian analysis should have
yielded for the mass profile in a MOND gravity field without DM.  Solving
Equation~(\ref{eqn.mond}) for $g_{\rm HE}$ gives,
\begin{equation}
g_{\rm HE} = \frac{GM_{\rm M}}{2r^2}\left(1 + \sqrt{1 + 4\frac{a_0r^2}{GM_{\rm
      M}}} \, \, \, \, \right). \label{eqn.mond.other}
\end{equation}
(Note -- in \citealt{buot16a} we referred to $GM_{\rm M}(<r)/r^2$ in
this equation as $g_{\rm M}$.)  If we take $M_{\rm M}$ to be the sum
of the enclosed masses of the baryonic components (i.e., stellar and
gas) and identify $g_{\rm HE}$ with $GM_{\rm N}(<r)/r^2$ as
appropriate for a Newtonian analysis, then we may use
Equation~(\ref{eqn.mond.other}) to predict the Newtonian mass profile
that one would infer for a MONDian gravity field generated only by the
baryonic mass components. For this exercise, we take the stellar and
gas mass profiles obtained from our fiducial model (i.e., Newtonian
analysis), combine them to give $M_{\rm M}(<r)$, and then use
Equation~(\ref{eqn.mond.other}) to compute $M_{\rm N}(<r)$ that MOND
would predict. In Figure \ref{fig.mond} we compare this predicted
Newtonian total mass profile to the one actually measured. The profile
predicted by MOND without DM underestimates the true Newtonian profile
out to $\approx\rtwofiveh$ at which point it crosses over, and then
exceeds, the Newtonian profile for larger radii. The largest
underestmate occurs again near a radius $\approx 35$~kpc where
$M_{\rm M}\approx 0.48M_{\rm N}$.  Accounting for the non-central
baryons and using the upper limit for $R_e$ for the BCG increases this
value to $M_{\rm M}\approx 0.65M_{\rm N}$.

We conclude that MOND requires substantial amounts of DM to explain
the X-ray emission of \src, consistent with results obtained for more
massive groups and clusters.

\subsection{Super-Massive Black Hole}
\label{smbh}

We note the good fit to the centrally peaked temperature
profile\footnote{We also mention that centrally peaked temperature
  profiles in massive elliptical galaxies can be explained by
  classical wind models~\citep[e.g.,][]{davi91a,ciot91a} without
  needing to invoke additional energy input from an AGN.} by the
fiducial model that does not include the mass of any putative
super-massive black hole (SMBH). That is, the stellar mass density is
sufficiently centrally peaked to produce the temperature peak in \src,
as was shown previously by \citet{hump06b} for this system and two
other elliptical galaxies (NGC~720 and NGC~4125) with centrally peaked
temperature profiles.  For the early-type galaxy NGC~1332, which also
has a centrally peaked temperature profile, \citet{hump09c} showed
that adding an SMBH with a plausible mass did little to alter the
central temperature over that established by the stellar
potential. Only for the massive elliptical galaxy NGC~4649, for which
the sphere of influence is largely resolved by \chandra, is there a
strong case for a central X-ray temperature peak generated by the
potential of the SMBH~\citep{hump08a,brig09a,pagg14a}.

From analysis of a large cosmological simulation, \citet{raou16a} find
evidence that the BCGs in fossil systems should possess SMBHs with
above-average masses. Visual inspection of Figure~6 of \citet{raou16a}
indicates that \src\ would possess an SMBH with $M_{\rm bh}\sim
10^{9}\, M_{\odot}$. Using this $M_{\rm bh}$ along with the central
stellar velocity dispersion of the BCG ($\sigma=310.4\pm
11.5$~km~s$^{-1}$;~\citealt{ho09a}), the sphere, or radius, of
influence of the SMBH is $r_g = GM_{\rm bh}/\sigma^2\sim 0.05$~kpc;
i.e., $r_g$ is only $\sim 4\%$ of the radius of \chandra\ Annulus~1
(Table~\ref{tab.gas}). Consequently, when we add an SMBH with $M_{\rm
  bh}=10^{9}\, M_{\odot}$ to our hydrostatic models there is no
tangible effect. We find that our model fits are only affected
noticeably when $r_g$ is at least about half the radius of Annulus~1
corresponding to $M_{\rm bh}>10^{10}\, M_{\odot}$, though even then
the statistical significance is weak.  Hence, presently we are unable
to place interesting constraints on $M_{\rm bh}$.

\section{Error Budget}
\label{sys}

Here we describe various systematic errors and their impact on our
analysis. The detailed error budget is listed for the mass parameters
in Table~\ref{tab.mass} and gas  and baryon fractions in
Table~\ref{tab.fb}. 

\subsection{Spherical Symmetry}
\label{sys.sphere}

To estimate the magnitude of possible systematic error due to the
assumption of spherical symmetry, we use the results of
\citet{buot12c} who examined the orientation-averaged biases (mean
values and scatter) of several quantities derived from spherically
averaged hydrostatic equilibrium studies of hot gas in ellipsoidal
galaxy clusters.  We use the ``NFW-EMD'' results for the halo
concentration, total mass, and gas fraction from Table~1 of
\citet{buot12c} to estimate the average error arising from assuming
spherical symmetry within a radius of $r_{500}$ when the system is
really an ellipsoid viewed at a random orientation. We adopt an
intrinsic flattening of $q\approx 0.6$ based on cosmological
simulations of DM halos with mass similar to \src~\citep{schn12a}. In
all cases the derived systematic errors (``Spherical'') on the
concentration, mass, and gas fraction within $r_{500}$ listed in
Tables \ref{tab.mass} and \ref{tab.fb} are insignificant.

\subsection{Entropy Model}
\label{sys.entropy}

We considered the systematic effects (``Entropy'') of two modifications to
the default entropy profile having two break radii.  For one test we set $r_{\rm
  baseline}=400$~kpc; i.e., twice the default value. For the other test we
added a third break radius to the entropy profile. In each case the induced
systematic error is everywhere insignificant.

\begin{figure*}
\parbox{0.49\textwidth}{
\centerline{\includegraphics[scale=0.35,angle=0]{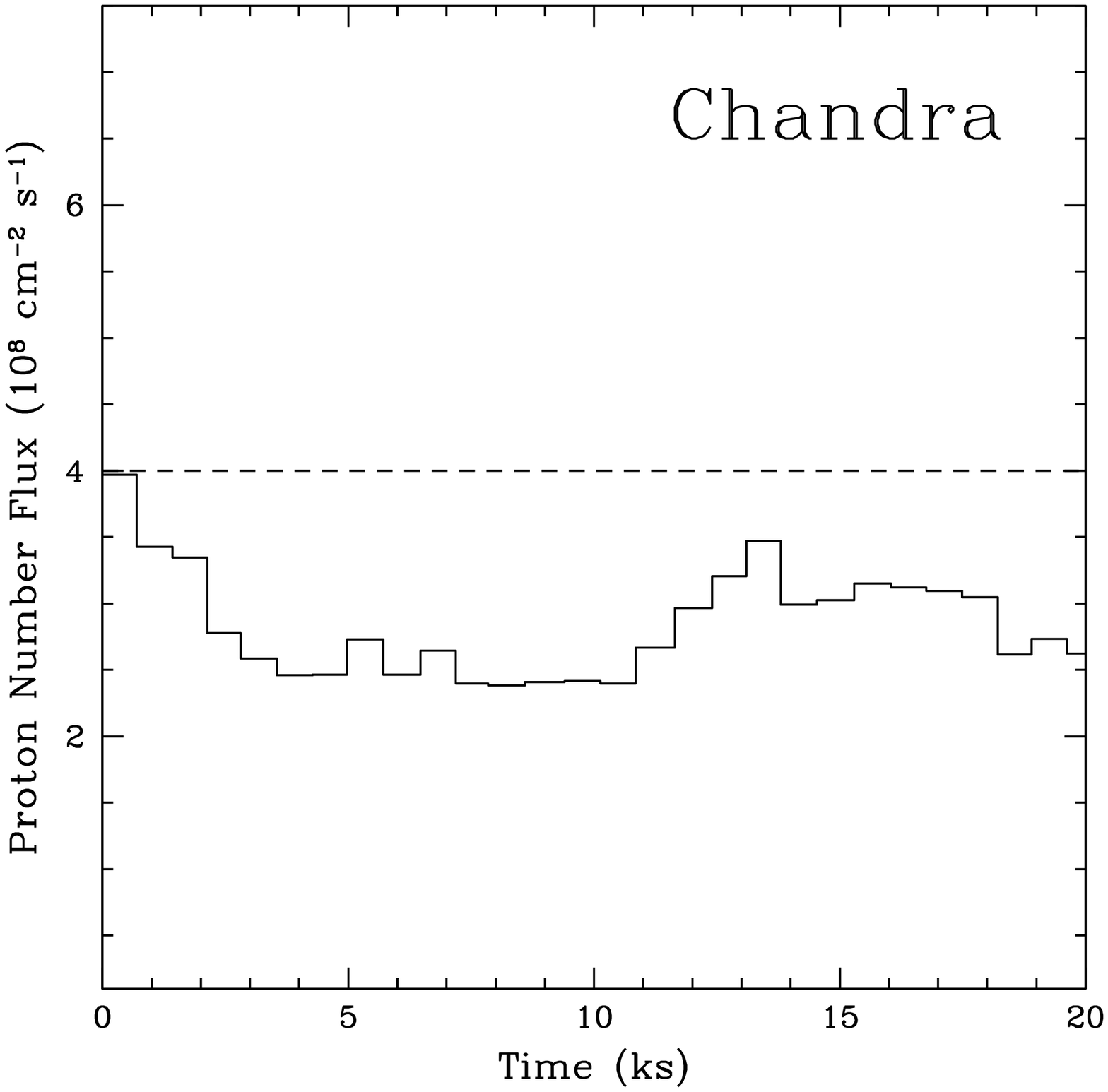}}}
\parbox{0.49\textwidth}{
\centerline{\includegraphics[scale=0.35,angle=0]{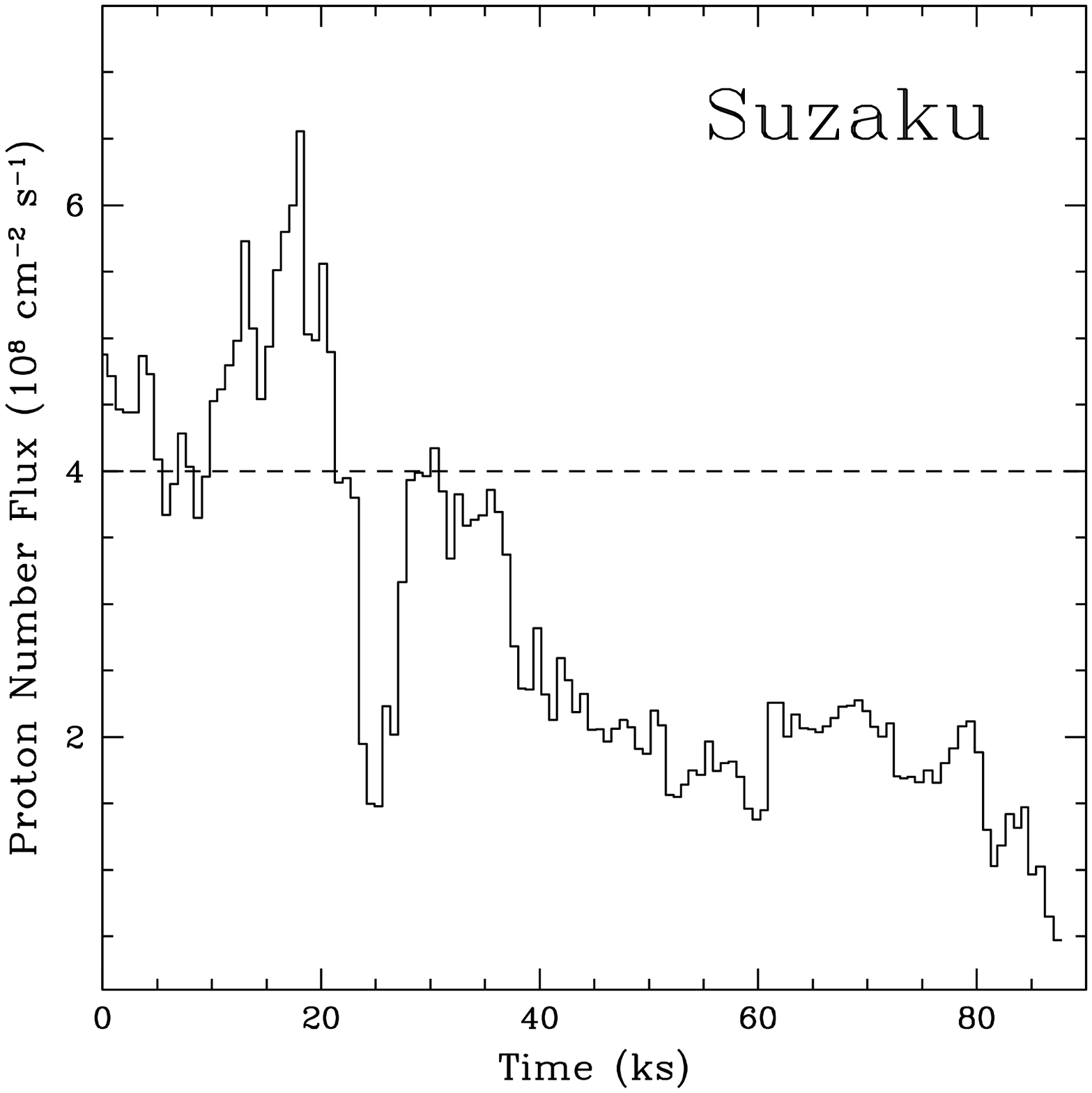}}}
\caption{\label{fig.swcx} Solar proton flux measured by SWEPAM/SWICS
  during the ({\sl Left Panel} ) \chandra\ and ({\sl
    Right Panel}) \suzaku\ observations of \src. The dashed line
  denotes the approximate level suggested by \citet{fuji07a} above
  which SWCX emission may significantly contaminate \suzaku\ data.}
\end{figure*}

\subsection{Stellar Mass}
\label{sys.stars}

We studied the full range of estimated uncertainty in $R_e$ for the
BCG stellar mass model (Table~\ref{tab.prop}; \S \ref{stars}) and its
impact on the derived hydrostatic equilibrium models (``BCG''). We
discussed the magnitudes of the systematic errors in \S \ref{mass} and
\S \ref{baryfrac}. The most important effect is on $\mlkband$ by
allowing larger values that bring the measured values into better
agreement with SPS models (\S \ref{mass}). 

The amounts and distributions of non-central stellar baryons (i.e.,
non-central galaxies and diffuse intracluster light (ICL)) are more
uncertain than for the BCG, and therefore we prefer to treat their
contributions to the total baryon fraction as a systematic error
similarly to our previous study of the fossil cluster
RXJ~1159+5531~\citep{buot16a}. \citet{lied13a} present a catalog of 48
group members of \src\ from deep $B$- and $R$-band imaging. Using
their measured $M_R$ for each galaxy, we find that the non-central
galaxies constitute 28\% of the $R$-band stellar light of the group.
For our calculation we assume this result also applies in the $K$ band
with the same $M_{\star}/L_K$ as the BCG. Since we lack an ICL
measurement for \src, we use the result of \citet{budz13a} that
typically the ICL in groups and clusters is $\approx 30\%$ of the
total stellar mass fraction. We assume that the non-central galaxies
and ICL follow the spatial distribution of the DM in our models.

We list the contribution of the non-central baryons to the baryon
fraction as ``$M_{\rm stellar}^{\rm other}$'' In Table
\ref{tab.fb}. This is one of the largest systematic effects (\S
\ref{baryfrac}) and lends support to the AC models which have smaller
baryon fractions. 

\subsection{DM Model}
\label{sys.dm}

The effects of using a DM profile different from NFW are indicated in
the rows ``Einasto'', ``Corelog'', and ``AC'' in Tables \ref{tab.mass}
and \ref{tab.fb}. The ``AC'' column includes the results of all 4 AC
models (Table~\ref{tab.ac.defs}) applied to the NFW DM profile, though
the results are dominated by AC1. We have discussed the magnitudes of
these systematic differences in \S \ref{mass} and \S\ref{baryfrac}.

\subsection{SWCX}
\label{swcx}

To investigate whether the \chandra\ and \suzaku\ observations may
have been impacted by enhanced SWCX emission we examined the solar
proton flux using the Level~3 data from
SWEPAM/SWICS\footnote{http://www.srl.caltech.edu/ACE/ASC/level2/sweswi\_l3desc.html}. In
Figure~\ref{fig.swcx} we plot the solar proton flux observed by
SWEPAM/SWICS during the \chandra\ and \suzaku\ observations for \src,
which includes the periods of time gaps over the duration of the
\suzaku\ observation, and we have accounted for the travel time from
the satellite to Earth. Almost all the \chandra\ observation has solar
proton flux below $4\times 10^8$~cm$^{-2}$~s$^{-1}$, which is
approximately the level suggested by \citet{fuji07a} above which
significant proton flare contamination of \suzaku\ observations can be
expected. However, the initial $\sim 25\%$ of the \suzaku\ observation
does exceed this flux level indicating possibly important SWCX flare
contamination.

As noted in \S \ref{chandra} and \S \ref{suzaku}, the light curves for
both the \chandra\ and \suzaku\ observations are very quiescent and
provide no evidence of significant flaring. In addition, following our
previous study of the fossil cluster RXJ~1159+5531~\citep{su15a} we
investigated including a spectral model for the SWCX emission in each
of the \chandra\ and \suzaku\ data and found no evidence for
significant SWCX emission.  Finally, when removing all the time
intervals with solar proton flux above
$4\times 10^8$~cm$^{-2}$~s$^{-1}$ in the \suzaku\ data, the total
cleaned exposure time is reduced substantially; i.e., by 25\% to
32.6~ks. For these reasons, and remembering that the key
$4\times 10^8$~cm$^{-2}$~s$^{-1}$ flux level does not guarantee strong
flare contamination, we do not use the nominally SWCX-cleaned data set
for \suzaku\ by default. Instead, we treat it as a systematic error
(``SWCX'') in Tables~\ref{tab.mass} and \ref{tab.fb}. For the mass
parameters (i.e., $\mlkband$, concentration, and mass) the effect is
insignificant in all cases. For the gas and baryon fractions, the
effect is comparable to the $1\sigma$ errors leading to slightly
larger values.

\subsection{Background}
\label{sys.bkg}

In addition to SWCX, we examined other systematic effects associated
with the characterization of the background. First, we varied the
\suzaku\ NXB model emission by $\pm 5\%$ to gauge how sensitive are
the results to the default level. We find that all induced changes are
statistically insignificant -- see row ``NXB'' in Tables
\ref{tab.mass} and \ref{tab.fb}. Second, for both the \chandra\ and
\suzaku\ data we examined the sensitivity of the results to the value
of the CXB power-law slope (i.e., $\Gamma=1.41$).  When changing the
slope to $\Gamma=1.3$ and $\Gamma=1.5$, representing
approximately the range of observed
values~\citep[e.g.,][]{tozz01b,delu04a,more09a}, we obtain the results
listed in row ``CXBSLOPE'' in the tables. As noted above in \S
\ref{mass} and \S \ref{baryfrac}, this is among the largest systematic
errors. In most cases the differences are $\sim 1\sigma$ or a little
less. The largest differences are $\sim 2\sigma$ and occur for the gas
and baryon fractions at $r_{200}$.

\subsection{Miscellaneous Spectral Fitting}
\label{misc.spec}

Here we summarize other tests involving the spectral analysis. We
varied the adopted value of Galactic $N_{\rm H}$ by $\pm 20\%$
(``$N_{\rm H}$'') finding the effect to be insignificant. Likewise,
minimizing $\chi^2$ instead of the C-statistic (``$\chi^2$'') did not
affect the parameters significantly. By default we use two
temperatures to model the gas emission within the central Suzaku
annulus (\S \ref{spec_suzaku}). If instead we use a single
temperature, we obtain the results listed in row ``$1T$ Ann1 Suzaku''
in Tables \ref{tab.mass} and \ref{tab.fb}. The induced changes are
generally insignificant for $\mlkband$, the concentrations, and total
masses, but are possibly significant ($\la 1\sigma$) for the gas and
baryon fractions.

We also examined how choices regarding the metal abundances affect the
results. First, we examined using different solar abundance
tables~\citep{grsa,lodd} than our default~\citep{aspl} and list the
results in the ``Solar Abun'' row in Tables \ref{tab.mass} and
\ref{tab.fb}. The induced changes are insignificant. The results of
using spectral fits allowing Ne, Mg, and Si to vary with non-solar
ratios with respect to Fe are listed in the row ``Other Abun.'' This
effect is insignificant in most cases, except for the gas fractions
where $\approx 1\sigma$ smaller gas fractions are indicated.

Finally, we considered various permutations of our treatment of the
LMXB component in the spectral fits. That is, for each observation we
performed fits both with the flux of the LMXB component set to a
nominal value and with it freely fitted (\S \ref{galmod}). We chose by
default to use the nominal LMXB flux for the \chandra\ spectral fits
and the freely fitted LMXB flux for the \suzaku\ fits. As a systematic
test, we used various permutations of these choices (i.e., nominal
LMXB flux for both, freely fitted for both) and list the results in
the row ``LMXB'' in the tables. The effect is only weakly
significant for the gas and baryon fractions leading to $\approx
1\sigma$ smaller values

\subsection{Miscellaneous Hydrostatic Modeling}
\label{mischm}

In this section we describe additional tests regarding details of the
hydrostatic equilibrium models.  First, we examined the sensitivity of
our models to the assumed distance to \src\ by instead using the
distance employed by~\citet{ma14a} but scaled to
$H_0=70$~km~s$^{-1}$~s$^{-1}$ (i.e., 62.3~Mpc). We list the results in
Tables \ref{tab.mass} and \ref{tab.fb} in row ``Distance'' and find
the effect to be insignificant in all cases. Second, we examined the
sensitivity of our results to the assumed maximum radius (1~Mpc) used
to define the hydrostatic equilibrium model.  In row ``Proj. Limit'' of
Tables \ref{tab.mass} and \ref{tab.fb} we show the results of using
0.75~Mpc or 1.5~Mpc for the maximum radius and find that in all
instances the effect is insignificant. Next we explored the result of
performing a standard $\chi^2$ frequentist analysis for the hydrostatic
equilibrium models (i.e., as opposed to the default nested sampling
bayesian approach) and list the results in row ``Frequentist'' in the
tables. Again, in all cases the effect is insignificant.

We also considered how the results changed by excluding some of the
data. The results obtained from excluding the outer annulus from each
of the \chandra\ and \suzaku\ data sets (i.e., annulus~10 and
annulus~3 respectively) are listed in row ``Exclude Last Bin'' in
Tables \ref{tab.mass} and \ref{tab.fb}. This is one of the largest
systematic effects investigated leading to differences of
$\sim 1.5\sigma$ significance in $M_{200}$ and $f_{\rm b, 200}$.
Unsurprisingly, the largest effects occur for the largest radius
considered (i.e., $\rtwoh$). Even larger differences are observed when
the \suzaku\ data, which extend out the largest radius in our study,
are omitted entirely (``No Suzaku''). As expected, the differences
manifested when omitting the \suzaku\ data increase with increasing
radius; e.g., $c_{200}$ increases and $M_{200}$ decreases at
$\sim 1.5\sigma$ significance. The baryon fraction changes even more,
such that $f_{\rm b, 200}$ increases by $\sim 4\sigma$ (with respect to
the error bar of the fiducial model) to a super-cosmic value of $\sim 0.21$. Note,
however, that omitting the \suzaku\ data increases the error bar so
that $f_{\rm b, 200}$ formerly exceeds the cosmic value at a lower significance
level ($\sim 2.5\sigma$).

Finally, we considered the effect of fitting the radial iron
abundances (Table~\ref{tab.gas}) with a projected, emission-weighted
parameterized model that is used in the evaluation of the plasma
emissivity $\Lambda_{\nu}(T,Z)$ (\S \ref{method}). For the parameterized
model we used a multi-component model related to that consisting of
two power-laws mediated by an exponential (eqn.\ 5 of
\citealt{gast07b}) and a constant floor. We list the results for this
test of the plasma emissivity in row ``$\Lambda_{\nu}(T,Z)$'' of
Tables \ref{tab.mass} and \ref{tab.fb}. This is one of the largest
systematic effects and induces parameter shifts very similar to the
``No Suzaku'' test. Apparently, the parameterized model fit for the iron
abundance is dominated by the \chandra\ measurements at the largest
radii because of the large \suzaku\ error bar on the iron abundance
(Table~\ref{tab.gas}). The smaller \chandra\ abundances lead to
smaller plasma emissivity and higher gas density at large radius. The
higher gas density translates to higher gas and baryon
fractions. Indeed, if we use the fitted model for the iron abundance,
but omit the \suzaku\ data entirely, then the parameter differences
with respect to the fiducial model are all insignificant. Thus, the ``No
Suzaku'' and ``$\Lambda_{\nu}(T,Z)$'' tests demonstrate the importance
of obtaining more accurate iron abundance measurements at large radius
($\ga r_{2500}$) to obtain robust measurements of the mass parameters,
especially the gas and baryon fractions (\S \ref{baryfrac.disc}).

\section{Discussion}
\label{disc}

\subsection{High Concentration and Adiabatic Contraction}
\label{highc}

\begin{figure}[t]
\begin{center}
\includegraphics[scale=0.42]{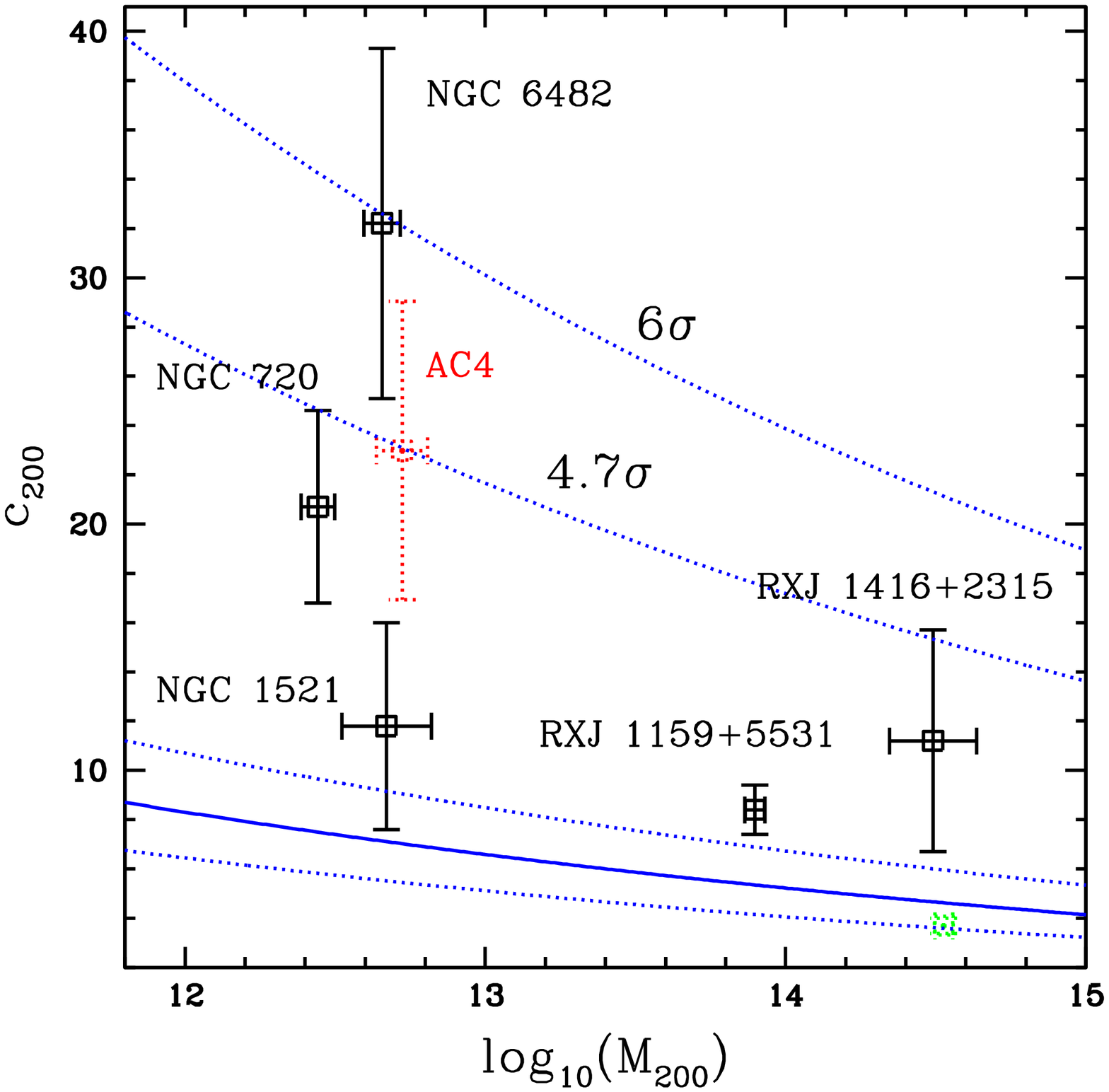}
\end{center}
\caption{\footnotesize Concentration and mass values for \src\ and
  other fossil systems (see \S \ref{highc}). The blue solid line is
  the \lcdm\ relation for relaxed halos from~\citet{dutt14a} evaluated
  at the distance of \src. The blue dotted lines indicate the
  intrinsic scatter in the \lcdm\ relation; i.e., the lines closest
  the mean relation are $\pm 1\sigma$ while the significances of the
  other lines are indicated. The data point indicated by red dots is
  obtained for the ``Forced Quenching'' AC NFW
  model~(i.e., AC4 -- Tables~\ref{tab.ac.defs} and
  \ref{tab.custom}). Finally, the green dotted data point is our
  preliminary measurement for RXJ~1416+2315 using more recent data.}
\label{fig.c-m}
\end{figure}

In \S \ref{mass} we found that the halo concentration for the fiducial
model with an NFW DM profile ($c_{200}=32.2\pm 7.1$,
$M_{200}=(4.5\pm 0.6)\times 10^{12}\, M_{\odot}$), though only about
half the original estimate of $c_{200}\sim 60$ obtained
by~\citet{khos04a} who did not include a separate BCG stellar
component, nevertheless significantly exceeds the value of
$c_{200}=7.1$ expected for relaxed \lcdm\ halos~\citep{dutt14a}; i.e.,
by $6\sigma$ with respect to the intrinsic scatter in the theoretical
\lcdm\ $c-M$ relation and by $3.5\sigma$ with respect to the
statistical measurement error for the observations.  Moreover, from
visual inspection of Figure~16 of \citet{dutt14a}, which shows
$c_{200}$ and $M_{200}$ obtained from NFW fits to simulated \lcdm\
halos, we notice that for the mass of \src\ there are no \lcdm\ halos
with $c_{200}>30$. While the observational error on $c_{200}$ permits
smaller values, the measured $c_{200}$ and $M_{200}$ values for the
fiducial model with an NFW DM halo represent an extremely rare system
formed in \lcdm.  (Note, however, that otherwise the DM profile does
not appear to be unusual; e.g., (1) we measure a DM fraction within
$5R_e$ of $74\% \pm 5\%$ for \src\ that is very consistent with other
massive elliptical galaxies~\citep[e.g.,][]{alab16a}; (2) the slope of
the total mass profile is consistent with the $\alpha$-$R_e$ scaling
relation within the observed scatter (\S \ref{slope}).)

We plot in Figure~\ref{fig.c-m} the $c_{200}$ and $M_{200}$ values for
\src\ and a few other fossil and nearly fossil systems with evidence
for above-average concentrations: NGC~720~\citep{hump11a},
NGC~1521~\citep{hump12b}, RXJ1159+5531~\citep{buot16a}, and
RXJ1416.4+2315~\citep{khos06a}. We designate NGC~720 and NGC~1521 as
``nearly'' fossil systems since they obey the fossil classification
within their projected virial radii, although typically they are
classified as members of larger groups owing to more distant galaxy
associations.  There is presently newer and deeper \chandra\ and
\suzaku\ data for the massive fossil cluster RXJ1416.4+2315 since the
time of the initial analysis of a shallow \chandra\ observation by
\citet{khos06a}. Our preliminary analysis incorporating all the
available \chandra\ and \suzaku\ data yields a smaller concentration
$\sim1\sigma$ below the mean $c-M$ relation (Buote et al.\ 2016, in
preparation). Our new measurement for \src\ places it as the most
extreme outlier in the $c-M$ relation for fossils, with the next
largest being NGC~720 ($\sim 4\sigma$ with respect to the intrinsic
scatter).

As noted above, if $c_{200}$ for \src\ is really $>30$ it would be a
very rare system formed in \lcdm, and perhaps its existence so near to
us (i.e., representing a small cosmological search volume) would be
difficult for \lcdm\ to explain.  The lower concentrations inferred by
the AC models would mitigate any such tension with \lcdm. While the
standard AC prescriptions (AC1, AC2) yield $\mlkband$ values for the
BCG that are uncomfortably small compared to the SPS models, both AC
models derived from cosmological simulations (AC3, AC4) are consistent
with the SPS models.  The ``Forced Quenching'' AC variant (AC4) gives
the highest $\mlkband$ of the AC models identical to that obtained for
the un-contracted NFW model.

Additional support for the AC models comes from their lower baryon
fractions in better agreement with the cosmic value once stellar
baryons not associated with the BCG are considered (\S \ref{baryfrac}
and \S \ref{baryfrac.disc}). Therefore, while AC4 still produces a
significant outlier in the $c-M$ relation (see Figure~\ref{fig.c-m}),
we believe its advantages over the un-contracted NFW model and other
AC models make it the preferred model of all those that we have
investigated. The support for AC4 as opposed to AC1 or AC2 provides
new evidence for ``weak'' AC and would seem to be consistent with
stellar dynamics and lensing measurements of the mass profiles of
early-type galaxies that favor no AC or weak AC~\citep[see \S
\ref{intro}; e.g.,][]{dutt14b}.  This is also generally consistent
with the X-ray studies of a small number of massive elliptical
galaxies and galaxy groups that somewhat disfavor AC1 and AC2 relative
to no AC ~\citep[see \S \ref{intro}; H06;][]{gast07b}.

The qualitative features and conclusions of the above discussion also
apply to when we use the Einasto profile for the DM. The principle
difference is that the outlier significance is lessened somewhat with
respect to NFW; e.g., the pure Einasto profile gives a $c_{200}$ value
that is $4.2\sigma$ above the mean \lcdm\ $c-M$ relation in terms of
the intrinsic scatter, and $3.2\sigma$ above the mean for AC4. The
lower outlier significance partially arises from the larger \lcdm\
intrinsic scatter for the Einasto profile; i.e., 0.13 dex for Einasto
vs.\ 0.11 dex for NFW~\citep{dutt14a}.

Finally, note that while the halo concentration of \src\ is extreme
compared to the other fossil systems, the $R$-band magnitude gap
between the BCG and the next brightest member is quite typical for
fossils (just over 2 magnitudes). In fact, the $R$-band luminosity
function of \src, as well as its color-magnitude relation, appear to
be typical of the general group and cluster
population~\citep{lied13a}.

\subsubsection{Theoretical Concentration-Mass Relation}
\label{c-m}

When comparing the $c_{200}$ we have measured from the X-ray data to
theory, we have used the theoretical concentration-mass relations
published by \citet{dutt14a} for the following reasons. (1) It is a
recent study using state-of-the-art numerical simulations with
up-to-date cosmological parameters from \planck. (2) The study
provides convenient power-law approximations to the $c_{200}-M_{200}$
relation as a function of redshift for relaxed halos. (3) Results for
both NFW and Einasto profiles are provided.

Moreover, the $z\approx0$ $c_{200}-M_{200}$ relations obtained by
recent studies are quite consistent for our needs. For example, the
double power-law approximation of \citet{klyp16a} for relaxed,
mass-selected, $z=0$ halos with $M_{200}=4.5\times 10^{12}\,
M_{\odot}$ gives $c_{200}=7.0$ compared to 7.2 which we obtain from
the relation of \citet{dutt14a} for $z=0$ halos. The relation of
\citet{klyp16a} gives a slightly larger value ($c_{200}=7.2$) for
halos selected according to the maximum circular velocity (see Tables
A1 and A2 of \citealt{klyp16a}). Another example is provided by the
recent study of \citet{ludl16a} who present the theoretical
$c_{200}-M_{200}$ relation for the Einasto profile as a function of
$z$. Using the redshift and best-fitting Einasto mass of \src\ (i.e.,
$M_{200}=4.9\times 10^{12}\, M_{\odot}$), we obtain $c_{200}=7.4$ for
the relation published by \citet{ludl16a} compared to the value of 8.1
we obtain from \citet{dutt14a}.

Therefore, the uncertainties in the theoretical $c_{200}-M_{200}$
relation at $z\approx 0$ are considerably less than the statistical
and systematic uncertainties associated with our measurement of
$c_{200}$ from the X-ray data. As a result, we do not discuss
systematic differences in the theoretical relations henceforth in this
paper.  

\subsection{Baryon Fraction}
\label{baryfrac.disc}

Whereas previously the global baryon fraction was effectively
unconstrained for \src, with the addition of new \suzaku\ data and an
improved hydrostatic modeling procedure we obtain
$f_{\rm b,200}=0.149\pm 0.014$ (i.e., $\pm 9\%$ precision) for the
fiducial hydrostatic model. As mentioned in \S \ref{baryfrac} and \S
\ref{sys.stars}, this value is very consistent with
$f_{\rm b,U}=0.155$, but it only includes the stellar baryons from the
BCG. Adding reasonable estimates for the non-BCG stellar baryons
increases the baryon fraction to $f_{\rm b,200}\approx 0.19$.  This
value marginally exceeds $f_{\rm b,U}$ ($\sim 2.5\sigma$), and various
systematic errors could lower the significance (Table~\ref{tab.fb}).
Nevertheless, the lower baryon fractions obtained for the AC models
(Table~\ref{tab.custom}) mitigate such tension with \lcdm. For
example, for AC4, considering only the BCG stellar bayons gives
$f_{\rm b,200}= 0.136\pm 0.016$ which rises to
$f_{\rm b,200}\approx 0.18$ when including the non-BCG stars; i.e.,
only $\sim 1.5\sigma$ above $f_{\rm b,U}$. We conclude that the baryon
fraction constraints favor the AC models. (To a lesser extent, they
favor the Einasto model as well -- see Table~\ref{tab.fb}).

This solid evidence for a baryonically closed
$\approx 4.5\times 10^{12}\, M_{\odot}$ halo associated with \src\ has important
implications for the ``Missing Baryons Problem'' at low
redshift~\citep{fuku98} as it suggests that, at least in massive
elliptical galaxy / small group halos, many of the ``missing'' baryons
can be located in the outer halo as part of the hot component --
consistent with results we obtained previously for
NGC~720~\citep{hump11a} and NGC~1521~\citep{hump12b}.  We emphasize,
however, that the baryon fraction at $r_{200}$ depends strongly on the
form of the assumed DM profile (Table~\ref{tab.fb}). For both the NFW
and Einasto models (i.e., profiles consistent with \lcdm\ simulations)
baryonic closure is clearly indicated. However, the pseudo-isothermal
CORELOG profile yields a much smaller value,
$f_{\rm b, 200} = 0.06\pm 0.01$. While the CORELOG profile is not
motivated by the standard cosmological paradigm, it is consistent with
the X-ray data of \src.

In addition to assumptions about the DM profile, the ability to
measure $f_{\rm b,200}$ accurately for $\ktemp\la 1$~keV systems like
\src\ is limited to a large extent by the accuracy of the iron
abundance profile measured within $r_{200}$. As discussed in \S
\ref{mischm}, for systems with $\ktemp\la 1$~keV the plasma
emissivity, and thus also the gas mass, are each very sensitive to the
iron abundance at large radius where most of the gas mass resides (see
the ``$\Lambda_{\nu}(T,Z)$'' systematic error in Table~\ref{tab.fb}).
This issue is particularly relevant when considering that
\citet{ande14a} found that $\beta$-model fits to the radial X-ray
surface brightness profile of NGC~720 gave a value for the gas mass at
$r_{200}$ only about half of that we obtained in \citet{hump11a} using
essentially the same methodology as in our present
investigation. While much of the difference undoubtedly can be
attributed to the different model types employed in the two studies
(e.g., the $\beta$ model is equivalent to the CORELOG potential
hosting isothermal hot gas -- see, e.g., \S 2.1.1
of~\citealt{buot12b}), another key difference is that \citet{ande14a}
assumed a constant metallicity $Z=0.6Z_{\odot}$ while \citet{hump11a}
assumed $Z\approx 0.3Z_{\odot}$ for $r\ga r_{2500}$. For
$\ktemp\sim 0.5$~keV the factor of 2 higher metallicity assumed by
\citet{ande14a} translates to a plasma emissivity higher by
essentially the same factor and a gas mass that is lower by
$\approx\sqrt{2}$.

\subsection{Hydrostatic Equilibrium and the Stellar IMF}
\label{he}

To directly quantify the accuracy of the hydrostatic equilibrium
approximation requires measurements of the gas kinematics. In the
central region of the Perseus cluster, which displays large cavities
and other asymmetric features in the X-ray image presumed to arise
from AGN feedback, {\sl Hitomi} found somewhat surprisingly that the
pressure from turbulence is only $4\%$ of the thermal gas
pressure~\citep{hit16a}. If, as the {\sl Hitomi} observation suggests, the hydrostatic equilibrium
approximation is very accurate in a system like Perseus that possesses
pronounced X-ray surface brightness irregularities, it also seems likely to
be very accurate in \src\ which exhibits a highly regular X-ray
morphology (see \S \ref{intro}). Consistent with an accurate
hydrostatic equilibrium approximation, we find that plausible
hydrostatic equilibrium models provide excellent fits to the X-ray
data (\S \ref{mass}) and the derived high halo concentration (\S
\ref{highc}) implies a system that is very old and is therefore highly
evolved and relaxed.

The value of $M_{\star}/L_{\rm K}$ that we obtain for the BCG from the
hydrostatic equilibrium analysis agrees with SPS models that assume a
Chabrier or Kroupa IMF and also favors a value for $R_e$ close to the
upper limit of the systematic range considered (see \S \ref{stars} and
\S \ref{mass}). Our measurement strongly disfavors a Salpeter IMF which
predicts an $M_{\star}/L_{\rm K}$ that is too high; i.e., even when
using the upper limit for $R_e$ (for the fiducial model and AC4) we
obtain $M_{\star}/L_{\rm K}\approx 0.89\pm 0.11$ solar which is
$\approx 6\sigma$ smaller than the SPS value with a Salpeter IMF (1.58
solar). The clear preference for a Chabrier or Kroupa IMF over Salpeter
is consistent with our previous X-ray studies of massive elliptical
galaxies, groups, and
clusters~\citep[H06,][]{gast07b,hump09c,hump12b,buot16a} with the
possible exception of NGC~720~\citep{hump11a}. However, many other
studies, particularly those that combine gravitational lensing and
stellar dynamics, clearly favor the Salpeter IMF in massive early-type
galaxies~\citep[e.g.,][]{auge10a,dutt12a,conr12a,newm13a,dutt14b,sonn15a}. 

The preference for different IMFs between the X-ray and lensing /
stellar dynamical studies likely indicates an unappreciated large
systematic error in one or both of the approaches. Since, other than
the case of Perseus noted above, it has not been possible to
accurately and precisely measure hot gas motions directly, it is
natural to consider deviations from hydrostatic equilibrium as a
leading candidate for a large systematic error. For \src, however, to
account for the almost factor of 2 difference in the stellar mass
implied by a Chabrier and Salpeter IMF would require a non-thermal gas
pressure comparable to the thermal pressure. Given the evidence
presented at the beginning of this section that \src\ is very old and
relaxed, we believe that such a large non-thermal pressure fraction is
extremely unlikely. Nevertheless, resolving this IMF discrepancy
provides additional motivation for obtaining measurements of hot gas
motions of similar quality to Perseus for massive early-type
galaxies.

\subsection{Entropy Profile and Feedback}
\label{entropy.disc}

In \S \ref{entropy} we found that for $r\ga r_{2500}$ the entropy
profile we obtain for the fiducial hydrostatic model, when rescaled by
$\propto f_{\rm gas}(<r)^{2/3}$, diverges from the baseline
$\sim r^{1.1}$ entropy profile produced by cosmological simulations
with only gravity, so that near $r_{500}$ it nearly equals the
original profile (see \S \ref{entropy}).  This result is generic to
all the DM models we investigated (NFW, Einasto, CORELOG, AC).
Whereas the good matching of the rescaled profile to the baseline
profile interior to $\sim r_{2500}$ suggests feedback spatially
rearranged the gas without heating it (as is found typically), the
fact that the rescaled entropy exterior to $\sim r_{2500}$ exceeds the
baseline profile suggests the feedback raised the temperature of
the gas in that region.

This behavior of the entropy profile is very similar to that of the
nearly fossil system NGC~1521~\citep{hump12b} which has a total mass
very close to \src\ and has gas temperature and density measurements
extending out to $r_{500}$. The entropy profiles of \src\ and NGC~1521
suggest possibly different feedback mechanisms shape the gas interior
and exterior to $\sim r_{2500}$ in those systems. But since other
systems like NGC~720 do not display the same behavior~\citep{hump11a},
the feedback history is not universal.  Measurements of the hot gas
properties extending out to at least $\sim r_{500}$ in more galaxies
are needed to assess the generality of this result and its possible
implications for galaxy formation.

\section{Conclusions}
\label{conc}

In 2010 \suzaku\ observed \src\ for a nominal 46.5~ks providing
motivation for us to revisit its uncertain, but possibly extremely
high, halo concentration inferred previously from a modest 20~ks
\chandra\ observation~(\citealt{khos04a}; H06). \src\ is an optimal
target for analysis of its mass profile with X-ray observations
because it is nearby and fairly bright with a highly regular X-ray
image that displays no evidence of a central AGN disturbance (see \S
\ref{intro}).  Whereas our \chandra\ density and temperature profiles
in H06 only extended to $\sim 0.5r_{2500}$, the \suzaku\ data allow
these profiles to be measured fully out to $r_{2500}$ leading to
tighter constraints on the mass profile. The improved constraints
partially arise from the addition of the \suzaku\ data but are also
the result of our hydrostatic modeling approach having advanced since
H06; i.e., we implement an ``entropy-based'' hydrostatic method that
allows us easily to enforce the additional constraint of convective
stability~\citep[e.g.,][]{hump08a,buot12a}.

The fiducial hydrostatic equilibrium model we employ for our study has
the following components: (1) reference pressure value at $r=1$~kpc;
(2) entropy proxy profile consisting of a constant plus a broken
power-law with two break radii; (3) stellar mass of the BCG
represented by a Sersic model; (4) an NFW profile for the DM halo. The
primary method we use to fit the hydrostatic equilibrium models to the
data is Bayesian nested sampling. In addition, we employ a standard
$\chi^2$ minimization frequentist approach both to compare
best-fitting parameters to the Bayesian values and for hypothesis
testing.

The main results of our analysis are the following.

\begin{itemize}
\item{\bf Extremely High Halo Concentration (\S \ref{mass}, \S
    \ref{highc})} We measure $c_{200}=32.2\pm 7.1$ and
  $M_{200}=(4.5\pm 0.6)\times 10^{12}\, M_{\odot}$ for the fiducial model. For
  a halo of this mass, the value we measure for $c_{200}$ exceeds the
  expected value of $c_{200}=7.1$ for relaxed \lcdm\
  halos~\citep{dutt14a} by $3.5\sigma$ with respect to the statistical
  observational error, and by $6\sigma$ considering the intrinsic
  scatter in the \lcdm\ $c-M$ relation. This measurement situates \src\
  as the most extreme outlier known for a fossil system.  A $6\sigma$
  outlier in the $c-M$ relation would represent an extremely rare
  object, possibly even too rare to form in \lcdm\ simulations when
  taking into account the small cosmological sampling volume implied
  by the low redshift of \src. We believe this possible tension with
  \lcdm\ favors AC models since they yield lower $c_{200}$ values that
  are modestly less significant outliers in the $c-M$ relation.  We
  reach the same conclusions when using Einasto DM profiles
  instead of NFW though with modestly lower significances.

\item{\bf Weak Adiabatic Contraction (\S \ref{ac}, \S \ref{mass}, \S
    \ref{highc}, \S \ref{baryfrac.disc})} We considered four variants
  of AC and applied them to the NFW and Einasto DM profiles. We found
  all the AC models fitted the data equally well in terms of a
  frequentist $\chi^2$ analysis. Generally, the AC models give smaller
  $\mlkband$, smaller $c_{200}$, larger $M_{200}$, and smaller
  $f_{\rm b,200}$ than the un-contracted models.  These parameter
  differences are strongest for the standard AC
  prescriptions~\citep{blum86a,gned04a}.  We argue the X-ray analysis
  favors the {\it ad hoc} ``Forced Quenching'' AC model of
  \citet{dutt15a}, since not only does it have the advantages of
  producing smaller $c_{200}$ and $f_{\rm b,200}$ than the
  un-contracted models, but it also yields the largest $\mlkband$ of
  the AC models (equal to the un-contracted models) which agrees best
  with the stellar mass predicted by the SPS models.  Support for AC
  that is weaker than the standard prescriptions is not inconsistent
  with other X-ray studies~\citep[H06;][]{gast07b} or stellar dynamics
  and lensing measurements~\citep{dutt14b} of the mass profiles of
  early-type galaxies~(see \S \ref{intro}).

\item{\bf Baryonically Closed (\S \ref{baryfrac}, \S \ref{baryfrac.disc})}
  We obtain $f_{\rm b,200}=0.149\pm 0.014$ very consistent with
  $f_{\rm b, U}$ for the fiducial model. When including estimates for
  the stellar baryons not associated with the BCG, $f_{\rm b,200}$
  increases and marginally exceeds $f_{\rm b, U}$ ($\sim
  2.5\sigma$). For AC models the discrepancy is weaker $(\la
  1.5\sigma)$. The solid evidence for $f_{\rm b,200}\approx f_{\rm b,
  }$ in \src\ indicates that at least some of the ``Missing
  Baryons'' at low redshift may be located in the outer regions of hot
  halos in massive elliptical galaxies.

\end{itemize}

Other notable results are as follows.

\begin{itemize}
\item{\bf Surface Brightness and Temperature (\S \ref{spec})} The
  $\xsurf$ and $\ktemp$ profiles obtained from \chandra\ and \suzaku\
  agree well in their overlap region. At the largest radii
  ($\sim\rtwofiveh$) there is weak evidence $(\sim 1\sigma)$ that the
  temperature profile turns around and begins to rise.  Our
  measurements for \chandra\ differ significantly from H06 due to
  improvements in the atomic physics database used by the {\sc vapec}
  plasma code.

\item{\bf Entropy (\S \ref{entropy}, \S \ref{entropy.disc})}
  Interior to $\sim r_{2500}$ the radial profile of the entropy proxy
  displays evidence for feedback that has spatially rearranged the hot
  gas without heating it, very consistent with measurements of other
  galaxies, groups, and clusters. This is not the case exterior to
  $\sim r_{2500}$, where the entropy profile suggests that feedback
  has raised the gas temperature.

\item{\bf Pressure (\S \ref{pressure})} The pressure profile of \src\
  broadly agrees with (considering the intrinsic scatter) the
  universal profile of galaxy clusters~\citep{arna10a} for radii
  approximately 0.1-0.7~$\rfiveh$, but exceeds it elsewhere.

\item{\bf BCG Stellar Mass and the IMF (\S \ref{mass}, \S \ref{he})}
  The stellar mass we infer for the BCG, when allowing for plausible
  uncertainty in the assumed $R_e$, agrees with SPS models having a
  Chabrier or Kroupa IMF and strongly disfavors a Salpeter IMF. This
  result agrees with previous X-ray studies of massive elliptical
  galaxies, groups, and clusters but disagrees with results from many
  lensing and stellar dynamical studies. We argue that it is very
  unlikely the preference for a Chabrier/Kroupa IMF in \src\ arises
  from a strong violation of the hydrostatic equilibrium approximation.

\item{\bf Dark Matter Profiles (\S \ref{mass})} The three DM profiles
  we investigated -- NFW, Einasto, and a pseudo-isothermal logarithmic
  potential with a core (``CORELOG'') -- all fit the data equally well
  in terms of a frequentist $\chi^2$ analysis.

\item{\bf Slope of the Total Mass Profile (\S \ref{slope})} The total
  mass profile interior to $\sim 2R_e$ is very close to a power-law
  with a mass-weighted density slope $\langle\alpha\rangle=2.00\pm
  0.04$. Exterior to this radius the density profile steepens so that
  $\langle\alpha\rangle=2.27\pm 0.09$ for $r=10R_e$, which is $\approx
  12\%$ larger than the mean value of the $\alpha-R_e$ scaling
  relation~\citep{hump10a} but is consistent within the observed
  scatter~\citep{auge10a}.

\item{\bf MOND (\S \ref{mond})} MOND is unable to explain the X-ray
  data without DM; e.g., at $r\approx 35$~kpc the MOND DM fraction is
  $0.68\pm 0.05$ considering only the stellar baryons in the BCG which
  decreases to $\approx 0.59$ when including plausible contributions
  from other stellar baryons.  If we use the upper limit on $R_e$ for
  the BCG, the DM fraction is reduced by an additional 0.09.

\end{itemize}

The properties we have measured for the dark and baryonic mass
profiles of \src, particularly the extremely high halo concentration
implying an ancient system, suggest it is a ``classical'' fossil group
of the kind envisioned when the first fossil was discovered
by~\citet{ponm94a}.

\acknowledgements 

I thank the anonymous referee for a constructively critical review
that led to improvements in the manuscript. I also thank Dr.\ Melania
Nynka for contributing software to assist with the spectral data
analysis. I gratefully acknowledge partial support from the National
Aeronautics and Space Administration (NASA) under Grant NNX15AM97G
issued through the Astrophysics Data Analysis Program.  Partial
support for this work was also provided by NASA through Chandra Award
Number GO4-15117X issued by the Chandra X-ray Observatory Center,
which is operated by the Smithsonian Astrophysical Observatory for and
on behalf of NASA under contract NAS8-03060.  The scientific results
reported in this article are based in part on observations made by the
Chandra X-ray Observatory and by the Suzaku satellite, a collaborative
mission between the space agencies of Japan (JAXA) and the USA
(NASA). This research has made use of the NASA/IPAC Extragalactic
Database (NED) which is operated by the Jet Propulsion Laboratory,
California Institute of Technology, under contract with the National
Aeronautics and Space Administration.  We acknowledge the usage of the
HyperLeda database.

\bibliographystyle{apj}

\end{document}